\documentclass[twocolumn, twocolappendix]{aastex631}

\usepackage[version=4]{mhchem}
\usepackage{enumitem}

\usepackage{float}

\shorttitle{Quantifying the C/O ratio in the planet-forming environments around very low mass stars}
\shortauthors{Díaz-Berríos, Walsh \& van Dishoeck}

\graphicspath{{./}{}}

\begin{document}

\title{Quantifying the C/O Ratio in the Planet-forming Environments around Very Low Mass stars}

\received{February 18, 2025}
\revised{February 5, 2026}
\accepted{February 5, 2026}

%\correspondingauthor{Javiera K. Díaz-Berríos}
%\email{pyjkd@leeds.ac.uk}

\author[0000-0003-0771-5343]{Javiera K. Díaz-Berríos}
\affiliation{School of Physics and Astronomy, University of Leeds, Leeds, LS2 9JT, UK}

\author[0000-0001-6078-786X]{Catherine Walsh}
\affiliation{School of Physics and Astronomy, University of Leeds, Leeds, LS2 9JT, UK}

\author[0000-0001-7591-1907]{Ewine F. van Dishoeck}
\affiliation{Leiden Observatory, Leiden University, P. O. Box 9513, 2300 RA Leiden, The Netherlands}
\affiliation{Max-Planck Institut für Extraterrestrische Physik (MPE), Giessenbachstrasse 1, 85748, Garching, Germany}

%\collaboration{6}{(AAS Journals Data Editors)}

\begin{abstract}
\noindent
The material in planet-forming disks determines the composition of planets; hence, it is crucial to understand the physical and chemical processes that set the abundance and distribution of key volatiles. 
James Webb Space Telescope observations of disks around very low mass ($\sim0.1~\rm{M}_\odot$) stars (VLMSs) have revealed their hydrocarbon-rich inner regions (e.g., \ce{C2H2}), with column densities significantly higher than predicted. 
We employ chemical kinetics models using the physical structure of the inner disk around an M~dwarf star with an X-ray luminosity of $L_\mathrm{X}\sim10^{29}~\mathrm{erg~s^{-1}}$.  
We adopt initial abundances that mimic the effects of carbon enhancement and oxygen depletion (C/O from 0.44 to 87.47) and quantify how the abundances and distributions of key volatiles respond.  
The column density and number of molecules ($\mathcal{N}$) of hydrocarbons and oxygen-bearing species are highly sensitive to the C/O ratio, with the largest increases in hydrocarbons occurring when carbon increases by a factor of 2, and/or oxygen decreases by a factor of 10, relative to solar.  
In the IR-emitting region ($T_\mathrm{gas}>200~\mathrm{K}$), a range of C/O ratios can reproduce the observed $\mathcal{N}$ and ratios relative to \ce{CO2}. 
The disk-integrated molecular ratio with respect to \ce{CO2} is highly sensitive to the underlying C/O ratio. 
However, our results apply only to a source with a single X-ray luminosity value at the middle of that observed for VLMSs; hence, a degeneracy between the stellar $L_\mathrm{X}$ and the C/O ratio cannot be discarded. 
Nonetheless, our findings support that an enhanced C/O is required to drive the hydrocarbon-rich chemistry observed in the inner disks around VLMSs. 
\end{abstract}

\keywords{Astrochemistry --- Protoplanetary disks --- ISM: molecules}

%% We recommend that authors also use the natbib \citep
%% and \citet commands to identify citations.  The citations are
%% tied to the reference list via symbolic KEYs. The KEY corresponds
%% to the KEY in the \bibitem in the reference list below. 

%%%%%%%%%%%%%%%%%%%%%%%%%%%%%%%%%%%%%%%%%%%%%%%%%%%%%%%%%%%%%%%%%%%%%%

\section{Introduction} \label{sec:introduction}

As the most populous stars, M~dwarf stars are likely the most common hosts of exoplanets \citep[e.g.,][]{henry_character_2024}.
Thus, the protoplanetary disks around these very low mass stars (VLMSs) are interesting targets. 
In particular, studying the distribution and abundance of different species reveals the composition of the material available to form planets, and understanding the chemical processes dominating and controlling the detected abundances can illuminate the physics taking place within the planet-forming regions of these sources \citep[e.g.,][]{oberg_protoplanetary_2023}.

The inner disk (less than a few au), characterized by its high densities and temperatures, is where terrestrial planets form \citep{dullemond_inner_2010}.
Important results characterizing the inner disk came from previous works using the Spitzer Space Telescope. 
Studies such as those by \citet{salyk_h2co_2008}, \citet{pontoppidan_rates_2010}, and \citet{carr_organic_2008, carr_organic_2011} reported detections of \ce{OH}, \ce{CO2}, \ce{HCN}, \ce{C2H2}, and \ce{H2O} emission in the innermost disk regions around T~Tauri stars, indicating that the inner disk is a chemically rich environment for planet formation (see \citealt{pontoppidan_volatiles_2014} for review). 
However, these results are not only applicable to T~Tauri stars. 
\citet{pascucci_atomic_2013} reported the detection of \ce{HCN} and \ce{C2H2} (as well as \ce{CO2} and \ce{H2O}) toward disks around M~dwarf and brown dwarf stars, suggesting that the inner regions of disks around VLMSs are particularly rich in hydrocarbon species. 
In summary, observations of the inner disk with Spitzer gave us the first indicators of the rich chemistry in the inner disk and revealed the abundances of key carriers of important elements (O, C, and N) for planet formation. 

Recent James Webb Space Telescope (JWST) observations have confirmed the results suggested by Spitzer, but have also provided new insights about the rich chemical environment that is the inner region of planet-forming disks \citep{kamp_chemical_2023}. 
Water was previously detected in the inner regions of disks by Spitzer and thanks to the higher spectral resolution of JWST, it is now possible to study the water reservoirs in disks in more detail.
These studies have focused on the water enrichment within the water snowline and its relation to the presence of substructures \citep{banzatti_depletion_2017, kalyaan_linkin_2021, banzatti_JWST_2023, kalyaan_effect_2023, gasman_abundant_2023, temmink_characterising_2024}.
Other studies, such as those by \citet{grant_minds_2023} and \citet{vlasbom_mid_2024}, have focused on the structure of the inner disk. 
\citet{grant_minds_2023} presented JWST observations of the disk around a low-mass T~Tauri star, detecting \ce{^12CO2}, \ce{^13CO2}, \ce{H2O}, \ce{HCN}, \ce{C2H2}, and \ce{OH} emission. 
In addition, a high \ce{CO2}/\ce{H2O} ratio was derived, suggesting not only a chemically rich inner disk but also a structured one \citep{grant_minds_2023, vlasbom_mid_2024}.
These studies show how spatially unresolved observations of molecular line emission with JWST can reveal the presence of structure in the inner regions of planet-forming disks.

JWST has also enabled the increased detection of carbon-bearing species in inner-disk regions. 
Previous studies using Spitzer reported the detection of \ce{CO}, \ce{CO2}, and \ce{C2H2} \citep{salyk_h2co_2008,pascucci_atomic_2013, carr_organic_2011}, which are thought to constitute the most abundant carbon carriers. 
JWST has also allowed the first detection of additional, and more complex, hydrocarbon species such as \ce{C4H2}, \ce{C6H6}, \ce{C2H4}, and \ce{C3H4} toward several disks around VLMSs, indicating that the inner-disk region may be carbon-rich \citep{tabone_rich_2023, arabhavi_abundant_2024, kanwar_minds_2024, kaeufer_disentangling_2024}. 

Motivated by the Spitzer results, \cite{walsh_molecular_2015} investigated the chemical composition of the inner planet-forming regions of disks for a range of stellar spectral types: M~dwarf, T~Tauri, and Herbig~Ae. 
They concluded that the inner-disk region is influenced by the relative UV and X-ray luminosities of the stars, i.e., the photochemistry and X-ray chemistry play an important role in setting the chemical composition. 
\cite{walsh_molecular_2015} suggested that a higher abundance of hydrocarbons is expected in disks around low-mass stars since the inner regions of disks around cooler stars (i.e., M~dwarf stars) are more carbon-rich owing to the release of carbon from CO via X-ray-induced processes. 
However, these models did not predict the large column densities of all the hydrocarbons reported in very recent work by \citet{tabone_rich_2023},  \cite{arabhavi_abundant_2024}, \citet{kanwar_minds_2024}, and \citet{kaeufer_disentangling_2024}, with one reason being that those models assumed the solar abundance of C and O, i.e., oxygen-rich conditions.
\citet{kanwar_minds_2024} and \citet{colmenares_jwst_2024} support this idea, suggesting that a higher-than-solar C/O ratio could explain the high abundance of hydrocarbons in the disks around the low-mass star Sz28 and the low accretion rate ($\sim 10^{-10}~\mathrm{M_\odot~yr^{-1}}$) T~Tauri star DoAr~33  \citep{cieza_nature_2010}, respectively.

In a recent study, \citet{raul_tracking_2025} found that the abundance of hydrocarbons has a strong dependence on the mechanisms that lead to an X-ray- or UV-driven chemistry, suggesting that X-ray-dominated chemistry can result in a higher production of \ce{C2H2}, CO, HCN, and long carbon chains (LCCs), similar to the conclusions reached by \citet{walsh_molecular_2015}. 
\citet{raul_tracking_2025} also suggest that hydrocarbons such as \ce{C2H2} are more abundant in a water-rich disk, while species such as \ce{CH4} and \ce{CH3} reach higher abundances in a disk where the water is depleted.
Other studies, such as \citet{sellek_chemical_2025}, have found that the destruction of gas-phase CO can deplete oxygen while producing hydrocarbons to explain observations, indicating a high C/O ratio.
\citet{kanwar_strong_2025} modeled the disk structure of J160532, testing the effect of variation in the C/O ratio, and found that the presence of a gap better explains the observations.
\citet{houge_burned_2025} studied the effect of the variation in C/H and C/O ratios, finding that the thermal decomposition process of refractory organics allows the carbon-rich gas to survive throughout the disk and over time.
Given that each of the models discussed so far uses a distinct chemical network, cross-checks of the main results with an independent chemical network are warranted.
More recently, the survey of 10 VLMS disks by \citet{arabhavi_very_2025} showed that high columns of hydrocarbon molecules are common. 
In addition, \citet{grant_transition_2025} found an anticorrelation between the flux ratio of \ce{C2H2}/\ce{H2O} and the stellar luminosity, with disks around VLMSs showing a spectra dominated by \ce{C2H2} emission.

In light of the hydrocarbon-rich inner disks around VLMSs revealed by JWST, we revisit the M~dwarf models of \cite{walsh_molecular_2015}  to quantify the changes in chemical structure and abundances expected for cases where we have an increased relative abundance of carbon to oxygen.
We compare the predicted abundances, column densities, total number of molecules, and molecular ratios of the key species observed in the inner regions of protoplanetary disks, paying special attention to those reported by \citet{tabone_rich_2023}, \citet{arabhavi_abundant_2024}, \cite{kanwar_minds_2024}, and \citet{arabhavi_minds_2025}. 
We investigate possible explanations for the high column densities of hydrocarbons detected around VLMSs reported in the listed works.
We focus solely on the response of the chemistry and trends in composition to variations in C/O ratio for a generic model of a disk around an M~dwarf star, i.e., we do not aim to reproduce absolutely the observations toward any specific source.
Note that in this work we test a wide range of C/O ratio values, investigating carbon enrichment, oxygen depletion, and the combination of both.

Section~\ref{sec:model} describes the chemical model used, including the chemical network. 
Section~\ref{sec:results} presents the results from the different scenarios investigated. 
In Section~\ref{sec:discussion} we analyze the results, compare the different cases, and identify trends. 
We also compare these trends with the results from \citet{tabone_rich_2023}, \citet{arabhavi_abundant_2024}, and \citet{kanwar_minds_2024} and present possible explanations for the observed trends. 
Finally, Section~\ref{sec:conclusion} presents a summary of the key points and the conclusions of this work.

%%%%%%%%%%%%%%%%%%%%%%%%%%%%%%%%%%%%%%%%%%%%%%%%%%%%%%%%%%%%%%%%%%%%%%

\section{Disk Model} \label{sec:model}
In this section, we describe the physical model and chemical network used. 
We also include details about the parameters adopted in the different models used to motivate the initial abundances that mimic the proposed physical mechanisms responsible for the high abundance of hydrocarbons in the inner disks around VLMSs.

%%%%%%%%%%%%%%%%%%%%%%%%%%%%%%%%%%%%%%%%%%%%%%%%%%%%%%%%%%%%%%%%%%%%%%

\subsection{Physical Model}
We use the physical structure for a disk of mass 0.6~$M_\mathrm{Jup}$ and radius 10~au around an M~dwarf star from \citet{walsh_molecular_2015}, which is calculated using the methods from \citet{nomura_molecular_2005} with the addition of X-ray heating as outlined in \citet{nomura_molecular_2007}. 
Here we outline the most important parameters of the system only, and we refer to \citet{walsh_molecular_2015} for further details on the disk physical model.
The star-disk system properties can be found in Table~\ref{table:model-properties}. 
The radiation field of the star (with mass and radius of 0.1~M$_\odot$ and 0.7~R$_\odot$, respectively) is simulated as a blackbody at the stellar effective temperature (3000~K).
UV excess emission is included, which has two components: a diluted blackbody spectrum to simulate bremsstrahlung emission ($T_\mathrm{br} \sim 25,000~\mathrm{K}$) and $\mathrm{Ly}\alpha$ line emission (see Fig.~1 in \citealt{walsh_molecular_2015}).
We assume that the X-ray luminosity of the star is $L_\mathrm{X}\sim 10^{29}~\mathrm{erg~s^{-1}}$, which is at the middle of the observed values for VLMSs \citep[e.g.,][]{preibisch_evolution_2005}. 
The far-UV flux through the disk has contributions from both the star and interstellar medium and is calculated over the range $6~\mathrm{eV} < h\nu < 13~\mathrm{eV}$ \citep{nomura_molecular_2005}.
The dust grain size distribution adopted is that which reproduces the extinction curve observed in dense clouds \citep{weingartner_dust_2001}.
Additionally, we assume that the gas and small dust grains are fully mixed.
The gas temperature is calculated assuming thermal balance between the heating and cooling of the gas.
The initial abundances assumed to compute the gas temperature for carbon and oxygen were $7.86\times10^{-5}$ and $1.8\times 10^{-4}$, respectively, which are the same as those used by \citet{nomura_molecular_2005} and \citet{walsh_molecular_2015}.

\begin{figure*}[ht]
\centering
\includegraphics[width=1\linewidth]{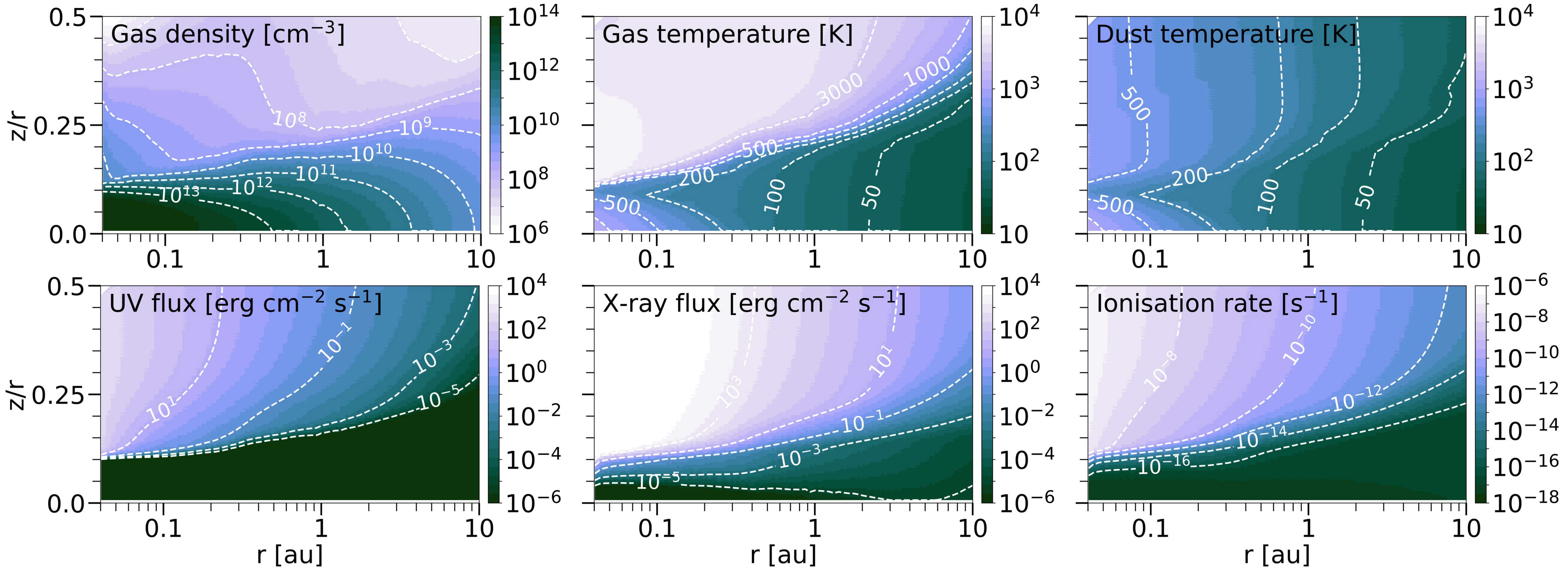}
\caption{Physical structure of the M~dwarf disk model used in this work, which is the same as that used in \cite{walsh_molecular_2015}. Presented are the gas density, gas temperature, and dust temperature (top row) and the UV flux, X-ray flux, and total ionization rate (bottom row). 
Each panel shows the dependence of each parameter as a function of disk radius (up to $10~\mathrm{au}$) and disk height (scaled by the radius; $z/r$). 
White contours on each panel represent the same information as the color bar. 
For ease of comparison, the gas and dust temperatures, as well as the UV and X-ray fluxes, share the same color scales.}
\label{fig:disk-structure}
\end{figure*}

Figure~\ref{fig:disk-structure} shows the gas density (cm$^{-3}$), gas and dust temperature (K), UV and X-ray fluxes (erg~cm$^{-2}$~s$^{-1}$), and total ionization rate (s$^{-1}$) for the M~dwarf disk model. 
The ionization rate includes contributions from X-rays and cosmic rays. 
The cosmic-ray ionization rate adopted is $1 \times 10^{-17}~\mathrm{s}^{-1}$.
The model follows the common protoplanetary physical structure; the gas temperature and density show a radial and vertical gradient with cooler temperatures near the midplane and in the outer disk ($\sim 50-500~\mathrm{K}$), and higher densities near the midplane and close to the star ($\sim 10^{9}-10^{13}~\mathrm{cm}^{-3}$).
Note that the disk structure used in our model includes accretion heating, which produces a temperature inversion in the midplane (for further discussion see Appendix~\ref{appendix:caveats}).
The dust temperature follows a similar gradient to the gas temperature but with lower temperatures in the disk atmosphere due to inefficient cooling of the gas in this region.
The UV and X-ray fluxes are set by the surface density of the disk and the relative penetration strengths of UV and X-ray photons.  
Note that the more energetic X-rays are able to penetrate to greater depths toward the disk midplane such that the X-ray flux is larger than that for UV photons at $z/r \sim 0.1 - 0.2$ and is an important driver of chemistry in this region.

\begin{deluxetable}{@{\extracolsep{0pt}}lcc}
\tablecaption{Physical model parameters}
\tablehead{
    \colhead{Parameter} &  \colhead{Description}  & \colhead{Value} 
    }
    \startdata
    \hline  
     $M_\star$ & Stellar mass & $0.1~\mathrm{M}_\odot$ \\
     $R_\star$ & Stellar radius & $0.7~\mathrm{R}_\odot$ \\
     $T_\star$ & Effective temperature & $3000~\mathrm{K}$ \\
     $\dot{M}$ & Accretion mass rate & $10^{-9}~\mathrm{M}_\odot~\mathrm{yr^{-1}}$ \\
     $\Sigma_{0.04\mathrm{au}}$ & Surface density at 0.04~au & $1.9\times 10^{25}~\mathrm{cm}^{-2}$ \\ 
     $\Sigma_{10\mathrm{au}}$ & Surface density at 10~au & $3.0\times 10^{23}~\mathrm{cm}^{-2}$ \\
     $M_\mathrm{disk,10~au}$ & Disk mass out to 10~au & $0.6~{M_\mathrm{Jup}}$ \\
    \hline
    \enddata
\label{table:model-properties}
\tablecomments{Star-disk system physical properties \citep{walsh_molecular_2015}.}
\end{deluxetable}

%%%%%%%%%%%%%%%%%%%%%%%%%%%%%%%%%%%%%%%%%%%%%%%%%%%%%%%%%%%%%%%%%%%%%%

\subsection{Chemical Network}
The disk model uses the \texttt{RATE12} version of the UMIST Database for Astrochemistry (UDfA) chemical network, which includes 6173 gas-phase reactions involving 467 species \citep{mcelroy_umist_2013}. 
The gas-phase network includes two-body reactions, direct cosmic-ray ionization, cosmic-ray-induced photodissociation and ionization, photodissociation and photoionization, direct X-ray ionization, and X-ray-induced ionization and dissociation. 
This network is supplemented with additional dissociative recombination reactions and photoreactions, collisional dissociation reactions, and three-body reactions necessary because of the higher densities and temperatures in the inner disk. 
Also included are reactions with excited \ce{H2} which are important for setting the composition in the hot atmosphere of the inner disk \citep{bruderer_survival_2013}.
Gas-grain chemistry is also included, allowing freeze-out to form ices and thermal and nonthermal desorption for release back into the gas phase. 
A grain-surface chemical network is included with all associated reactions taken from the \texttt{OSU2008} database \citep{garrod_osu_2008}. 
Additional glyoxal (\ce{(HCO)2}) reactions are adopted from \cite{woods_glycolaldehyde_2013}. 
The Langmuir-Hinshelwood mechanism is assumed to set the grain-surface reaction rates.
Table \ref{table:binding-energies} presents the binding energies that we adopt for species of interest in this study.

\begin{deluxetable}{@{\extracolsep{0pt}}lclc}
\tablecaption{Binding energies for key species}
\tablehead{
    \colhead{Species} &  \colhead{${E_\mathrm{bind}}$ [K]} & \colhead{Species} &  \colhead{${E_\mathrm{bind}}$ [K]}
    }
    \startdata
    \hline  
    \ce{C2H2} & 2090 & \ce{OH}   & 3210 \\
    \ce{CH4}  & 1252 & \ce{CO}   & 855 \\
    \ce{C6H6} & 7587 & \ce{CO2}  & 2267 \\
    \ce{CH3}  & 1040 & \ce{H2O}  & 4880 \\
    \ce{C4H2} & 4187 & \ce{HCN}  & 3610 \\
    \ce{C2H4} & 2010 & \ce{NH3}  & 2715 \\
    \ce{C3H4} & 4287 & \ce{HC3N} & 4580 \\
    \ce{C2H6} & 2320 & \ce{N2}   & 790 \\
    \hline
    \enddata
\label{table:binding-energies}
\tablecomments{Binding energies used here are mostly from the compilations by \citet{penteado_sensivity_2017} and \citet{mcelroy_umist_2013}. We adopt the values for CO, \ce{N2}, and HCN from the experimental work by \citet{oberg_competition_2005} and \citet{noble_thermal_2013}.}
\end{deluxetable}

%%%%%%%%%%%%%%%%%%%%%%%%%%%%%%%%%%%%%%%%%%%%%%%%%%%%%%%%%%%%%%%%%%%%%%

\subsection{Initial Abundances}
We use the same initial abundances adopted in \citet{walsh_molecular_2015} as our fiducial model (M0.44F). 
This model has a C/O ratio of $\sim0.44$, and C/H and O/H ratios of $1.4\times10^{-4}$ \citep{cardelli_abundance_1996} and $3.2\times10^{-4}$ \citep{meyer_definitive_1998}, respectively (see Table~\ref{table:model-versions}).
We also assume that He/H and N/H  have values of $9.75\times10^{-2}$ and $7.5\times10^{-5}$, respectively \citep{meyer_solar_1985, cardelli_interstellar_1991}.
To generate the suite of initial abundances for the disk model, we run a dark cloud model with $T_\mathrm{gas} = T_\mathrm{dust} = 10~\mathrm{K}$, $n_\mathrm{gas} = 10^4~\mathrm{cm^{-3}}$, and $A_\mathrm{v} = 10~\mathrm{mag}$, and adopt the abundances at a time of $3.2\times10^5$~yrs. 
The initial atomic abundances used for the dark cloud model are the low-metallicity (i.e., depleted) values from \citet{graedel_kinetic_1982} for a diffuse cloud.
Thus, we begin our disk chemical model with molecular material and interstellar abundances of ice species, thus mimicking the scenario where the disk has inherited interstellar material.
Whilst the chemistry through much of the inner disk is expected to be at or close to steady state, this is not true for the colder midplane, where ice formation can be efficient \citep{eistrup_setting_2016}.

To investigate different scenarios for the observed high abundances of hydrocarbons in the inner regions of VLMSs, we then run variations of the model in which we modify our initial abundances by increasing the carbon abundance (C/H) and/or decreasing the oxygen abundance (O/H).  
This is done to mimic elemental redistribution that might take place within the disk following the dark cloud stage. 
To increase C/H, we simply add additional carbon to the initial atomic carbon abundance, increasing that adopted in the fiducial model. 
On the other hand, when decreasing oxygen, we homogeneously deplete all oxygen-bearing species by the stated factor. 
Because this also depletes any species with both carbon and oxygen, we set the final C/O ratio by adding any depleted carbon to the initial atomic carbon abundance.
This leads to the final C/H, O/H, and thus C/O ratios shown in Table~\ref{table:model-versions}. 
Note that rounding errors are responsible for small variations in C/H and O/H across models.

The scenarios that we explore are the following and are based on those proposed by \citet{tabone_rich_2023}, \citet{van_dishoeck_diverse_2023}, and \citet{arabhavi_abundant_2024}.
\begin{enumerate}[label=(\roman*)]
    \item {\em{Carbon grain destruction releasing carbon into the gas phase.}} We mimic this by increasing the elemental carbon abundance everywhere in the disk by a factor of 2.
    \item {\em{Oxygen depletion due to icy pebble trapping in the outer disk.}} To mimic this, we retain the fiducial carbon elemental abundance and decrease the oxygen abundance everywhere in the disk by factors of 10 and 100.
    \item {\em{Carbon enrichment and oxygen depletion.}} To simulate the combined case, we increase the carbon elemental abundance by a factor of 2  and decrease the oxygen abundance everywhere by factors of 10 and 100. 
\end{enumerate}
In the first case, the choice of the factor of 2 was based on \cite{wei_effect_2019} and \cite{asplund_chemical_2021}. 
In \cite{wei_effect_2019} they simulated the effects of carbon grain destruction in disks by increasing the elemental abundance of carbon from $7.30\times10^{-5}$ \citep[the abundance in diffuse clouds;][]{woodall_umist_2007} to $2.95\times 10^{-4}$ which is the solar abundance of carbon \citep{asplund_chemical_2009}. 
As we have adopted more recent estimates for the carbon abundance as measured in the diffuse interstellar medium, this translates to increasing the carbon abundance by a factor of $\sim 2$ for our model when using the updated solar C/H reported by \citet{asplund_chemical_2021}, $2.88\times 10^{-4}$. 
This is considered an extreme case that assumes that all the carbon in the grains has been destroyed and released into the gas phase. 
The oxygen depletion factor in the disk has a broader range of values motivated by the depletion factors proposed for water and \ce{CO} in the outer disks of nearby stars \citep[e.g.,][]{schwarz_radial_2016,du_survey_2017,zhang_molecules_2021}.
See Section~\ref{sec:physical-scenarios} for further discussion on physical scenarios that can (re)set the C/O ratio in disks.
We adopt two levels of oxygen depletion: $10$ and $100$. 
Table~\ref{table:model-versions} shows the different models we investigated with the model names used in this work.
Note that hereafter we also refer to the models as carbon-poor/rich or oxygen-poor/rich. 
This criterion is based on the C/O ratio being higher than or lower than 1 (i.e., carbon-rich means C/O $>1$).
We extract and analyze the abundances at a time of $10^{6}$~yr.

\begin{deluxetable*}{@{\extracolsep{25pt}}lcccc}[ht]
\tablecaption{Initial abundances and models}
\tablehead{
    \colhead{Model} &  \colhead{Description}  & \colhead{C/O ratio} & \colhead{C/H} & \colhead{O/H}  
    }
    \startdata
    \hline  
     M0.44F & Fiducial model\tablenotemark{a}                                           & 0.44  & $1.4\times 10^{-4}$   & $3.2\times 10^{-4}$ \\
     M0.88C & $\uparrow ~\mathrm{C} \times 2$                                           & 0.87  & $2.8\times 10^{-4}$   & $3.2\times 10^{-4}$ \\
     M4.4   & $\downarrow ~\mathrm{O} \times 10$                                        & 4.37  & $1.5\times 10^{-4}$   & $3.3\times 10^{-5}$ \\
     M8.8C & $\uparrow ~\mathrm{C} \times 2$ ~~ $\downarrow ~\mathrm{O} \times 10$     & 8.75  & $2.9\times 10^{-4}$   & $3.3\times 10^{-5}$ \\
     M44  & $\downarrow ~\mathrm{O} \times 100$                                       & 43.72  & $1.5\times 10^{-4}$   & $3.3\times 10^{-6}$ \\
     M88C  & $\uparrow ~\mathrm{C} \times 2$ ~~ $\downarrow ~\mathrm{O} \times 100$    & 87.47 & $2.9\times 10^{-4}$   & $3.3\times 10^{-6}$ \\
    \hline
    \enddata
\label{table:model-versions}
\tablenotetext{a}{Same initial abundances used in \cite{walsh_molecular_2015}.}
\tablecomments{Initial abundances for the different models. 
The up and down arrows represent enrichment and depletion, respectively. 
All of the initial abundances were made with respect to the fiducial model M0.44F.}
\end{deluxetable*}

%%%%%%%%%%%%%%%%%%%%%%%%%%%%%%%%%%%%%%%%%%%%%%%%%%%%%%%%%%%%%%%%%%%%%%

\section{Results} \label{sec:results}

\subsection{Abundance maps and column density profiles}
\begin{figure*}[ht]
\centering
\includegraphics[width=1\linewidth]{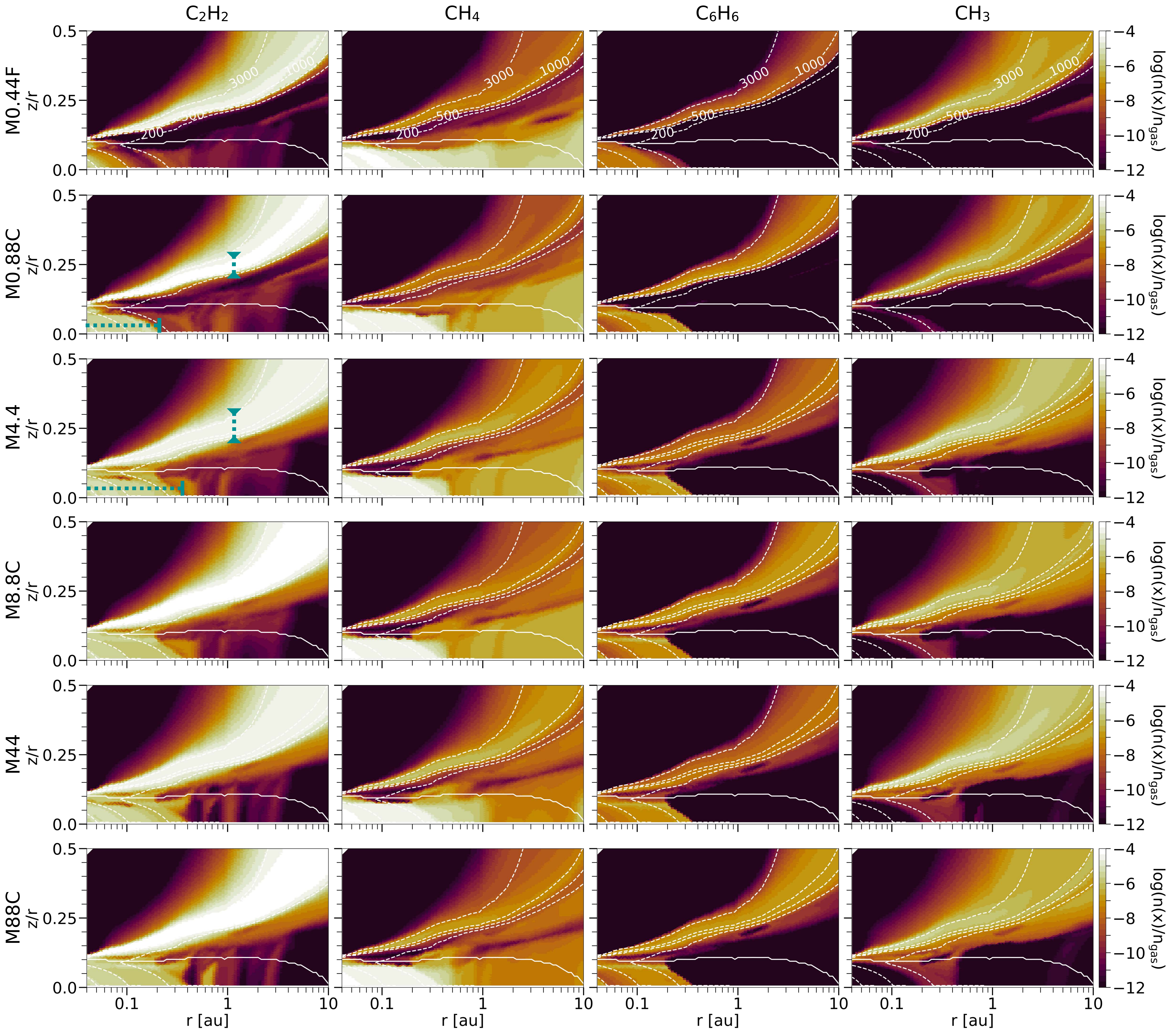}
\caption{Fractional abundance with respect to the gas number density $n_\mathrm{gas}$ of the carbon-bearing species (hydrocarbons), \ce{C2H2}, \ce{CH4}, \ce{C6H6}, and \ce{CH3} (left to right) as a function of disk radius $r$ and disk height scaled by the radius $z/r$. 
{First row:} fiducial model (M0.44F), following the same initial conditions of \cite{walsh_molecular_2015}. The C/O ratio in this case is $0.44$.
{Second row:} model M0.88C; increasing initial carbon abundance by a factor of $2$ ($\mathrm{C/O} = 0.87$), this is the extreme case of carbon enrichment.
{Third row:} model M4.4; decreasing the initial oxygen abundance by a factor of $10$ ($\mathrm{C/O} = 4.37$).
{Fourth row:} model M8.8C; the combination of oxygen depletion by a factor of $10$ and carbon enrichment by a factor of $2$, resulting in a C/O ratio of $8.75$.
{Fifth row:} model M44; strong oxygen depletion by a factor of $100$ resulting in a $\mathrm{C/O} = 43.72$.
{Sixth row:} model M88C; oxygen depletion by a factor of $100$ and a carbon enrichment by a factor of $2$ resulting in a $\mathrm{C/O}=87.47$.
Light-green arrows in two panels illustrate the variations in vertical and radial extent with varying C/O ratio.}
\label{fig:FAmap_hydrocarbon_set1}
\end{figure*}

Figures~\ref{fig:FAmap_hydrocarbon_set1}, \ref{fig:FAmap_hydrocarbon_set2}, \ref{fig:FAmap_oxygen}, and \ref{fig:FAmap_nitrogen} present the fractional abundance with respect to the gas number density, $n_\mathrm{gas}$, $(n(X)/n_\mathrm{gas})$, as a function of disk radius, $r$, and height/radius, $z/r$, for the different carbon and oxygen initial abundances (see Table \ref{table:model-versions}). 
We present vertical distributions in Appendix~\ref{appendix:vertical_distribution}. 
Figures \ref{fig:TA_Tgas_ionizationRate_0.1au_hydrocarbons}, \ref{fig:TA_Tgas_ionizationRate_0.1au_nitrogen_oxygen},  \ref{fig:TA_Tgas_ionizationRate_1.0au_hydrocarbons}, and \ref{fig:TA_Tgas_ionizationRate_1.0au_nitrogen_oxygen} present the ionization rate, gas temperature, and number density of the same set of molecules as a function of height at radii of 0.1 and 1~au.

\begin{figure*}[ht]
\centering
\includegraphics[width=1\linewidth, page=1]{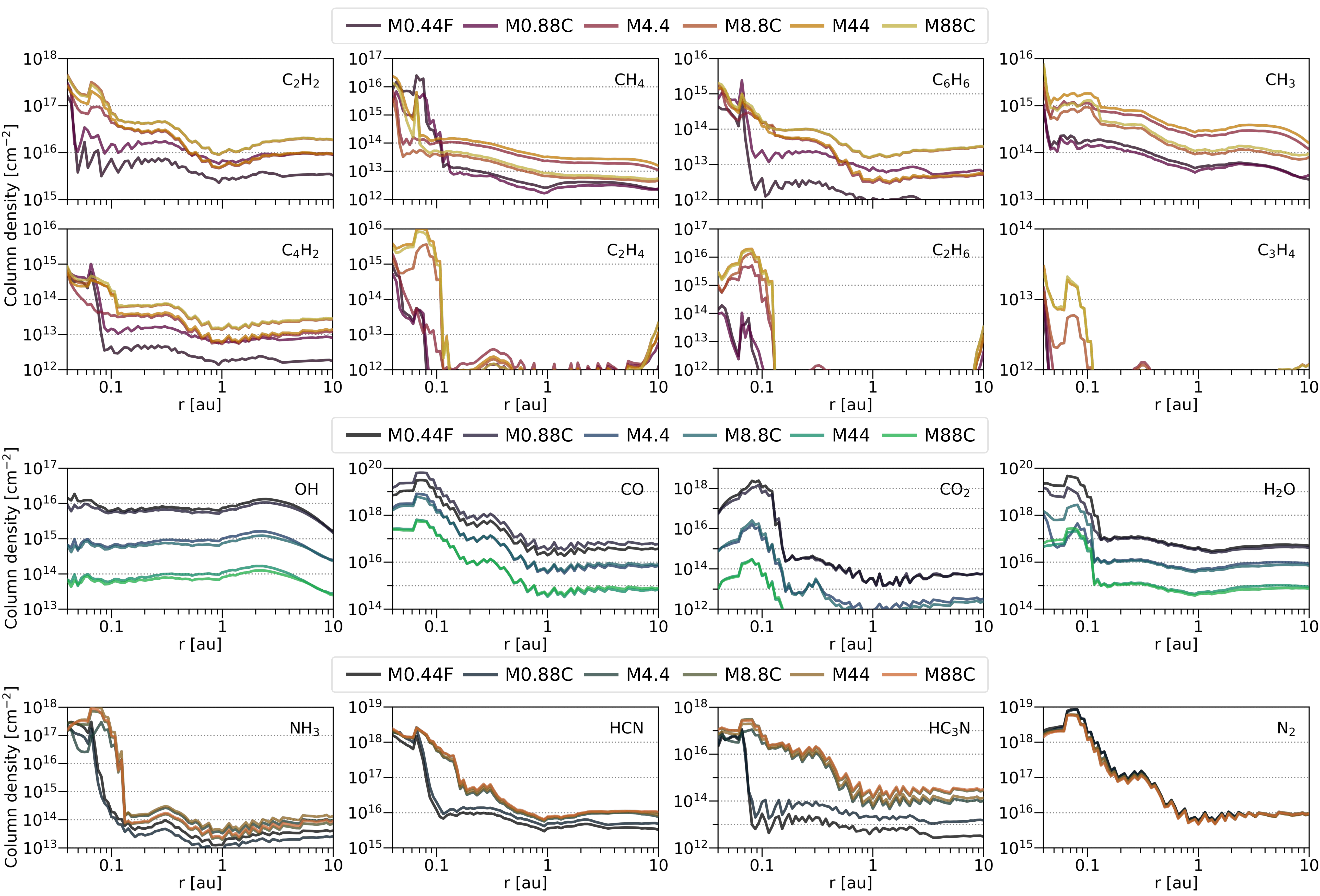}
\caption{Column density as a function of disk radius, $r$, for each species integrated vertically from the disk surface to the boundary defining the IR-emitting region (either down to $\tau=1$ surface at $14~\mu\mathrm{m}$ or to where the gas temperature reaches 200~K).
Equivalent figures showing the column density integrated vertically from the disk surface to the $\tau=1$ surface at $14~\mu\mathrm{m}$ (atmosphere component) and integrated vertically down to the midplane (midplane component) are available online. 
The line colors from dark to light have increasing values of C/O ratio.
(The complete figure set (2 images) is available in the online article.)
}
\label{fig:ColdensProfile_IRemitting}
\end{figure*}

\begin{figure*}[ht]
\centering
\includegraphics[width=1\linewidth]{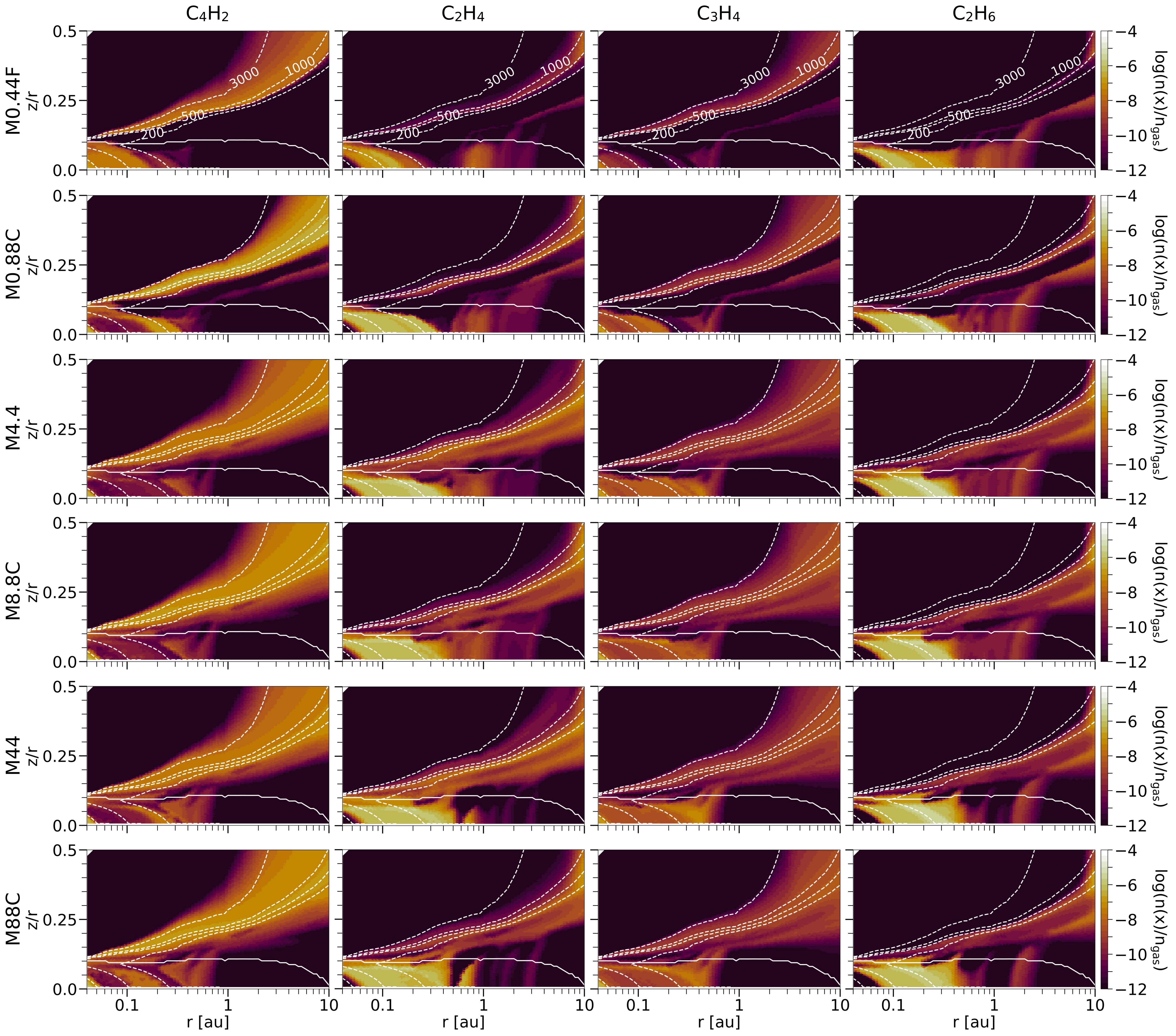}
\caption{Same as Fig.~\ref{fig:FAmap_hydrocarbon_set1}, but for \ce{C4H2}, \ce{C2H4}, \ce{C3H4} (propyne), and \ce{C2H6}.
}
\label{fig:FAmap_hydrocarbon_set2}
\end{figure*}

We focus here on a set of hydrocarbons (\ce{C2H2}, \ce{CH4}, \ce{C6H6}, \ce{CH3}, \ce{C4H2}, \ce{C2H4}, \ce{C3H4}, and \ce{C2H6}), oxygen- (\ce{OH}, \ce{CO}, \ce{CO2}, \ce{H2O}) and nitrogen-bearing species (\ce{HCN}, \ce{NH3}, \ce{HC3N}, and  \ce{N2}).
These species either have been previously detected in the inner region of protoplanetary disks \citep{najita_gas_2003, salyk_h2co_2008, pontoppidan_rates_2010, carr_organic_2008, carr_organic_2011, pascucci_atomic_2013, tabone_rich_2023,banzatti_JWST_2023,arabhavi_abundant_2024,kanwar_minds_2024,kaeufer_disentangling_2024,colmenares_jwst_2024} or are considered to be key carriers of carbon, oxygen, and nitrogen \citep{walsh_molecular_2015}. 
Figure~\ref{fig:ColdensProfile_IRemitting} shows the column densities as a function of disk radius, $r$, for the same set of species, for all six models. 
We show the column density integrated down to either the $\tau=1$ surface at 14~$\mu$m or where the gas temperature reaches 200~K.
This is what we define as the IR-emitting region, where the temperature boundary is consistent with the lowest excitation temperature derived for molecules from observations with JWST \citep[e.g., ][]{tabone_rich_2023, arabhavi_abundant_2024, kanwar_minds_2024, arabhavi_minds_2025, arulanantham_jdisc_2025}.
The adoption of this temperature boundary may slightly underestimate the number of molecules in the IR-emitting region for the case in which the reported excitation temperatures are an average over a distribution. We discuss the implications of this assumption in Section~\ref{sec:moredetailedcomparison}. 
For interest, versions of the same figure integrated down to the $\tau=1$ surface at $14~\mu\mathrm{m}$ and to the midplane are available online.

%%%%% Figure set for Figure 3 -- Begin
%\figsetstart
%\figsetnum{3}
%\figsettitle{Column density profiles for the region above the $\tau=1$ surface at $14~\mu\mathrm{m}$ (disk atmosphere component) and full vertical disk extent down to the midplane (midplane component).}

%\figsetgrpstart
%\figsetgrpnum{3.1}
%\figsetgrptitle{Region above the $\tau=1$ surface at $14~\mu\mathrm{m}$}
%\figsetplot{ColumnDensityProfiles_atmosphere14um.pdf}
%\figsetgrpnote{
%Same as Fig.~\ref{fig:ColdensProfile_IRemitting}, but for the integration down to the $\tau=1$ surface at $14~\mu\mathrm{m}$ (disk atmosphere component).}
%\figsetgrpend

%\figsetgrpstart
%\figsetgrpnum{3.2}
%\figsetgrptitle{Region above the midplane ($z=0$)}
%\figsetplot{ColumnDensityProfiles_midplane.pdf}
%\figsetgrpnote{
%Same as Fig.~\ref{fig:ColdensProfile_IRemitting}, but for the integration down to the midplane (midplane component).}
%\figsetgrpend

%\figsetend

%%%%% Figure set for Figure 3 -- End

Different species have different distributions through the disk, and their abundances can vary radially and vertically, which can affect the radial column density distributions. 
However, there is a general abundance map pattern for most species with two primary reservoirs: one reservoir in the disk atmosphere at elevated $z/r \gtrsim 0.1$ and generally elevated temperature $\gtrsim 300$~K, and a second in the inner-disk midplane $z/r \lesssim 0.1$. 
The specific region where the species resides will depend on what physical properties determine the appropriate conditions for its formation.
The radial extent of this midplane reservoir is dependent on the volatility of the molecule (more volatile species will generally have a greater radial extent) and whether or not efficient gas-phase synthesis is possible in the shielded midplane region. 
Exceptions to this pattern are the small radicals \ce{CH3} and \ce{OH}, which are only abundant in the disk atmosphere (see Figs.~\ref{fig:FAmap_hydrocarbon_set1} and \ref{fig:FAmap_oxygen}), and the hypervolatile species CO and \ce{N2}, which have a wider distribution through the disk, as they are depleted onto dust grains at low temperatures only ($\lesssim$~20~K; see Figs.~\ref{fig:disk-structure}, \ref{fig:FAmap_hydrocarbon_set1}, and \ref{fig:FAmap_nitrogen}).  
As another volatile species, \ce{CH4} also has a more widespread distribution through the disk than most other hydrocarbons, e.g., \ce{C2H2} (see~Fig.~\ref{fig:FAmap_hydrocarbon_set1}). 
\ce{CO2} also has a slightly different distribution in that its molecular layer in the disk atmosphere lies at slightly lower temperatures ($\gtrsim 50$~K) than most other species considered here (see~Fig.~\ref{fig:FAmap_oxygen}).
These general abundance patterns are reflected in the column density profiles and vertical distributions (see Appendix~\ref{appendix:vertical_distribution}).  

In the next subsections, we discuss how the abundance and distribution of each member of each family of species considered here (hydrocarbons, O-bearing, and N-bearing) respond to the imposed variations in the global C/O ratio. 
A more detailed description for each hydrocarbon species is available in Appendix~\ref{appendix:hydrocarbons_CxHy}, with the exception of \ce{CH4} and \ce{CH3} which have different distributions than the other hydrocarbons.

%%%%%%%%%%%%%%%%%%%%%%%%%%%%%%%%%%%%%%%%%%%%%%%%%%%%%%%%%%%%%%%%%%%%%%

\subsubsection{Hydrocarbons}

We find that most hydrocarbons are abundant in the atmosphere between 300 and 4000~K.
The upper boundary in the distribution of hydrocarbons (around 4000~K) is mainly determined by the gas temperature (see Figs.~\ref{fig:disk-structure}, \ref{fig:FAmap_hydrocarbon_set1}, and \ref{fig:FAmap_hydrocarbon_set2}) and is controlled by neutral-neutral and ion-molecule reactions \citep[e.g.,][]{walsh_molecular_2015, kanwar_hydrocarbon_2024}.
This is true for both carbon-poor and carbon-rich scenarios.
On the other hand, the lower boundary (around 300~K) is determined by different physical properties depending on the C/O ratio value.
This boundary moves to lower heights in the disk for some species and models; however, in most of those cases, the abundance reached toward the boundary is at least one order of magnitude lower than that in the peak abundance region.
For species such as \ce{C6H6} and \ce{C4H2} in the fiducial model, the lower boundary is shaped by gas temperatures higher than $200$~K ($\sim 500$~K). 
The upper boundary of the peak abundance region in the disk atmosphere for all the species mentioned is shaped by the gas temperature, X-ray flux, UV flux, and ionization rate depending on the radius and height.
For \ce{C2H4} and \ce{C2H6}, gas temperatures lower than $3000$~K ($\sim 1000$) shape the upper boundary.
For C/O~$<1$, the lower boundary of the peak abundance of \ce{C2H2}, \ce{CH3}, \ce{C4H2}, \ce{C2H4}, and \ce{C2H6} is shaped by the gas temperature (200~K) and ionization rate ($\zeta_\mathrm{XR+CR}\approx 10^{-12}~\mathrm{s^{-1}}$).
For \ce{C6H6} and \ce{C3H4}, only the gas temperature appears to determine the same region ($\sim500$~K).
The abundance of \ce{CH4} is shaped by the gas temperature in the region where this ranges between 500 and 4000~K, the X-ray flux in the region where $z/r<0.25$ ($10^{-3}$ to $10^{-4}~\mathrm{erg~cm^{-2}~s^{-1}}$), the UV flux ($10^{-5}~\mathrm{erg~cm^{-2}~s^{-1}}$), and ionization rates (for values $<10^{-12}~\mathrm{s^{-1}}$ within $\sim4$~au).
For the fiducial model, the fractional abundance map for some species (\ce{C2H2}, \ce{CH4}, \ce{CH3}, \ce{C2H4}, \ce{C3H4}, and \ce{C2H6}) shows a depletion layer below 500~K, which in all cases is then `filled in' as C/O increases. 
The lower boundary of this layer seems to be determined by the X-ray ($10^{-1}~\mathrm{erg~cm^{-2}~s^{-1}}$) and UV flux ($10^{-5}~\mathrm{erg~cm^{-2}~s^{-1}}$) within $\sim4$~au and by the gas temperature $(\sim 50$~K) beyond that radius.
For C/O~$>1$, the lower boundary of \ce{C2H2}, \ce{CH3}, \ce{C4H2}, \ce{C2H4}, and \ce{C2H6} is determined by the gas temperature (a lower temperature than those for C/O~$<1$, $\sim40-50$~K), the X-ray and UV fluxes ($10^{-1}$ and $10^{-5}~\mathrm{erg~cm^{-2}~s^{-1}}$, respectively), and the ionization rate ($\sim10^{-16}$ to $10^{-14}~\mathrm{s^{-1}}$), with species going into ices.
For \ce{C6H6} and \ce{CH4}, the gas temperature contours do not match with either of the fractional abundance maps within 3~au, which suggests that only the X-ray flux, UV flux, and the ionization rates shape its abundance in that region, with the same values as those for \ce{C2H2}, \ce{CH3}, \ce{C4H2}, \ce{C2H4}, and \ce{C2H6}.
The fractional abundances of other species such as \ce{C3H4} better fit with the contours of gas temperature ($\sim3000$~K) and ionization rates ($\sim10^{-16}~\mathrm{s^{-1}}$) for the upper and lower boundary, respectively, and within $\sim 4$~au.
The reason behind this is that each molecule has different routes of formation and destruction that can be enhanced not only by the individual different physical conditions but also by the combination of them.
Additionally, the reservoir in the inner-disk midplane seems to be determined by the high temperatures and densities in that region of the disk, where neutral-neutral and ion-molecule gas-phase chemistry dominates.

\paragraph{\ce{C_xH_y}}
All of the hydrocarbons,$\mathrm{C}_x\mathrm{H}_y$, follow the general abundance pattern discussed above (see Figs.~\ref{fig:FAmap_hydrocarbon_set1} and \ref{fig:FAmap_hydrocarbon_set2}). 
\ce{C2H2} reaches a higher peak abundance in the disk atmosphere than the disk midplane (i.e., $\sim 10^{-5}$ in the fiducial model).
On the other hand, \ce{C2H4} and \ce{C2H6} reach a higher peak abundance in the midplane than in the atmosphere.
In the fiducial model, the peak abundance in the midplane takes values of $\sim 10^{-7}$ and $\sim 10^{-6}$ for \ce{C2H4} and \ce{C2H6}, respectively.
\ce{C6H6}, \ce{C4H2}, and \ce{C3H4} reach a similar peak abundance ($\sim 10^{-8}$) in both components.

A general trend with increasing C/O is that for the cases with enhanced carbon (M0.88C, M8.8C, and M88C) the vertical width of the atmospheric component increases with respect to the equivalent reference model and models with oxygen depletion only (M4.4 and M44), with the peak abundance region expanding to denser and colder regions (see Figs. \ref{fig:TA_Tgas_ionizationRate_0.1au_hydrocarbons} and \ref{fig:TA_Tgas_ionizationRate_1.0au_hydrocarbons}).
Exceptions to this are the results for \ce{C3H4} and \ce{C2H6}, for which the fractional abundance maps show that the vertical width remains constant.
This is because the formation pathways of these species are associated not to \ce{C2H2} but instead with CH$_x$ ions, built up from \ce{C+} and \ce{C}.
This suggests that once C/O~$>1$, no matter the mechanism behind the increase in C/O, carbon will be more easily incorporated into large hydrocarbons, instead of CH$_x$ ions.
Another exception is \ce{C2H4}, for which the vertical width shrinks with respect to the models with oxygen depletion only, in particular, the upper boundary moves to colder layers (from $\sim 3000$ to $\sim 1000$~K).
Note that we consider a vertical or radial width expansion (or shrinkage) to be significant when the abundance changes by at least one order of magnitude.
Another general trend is that for several species the radial extent of the midplane component increases with increasing C/O (see Appendix~\ref{appendix:midplane}).
This is the case for \ce{C2H2}, \ce{C4H2}, \ce{C2H4}, and \ce{C3H4}.
We also see an increase in the vertical extent of the midplane component with the increase of C/O, which is most significant for \ce{C2H2}, \ce{C2H4}, and \ce{C2H6}.
A further general observation is that the biggest impact on the distribution and peak abundance reached for all species comes in the first two perturbations, i.e., in the enhancement of carbon (M0.88C) and/or the depletion of oxygen by a factor of 10 (M8.8C/M4.4).
Further oxygen depletion, i.e., by a factor of 100, does not result in any further significant increases in peak abundance of the hydrocarbons.
This suggests that the chemistry becomes limited by the availability of carbon, i.e., C/H, rather than depletion of oxygen.

The described changes in abundance pattern are reflected in column densities presented in Fig.~\ref{fig:ColdensProfile_IRemitting}.
We provide a more detailed and quantitative description of the response to C/O of each individual hydrocarbon considered in Appendix~\ref{appendix:hydrocarbons_CxHy}.  
In the remainder of the results we provide this quantitative information for \ce{CH4} and \ce{CH3}, which have a different behavior, and for the O-bearing and N-bearing species, which also have more individual behaviors with increasing C/O. 

\paragraph{\ce{CH4}}
\ce{CH4} has a broader distribution over the disk than that for the hydrocarbons discussed thus far. 
It is more abundant in the midplane than in the disk atmosphere, reaching a peak fractional abundance of $10^{-4}$ across all models within the midplane ($z/r\lesssim0.1$), and out to $\sim 0.06$ or $0.1$~au, depending on the model. 
There is a 10 times increase in the abundance of \ce{CH4}  where the gas temperature is $\sim100-500$~K and the abundance of \ce{CH4} is lower than $\sim10^{-10}$ in the fiducial model (M0.44F) when the C/O ratio increases to $0.87$ (due to carbon enrichment). 
However, oxygen depletion has a bigger effect on the abundance of \ce{CH4} in the upper layers of the disk atmosphere ($T_\mathrm{gas}\sim 1000-3000$~K), with models M4.4 and M44 showing a higher abundance than those that include carbon enhancement. 
Hence, the abundance of \ce{CH4} is more sensitive to the oxygen abundance in the atmosphere, with a reduced rate of \ce{CH4} destruction when oxygen is depleted \citep[see also][where similar results are reported]{raul_tracking_2025}. 
For C/O ratios greater than $1$, the drop in abundance in the layer in the outer disk at $z/r \sim 0.1 - 0.2$, and beyond a radius of $\sim 0.8 - 0.9$~au becomes more prominent. 
This could be related to this region being a sweet spot for grain-surface chemistry ($\sim 20-70~\mathrm{K}$).  
The behavior of the radial extent of the midplane component is somewhat sensitive to C/O.
The largest radial extent is found for the fiducial model M0.44F ($\sim 1$~au), with peak abundance values between $10^{-5}$ and $10^{-4}$.
An increase in C/O via carbon enhancement (M0.88C) shrinks this region to $\lesssim 0.4$~au (toward higher temperatures), which stays similar for all subsequent increases in C/O. 
This is likely to be due to other reactions in which atomic carbon is a reactant that are more efficient than those forming \ce{CH4}, such as those forming hydrocarbon ices. 
Also note that in our model the snowlines for hydrocarbons all lie internal to that for \ce{CH4} (see~Table~\ref{table:binding-energies}), which means that if \ce{CH4} formation reactions are not as efficient, carbon can be readily incorporated into hydrocarbon ices.

In general, the column density of \ce{CH4} in the IR-emitting region monotonically decreases with radius from $\sim 10^{16}~\mathrm{cm}^{-2}$ at 0.04~au to $\sim 10^{13}~\mathrm{cm}^{-2}$ at 10~au in all models.
The column density of \ce{CH4} in the fiducial model peaks at $\sim 10^{16}~\mathrm{cm}^{-2}$ within $\sim 0.1$~au and then drops to $\sim 10^{12}-10^{13}~\mathrm{cm}^{-2}$.
An increase in C/O does not necessarily lead to an overall increase in column density of \ce{CH4} in the IR-emitting region. 
In fact, the opposite is true within 0.1~au: when the C/O ratio increases, the column density of \ce{CH4} drops up to two orders of magnitude within 0.1~au, depending on the model.
However, beyond that radius, the column density of \ce{CH4} in oxygen-rich cases (C/O~$<1$) is generally at least a factor of three lower than that in the carbon-rich (C/O~$>1$) models.
The drop in the column density within 0.2~au is a result of the depleted region between 0.1 and 0.2~au expanding radially when C/O increases. 
When C/O increases, the main destruction pathway of \ce{CH4} changes from X-ray-induced photodissociation to the ion-molecule reaction with \ce{C8H2+}.
As long chains of hydrocarbons are more abundant in the carbon-rich scenario, \ce{CH4} is destroyed more efficiently.
As a result, carbon goes into nitriles, hydrocarbons, and LCCs.
Note that this is a feature of our adopted chemical network, where we observe a carbon sink effect in the largest unsaturated hydrocarbons in the network, such as \ce{C8H2+}.

\paragraph{\ce{CH3}}
As a radical, \ce{CH3} has an appreciable abundance only in the disk atmosphere (see Fig.~\ref{fig:FAmap_hydrocarbon_set1}).
The abundance in the fiducial model reaches high values of $\sim10^{-6}$ in this layer, where the temperature ranges from 1000 to 3000~K. 
Fig.~\ref{fig:FAmap_hydrocarbon_set1} shows that when increasing the C/O ratio the region over which \ce{CH3} reaches its peak value expands vertically to colder and denser regions, and the peak abundance also increases (up to $\sim10^{-5}$ in models M44 and M88C). 

This enhancement is also visible in Fig.~\ref{fig:ColdensProfile_IRemitting}, where the column density increases by at least a factor of two at all radii when the C/O increases from oxygen-rich to carbon-rich values.
However, again, the biggest effect (i.e., relative increase) comes from combining the first two perturbations (M8.8C, where we have both enhanced carbon and depleted oxygen by a factor of 10). 
Further increases in C/O do not significantly boost the column density of \ce{CH3} in the IR-emitting region.

%%%%%%%%%%%%%%%%%%%%%%%%%%%%%%%%%%%%%%%%%%%%%%%%%%%%%%%%%%%%%%%%%%%%%%

\subsubsection{Oxygen-bearing Species}

\begin{figure*}[ht]
\centering
\includegraphics[width=1\linewidth]{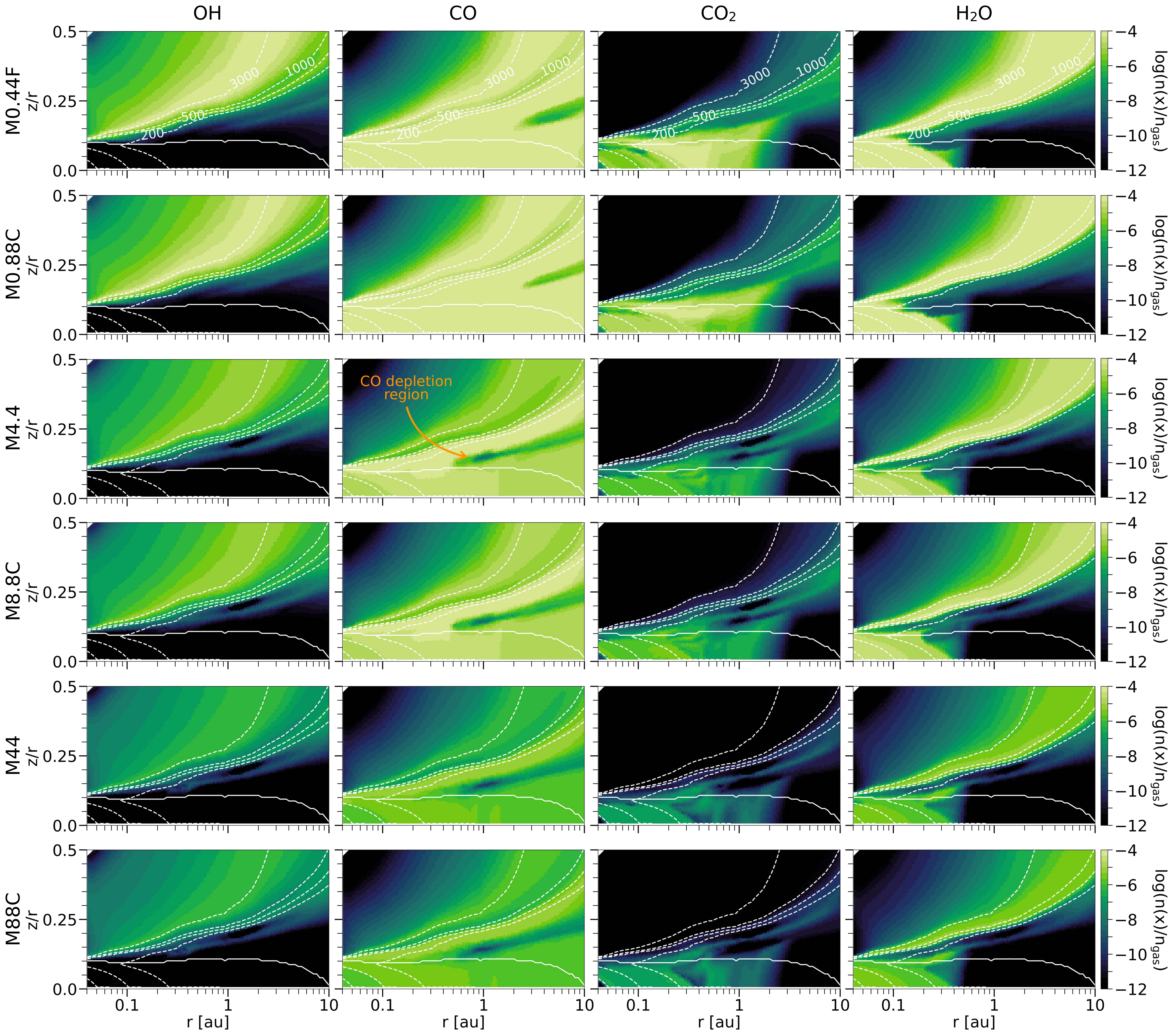}
\caption{Same as Fig.~\ref{fig:FAmap_hydrocarbon_set1}, but for the oxygen-bearing species: \ce{OH}, \ce{CO}, \ce{CO2}, and \ce{H2O}. 
}
\label{fig:FAmap_oxygen}
\end{figure*}

The behavior of oxygen-bearing species is somewhat different from that for the hydrocarbons, with a greater dependence on oxygen depletion, rather than carbon enhancement. 
There is an almost linear response in both peak abundance and column density to the oxygen depletion factor for most oxygen-bearing species. 

\paragraph{OH}
The fractional abundance of \ce{OH} is high in the upper layers, where the temperature ranges from $\sim 300$ to $\sim3000$~K, reaching a peak fractional abundance of $\sim10^{-4}$ in the fiducial model (M0.44F).
However, Fig.~\ref{fig:FAmap_oxygen} shows that when C/O increases the peak fractional abundance decreases (reaching a peak $\sim 10^{-6}$ when C/O~$\sim88$).
Similar to that found for the radical \ce{CH3}, OH is also only abundant in the disk atmosphere above the dust photosphere at $14~\mu$m. 

The column density profiles (Fig.~\ref{fig:ColdensProfile_IRemitting}) of OH across the disk have a close-to-linear response to the oxygen depletion factor. 
The peak value is $\sim 10^{16}$~cm$^{-2}$ in the fiducial model and drops by a factor of $\sim 10$ in the M4.4 model and by a factor of $\sim 100$ in the M44 model.
Carbon enhancement (models M0.88C, M8.8C, and M88C), on the other hand, affects the OH column density very little, with almost indistinguishable results for models M4.4 and M8.8C and models M44 and M88C.

\paragraph{CO}
The abundance of \ce{CO} in the fiducial model (M0.44F) reaches a peak of $\sim10^{-4}$ from the midplane to the layers where the temperature reaches 3000~K, with two layers of CO depletion (drop of the abundance by a factor of 10 to 100, depending on the model) at gas temperatures of $\sim1000$ and $\lesssim40$~K.
As a hypervolatile, CO has one of the more expansive distributions across the disk, as it only freezes out at temperature $\lesssim 20$~K. 
The \ce{CO}-depleted region in the layer between $z/r\sim0.15$ and $0.25$ beyond $\sim2~\mathrm{au}$ is likely related to grain-surface chemistry transforming CO into other molecules, mostly into \ce{CO2} ice.
On the other hand, the slightly depleted layer at $\sim1000$~K is more related to photodissociation reactions.
When the C/O ratio increases, the size of the region over which \ce{CO} reaches its peak abundance in the disk atmosphere shrinks vertically, with the \ce{CO} layer moving to lower gas temperature regions (below 200~K). 
The lower CO abundance in the atmosphere for these cases weakens the self-shielding of CO to photodissociation and causes the layer to shift downward. 
The CO-depleted region closer to the midplane also extends inward down to $\sim0.6~\mathrm{au}$.
This is due to CO gas being efficiently converted to ices composed of CO, \ce{CO2}, and LCCs.
For models M44 and M88C, the peak abundance decreases by a factor of $\sim 10$ to $\sim10^{-5}$.

This is reflected in the column density profiles in Fig.~\ref{fig:ColdensProfile_IRemitting}.
The column density in the IR-emitting region within 0.1~au drops from $\sim10^{19}~\mathrm{cm^{-2}}$ (M0.44F) to $\sim10^{17}~\mathrm{cm^{-2}}$ (M88C).
Similar to OH, carbon enhancement does not significantly increase the column density of CO relative to the M0.44F, M4.4, and M44 models.

\paragraph{\ce{CO2}}
\ce{CO2} is an interesting diagnostic, as it has been detected in the three sources discussed here and appears to be a prevalent O-bearing species that can survive and be emissive in an otherwise apparently carbon-rich environment.
The fractional abundance of \ce{CO2} in the model M0.44F reaches a peak value of $\sim10^{-4}$ in the region near the midplane $(z/r<0.2)$ between 0.3 and 0.5~au (see Fig.~\ref{fig:FAmap_oxygen}).
The peak decreases by a factor of 10 and expands inward radially to 0.05~au when C/O increases to 0.87.
The region over which \ce{CO2} reaches its peak abundance in the disk atmosphere is also generally at a lower temperature than that for other O-bearing species such as OH and \ce{H2O}. 
Above $z/r \sim 0.1$ and beyond $r\approx 0.5$~au, \ce{CO2} generally resides in the region between 50 and 200~K with a peak abundance of $\sim 10^{-5}$ in the fiducial model (M0.44F).
This lower temperature bound is likely tracing the location of the snow surface for \ce{CO2}.
The upper boundary of 200~K is shaped by the high abundance of ions that destroy \ce{CO2}, such as \ce{H+} and \ce{H3+}, and \ce{OH} being driven into \ce{H2O} rather than \ce{CO2}.
Additionally, the ionization rate ($\sim10^{-14}~\mathrm{s^{-1}}$), X-ray flux ($\sim 10^{-1}~\mathrm{erg~cm^{-2}~s^{-1}}$), and UV flux ($\sim 10^{-5}~\mathrm{erg~cm^{-2}~s^{-1}}$) also shape that region (see Fig.~\ref{fig:disk-structure} and \ref{fig:FAmap_oxygen}).

When the C/O ratio increases to $> 1$ (models M4.4 and M8.8C), the abundance decreases by around a factor of 10 in the disk midplane.
In particular, the peak abundance within 0.5~au decreases to $\sim10^{-6}$.
There is a similar drop in the atmosphere by almost an order of magnitude with respect to M0.44F and M0.88C.
For the two higher values of the C/O ratio (models M44 and M88C), the peak fractional abundance drops to $\sim10^{-7}$ in the midplane and $\sim 10^{-9}$ in the atmosphere, again demonstrating an almost linear response to the oxygen depletion factor.

This is also somewhat reflected in the column density profiles in Fig.~\ref{fig:ColdensProfile_IRemitting}, where the peak column density in the IR-emitting region reduces from 
$\sim 10^{18}$~cm$^{-2}$ within $0.1 - 0.2$~au in the fiducial and oxygen-rich models (M0.44F and M0.88C) to values lower than $10^{12}$~cm$^{-2}$ at $0.2 - 0.3$~au in models M44 and M88C. 
The column density profiles beyond 1~au are also affected by the \ce{CO2} depletion in the middle layers of the disk (similar to that for CO), with the drop beyond this radius increasing in magnitude with increasing C/O.
Again, oxygen depletion has the biggest effect on the \ce{CO2} column densities rather than carbon enhancement. 
This is the result of the linear response of \ce{CO2} to the C/O ratio; when C/O increases, less oxygen is available to form \ce{CO2} in the first place.
% START HERE
\paragraph{\ce{H2O}}
The abundance of \ce{H2O} is high in the region where the temperature ranges from 100 to 3000~K, reaching a peak fractional abundance of $\sim10^{-4}$ in the fiducial model (M0.44F), and its distribution has both the midplane component and the atmospheric component. 
The response of the abundance of \ce{H2O} to the C/O ratio is similar to that of the other oxygen-bearing species, e.g., \ce{CO2}: the peak abundance drops to $\sim10^{-6}$ in model M88C (see Fig.~\ref{fig:FAmap_oxygen}) and is much more sensitive to oxygen depletion than carbon enhancement. 

This is also reflected in the calculated column density shown in Fig.~\ref{fig:ColdensProfile_IRemitting}, where the profiles show a steplike structure. 
The column density of \ce{H2O} in the IR-emitting region when C/O~$= 0.44$ reaches a peak value of $\sim10^{19}~\mathrm{cm^{-2}}$ within 0.1~au. 
When the C/O ratio increases, this drops by up to two orders of magnitude, to $\sim 10^{17}~\mathrm{cm^{-2}}$ in the same region of the disk.
Finally, as found for the other oxygen-bearing species, oxygen depletion has the biggest effect.
Carbon enrichment has an effect only in the innermost region within 0.1~au (a variation by a factor up to 100 depending on the model).
Beyond that radius, there is no effect of carbon enrichment on the column density profile of water, with the peak value of the column density showing a linear response to the oxygen depletion factor.

%%%%%%%%%%%%%%%%%%%%%%%%%%%%%%%%%%%%%%%%%%%%%%%%%%%%%%%%%%%%%%%%%%%%%%

\subsubsection{Nitrogen-bearing Species}

\begin{figure*}[ht]
\centering
\includegraphics[width=1\linewidth]{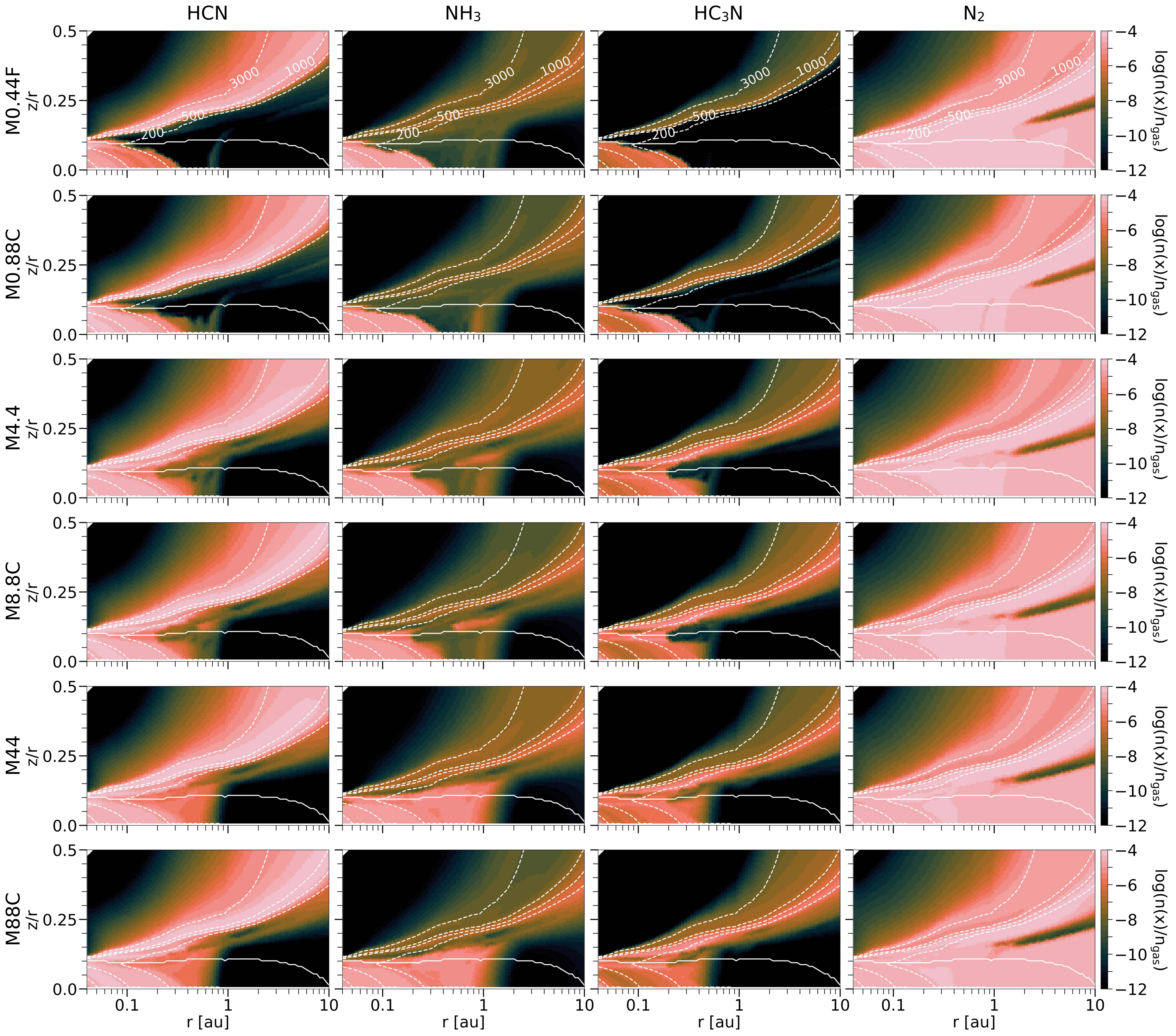}
\caption{Same as Fig.~\ref{fig:FAmap_hydrocarbon_set1}, but for the nitrogen-bearing species: \ce{HCN}, \ce{NH3}, \ce{HC3N}, and \ce{N2}.
}
\label{fig:FAmap_nitrogen}
\end{figure*}

\paragraph{HCN}
The abundance distribution of HCN exhibits the same two-component morphology as most other species: a midplane component and an atmospheric component, with the peak fractional abundance in the atmosphere up to 10 times higher than in the midplane. 
In the fiducial model, the fractional abundance of \ce{HCN} is high in the region where the temperature is around 1000~K, reaching a peak abundance of $\sim10^{-4}$. 
A peak abundance of $\sim 10^{-5}$ is reached in the midplane within $\approx 0.3$~au. 
The size of the disk atmosphere component that reaches values higher than $10^{-8}$ expands vertically when the C/O ratio increases, with the upper and lower boundaries moving to higher ($\gtrsim3000$~K) and lower ($\sim50$~K) gas temperatures, respectively. 
The models with carbon enhancement have an increased vertical extent when compared with the models with oxygen depletion only (e.g., M88C vs. M44); hence, excess carbon is important for boosting the abundance of HCN in the disk atmosphere.
However, the peak fractional abundance reached in the IR-emitting region does not vary with C/O (similar to, e.g., \ce{C2H2}).
The peak abundance region in the midplane component also expands radially and vertically, with increasing C/O ``filling in'' the region between the midplane and $z/r\approx 0.1$ within $\approx 0.3$~au. 
This radial expansion is also observable beyond the mentioned radius for each model, but with a lower abundance.

The column density profiles show a similar shape to those for \ce{H2O} (i.e., a steplike behavior).
The peak column density is the same in all models, with values of $\sim 10^{18}$~cm$^{-2}$ at 0.04~au in the IR-emitting region.
The column density drops to $\sim 10^{16}$~cm$^{-2}$ beyond $\approx$~0.1~au in the oxygen-rich models (M0.44F and M0.88C).
The carbon-rich models reach a higher column density beyond $\sim0.1$~au, which ranges from more than one order of magnitude higher within 0.2~au to a factor of a few higher beyond.
The effects of the ``filling in'' of the HCN reservoir toward the midplane are evident in the smoothing of the column density profiles within $\approx$~1~au as C/O increases.
The reason behind this effect is that when C/O increases, nitrogen that was in form of \ce{N}, \ce{N2}, \ce{NH3}, and \ce{NO} is now incorporated into nitriles (e.g., \ce{NCCN}, \ce{HC3N}, \ce{HC5N}, \ce{HC9N}, and \ce{HCN}).

\paragraph{\ce{NH3}}
The distribution of \ce{NH3} shows the same two-component morphology as HCN; however, \ce{NH3} is significantly more abundant in the disk midplane than in the disk atmosphere in the fiducial model (M0.44F).
The abundance of \ce{NH3} reaches a peak value of $\sim10^{-5}$ in the midplane within 0.3~au and a value of $\sim 10^{-7}$ in the disk atmosphere, where the temperature is $\sim1000$~K.
\ce{NH3} is more sensitive to oxygen depletion than to carbon enhancement.
When oxygen is decreased by a factor of 10 (M4.4 and M8.8C), the peak abundance in the atmosphere increases by around the same factor, and the midplane component increases in vertical extent (similar to HCN).
Further oxygen depletion (M44 and M88C) does not significantly boost the abundance in the atmospheric component, but the midplane region expands radially outward to 1~au to colder and less dense regions. 

This behavior is also reflected in the column density profiles (Fig.~\ref{fig:ColdensProfile_IRemitting}).  
The column density at 0.04~au in the IR-emitting region is $\sim 10^{17}$~cm$^{-2}$ in all models.
When the C/O ratio increases to carbon-rich values (M4.4, M8.8C, M44, and M88C) there is a boost in the column density of \ce{NH3} within $\sim0.1$~au by almost one order of magnitude.  
Beyond that radius, models with a C/O$>1$ have a higher column density by a factor of a few, with all models ranging between $\sim10^{13}$ and $\sim10^{14}~\mathrm{cm^{-2}}$.

\paragraph{\ce{HC3N}}
The behavior of \ce{HC3N} is very similar to that for \ce{NH3}.
The fractional abundance of \ce{HC3N} reaches a peak value of $\sim 10^{-6}$ in the innermost regions of the disk and a value of $\sim 10^{-8}$ in layers where the temperature ranges from 500 to 1000~K in the fiducial model.
The response of the abundance of \ce{HC3N} to the C/O variations is somewhat similar to that for \ce{HCN}, in that the region size where the abundance peaks expands radially (in the disk midplane to lower temperatures and less dense regions) and vertically (in both components, filling up the region between these two components) when C/O increases (see Fig \ref{fig:FAmap_nitrogen}). 
For the first perturbation, there is an increase in the peak abundance reached in the disk midplane by a factor of a few between $\sim0.1$ and $\sim 0.2$~au.
For higher values of C/O ratio, there is no increase in the peak abundance; however, the peak in the atmospheric component does increase to $\sim 10^{-6}$ in models with C/O $\ge 4.4$.  
The distribution of \ce{HC3N} is also sensitive to the enhancement of carbon, with the carbon-enriched models having a slightly greater vertical extent of atmospheric \ce{HC3N} (moving to higher gas temperatures and less dense regions) than the equivalent models with oxygen depletion only; however, this effect is very subtle.

This behavior with C/O is also seen in the column density profiles in Fig.~\ref{fig:ColdensProfile_IRemitting}.
The peak column density in the fiducial model is $\sim~ 10^{17}$~cm$^{-2}$ within 0.07~au and reduces to $\sim 10^{13}$~cm$^{-2}$ beyond. 
The peak column density in the inner disk does not change with C/O; only the region over which this is reached increases in radial extent.
Again, the behavior of carbon-rich models differs from the oxygen-rich cases.
Beyond 0.1~au, the column density for M0.44F and M0.88C is lower than $\sim 10^{14}~\mathrm{cm}^{-2}$. 
The column density in the carbon-rich models, on the contrary, decreases monotonically out to 1~au and thereafter remains flat at $\sim 10^{14}~\mathrm{cm}^{-2}$.

\paragraph{\ce{N2}}
The fractional abundance of \ce{N2} reaches peak values ($\sim10^{-4}$) in the region from the midplane to the layers where the temperature reaches 500~K (see Fig.~\ref{fig:FAmap_nitrogen}).
Similar to \ce{CO}, \ce{N2} shows a depleted region that expands inward (to denser and colder regions) when the C/O ratio increases.
This is the only major effect of increasing C/O. 
This is because the chemistry is driving nitrogen into N-bearing ices, such as \ce{NH3}, \ce{HCN}, \ce{CH3NH2}, and \ce{HC3N}.
The column density for the IR-emitting region has similar profiles for all C/O values. 
In general, the column density profile decreases with radius, reaching a constant value beyond $\sim0.8$~au ($\sim 10^{16}~\mathrm{cm}^{-2}$).

%%%%%%%%%%%%%%%%%%%%%%%%%%%%%%%%%%%%%%%%%%%%%%%%%%%%%%%%%%%%%%%%%%%%%%

\subsection{Potentially Observable Species}
Other molecules have been previously predicted to be abundant enough to be detected in the inner disk \citep{bast_exploring_2013}. 
This is the case of the nitriles \ce{CH3CN} and \ce{HNC}, which are also commonly detected at millimeter wavelengths. 
\citet{kanwar_minds_2024} also suggest that the abundance of \ce{C2}, \ce{C2H}, \ce{C3}, \ce{C3H}, and \ce{CH2CCH} may reach observable abundances in the IR-emitting region at elevated C/O.

We present the fractional abundance maps and column density profiles predicted by our models for the different C/O scenarios for these potentially observable species in Figs.~\ref{fig:FAmap_potentially_observed_set1}, \ref{fig:ColdensProf_potentially_observed_IRemitting}, and \ref{fig:FAmap_potentially_observed_set2} in Appendix~\ref{appendix:potentially_observable}. 
As shown in these figures and in Fig.~\ref{fig:Nmol-C2Oratio-others}, the abundance and number of molecules of \ce{C2}, \ce{C2H}, \ce{C3}, \ce{C3H}, \ce{CH2CCH}, \ce{CH3CN}, and \ce{HNC} increase when C/O increases with respect to the fiducial model (M0.44F).
However, similar to the case of \ce{C2H2} and most of the hydrocarbons mentioned so far in this work, the abundance does not increase linearly with the C/H and O/H variations.

%%%%%%%%%%%%%%%%%%%%%%%%%%%%%%%%%%%%%%%%%%%%%%%%%%%%%%%%%%%%%%%%%%%%%%

\section{Discussion} \label{sec:discussion}

Here we more closely compare the model results with the observations in order to determine whether any conclusions can be drawn on the C/O ratio needed to explain the observations of the sources considered here.  
Because we have adopted a generic model, rather than one tailored to a particular source, we also compare the total number of emitting molecules, as well as their ratios.
This is because this removes, to some extent, the dependence on physical properties such as the disk gas mass and radial distribution, as well as degeneracies that are often met when trying to fit both column densities and emitting areas for optically thin emission (see discussion in \citealt{kamp_chemical_2023} and \citealt {arulanantham_jdisc_2025}).

%%%%%%%%%%%%%%%%%%%%%%%%%%%%%%%%%%%%%%%%%%%%%%%%%%%%%%%%%%%%%%%%%%%%%%

\subsection{Observational Trends}\label{sec:observational_trends}
\begin{figure}[ht]
\centering
\includegraphics[width=\linewidth]{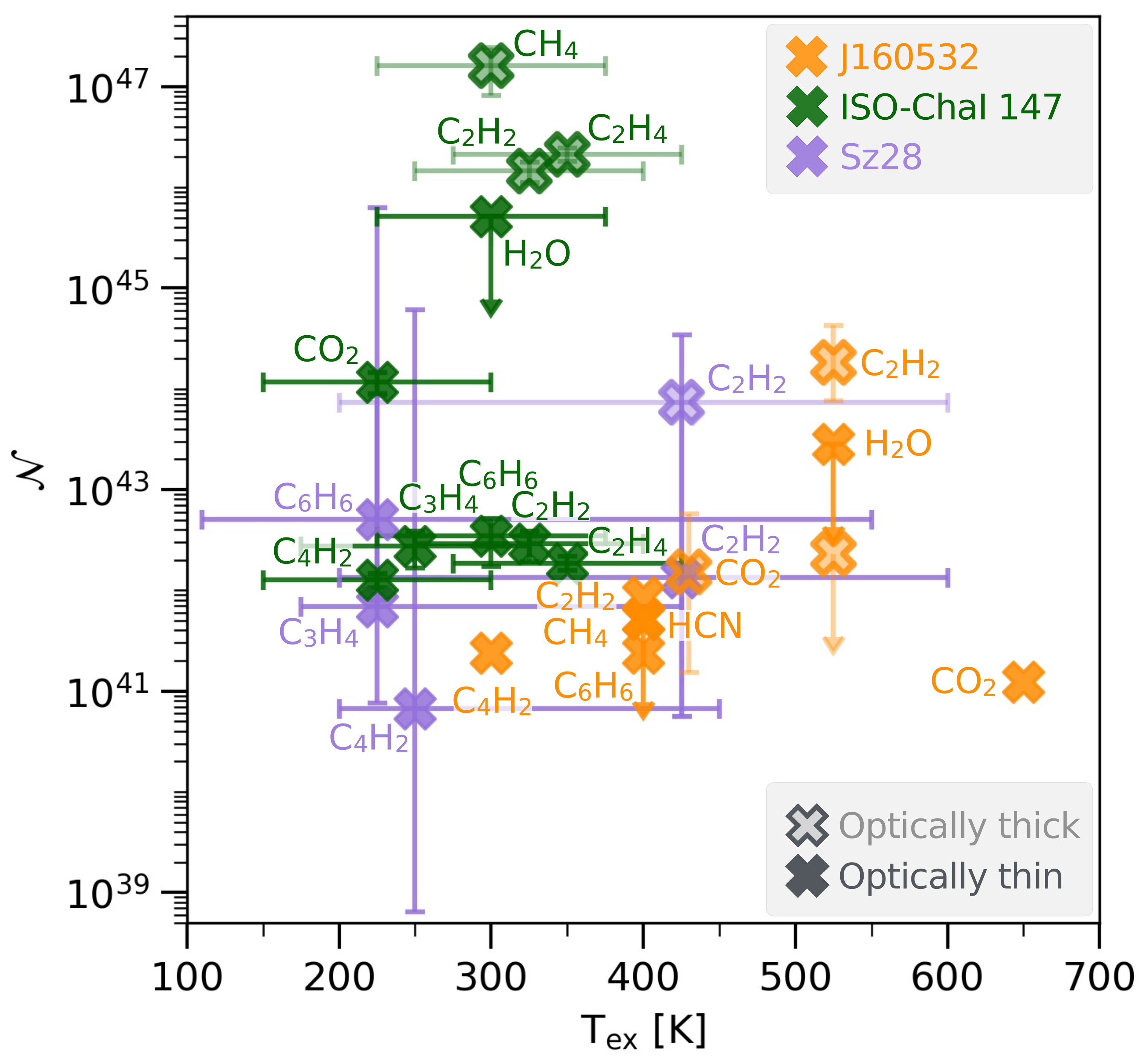}
\caption{
Summary of the results available in the literature for J160532 \citep{tabone_rich_2023, arabhavi_minds_2025}, ISO-ChaI~147 \citep{arabhavi_abundant_2024}, and Sz28 \citep{kanwar_minds_2024}, using a 0D slab model to estimate the total number of emitting molecules, $\mathcal{N}$, and excitation temperature, $T_\mathrm{ex}$.
The different colors represent the different disks.
The filled and open crosses represent the optically thin and thick components, respectively, for each species when available (\ce{C2H2} and \ce{CO2} for J160532; \ce{C2H2}, \ce{CH4}; and \ce{C2H4} for ISO-ChaI~147, and \ce{C2H2} for Sz28.)
Note that some error bars for different species and/or sources might superpose.
For more information see Table~\ref{table:Nmol_literature}.}
\label{fig:Nmol_Tex_literature}
\end{figure}

In Table \ref{table:Nmol_literature} we show the total number of molecules reported by \citet{tabone_rich_2023}, \citet{arabhavi_abundant_2024}, \citet{kanwar_minds_2024}, and \citet{arabhavi_minds_2025} for the species detected toward the hydrocarbon-rich disks of J160532, ISO-ChaI~147, and Sz28, respectively. 
We do not include estimates by \citet{kaeufer_disentangling_2024} for Sz28, which are up to eight orders of magnitude higher than those reported by \citet{kanwar_minds_2024}.
To better visualize the range of molecular parameters derived from the observations, we present the data reported by the papers above in Figure~\ref{fig:Nmol_Tex_literature}.
Presented are excitation temperatures $T_\mathrm{ex}$ versus the total number of molecules $\mathcal{N}$ for each source and each detected species.
This highlights two by-eye interesting trends. 
First, the excitation temperatures toward ISO-ChaI~147 are lower than those for J160532, with Sz28 somewhat intermediate between the two. 
A second trend is that ISO-ChaI~147 generally has the largest number of molecules, with the values for Sz28 and J160532 generally lower.

The number of molecules for hydrocarbons in the thin component ranges from $\sim10^{41}$ to $\sim 10^{43}$. 
The values for the thick component for the same set of species are, in general, higher ($\sim10^{44}-10^{47}$).
From Fig.~\ref{fig:Nmol_Tex_literature}, ISO-ChaI~147 is overall more rich in hydrocarbons and oxygen-bearing species than J160532 and Sz28.
For example, the number of molecules of \ce{CO2} is three orders of magnitude higher in ISO-ChaI~147 than in J160532.
The total numbers of molecules in the optically thin component for hydrocarbons are similar for both ISO-ChaI~147 and J160532, while $\mathcal{N}$ of \ce{H2O} is three orders of magnitude higher in ISO-ChaI~147 than in J160532.
Because the numbers of molecules in the optically thin component for the hydrocarbons in both J160532 and ISO-ChaI~147 are similar, this results in the latter being relatively more hydrocarbon-poor relative to \ce{CO2} than the former.

The results also suggest that hydrocarbons and oxygen-bearing species in ISO-ChaI~147 and Sz28 are emitting mostly from regions colder than those in J160532. 
The excitation temperature for the optically thick component of \ce{C2H2} is 525~K in J160532 and 300~K in ISO-ChaI~147. 
In addition, the excitation temperature of \ce{CO2} is $\sim200$ and $\sim650$~K for ISO-ChaI~147 and J160532, respectively.
We also see that \ce{C2H2} tends to have a temperature higher than or equal to those of other hydrocarbons, such as \ce{C6H6}, \ce{CH4}, \ce{C4H2}, and \ce{C3H4} in the same disk.

We can compare these excitation temperatures with the gas temperature distribution and average temperature for each species predicted by our models (see Appendix~\ref{appendix:temp_distribution}).
Our results show that larger hydrocarbons such as \ce{C3H4}, \ce{C4H2}, \ce{C2H4}, and \ce{C6H6} exhibit average temperatures in the IR-emitting region between 700 and 1200~K in the fiducial model (M0.44F).
The lower and upper ends of this range decrease to 400 and 1000~K, respectively, when C/O increases to $\sim 88$.
Note that the upper end decreases even more (to 800~K) when only oxygen depletion by a factor of 100 is considered.
\ce{CO2} remains comparatively cooler ($\lesssim 300~\mathrm{K}$) across all C/O values, whereas HCN, \ce{H2O}, and smaller hydrocarbons such as \ce{CH4} and \ce{C2H2} cover a larger temperature range of $1000-1450~\mathrm{K}$.
From the values derived from observations and shown in Fig.~\ref{fig:Nmol_Tex_literature}, larger hydrocarbons also cluster around intermediate temperatures, but significant differences between sources are observed for \ce{CO2}, HCN, and \ce{H2O}.
For example, the derived excitation temperature for \ce{CO2} is relatively low in ISO-ChaI~147 (225~K) but higher in J160532 ($430-650$~K).

%%%%%%%%%%%%%%%%%%%%%%%%%%%%%%%%%%%%%%%%%%%%%%%%%%%%%%%%%%%%%%%%%%%%%%

\subsection{Comparison with Observations}\label{sec:moredetailedcomparison}

In this subsection, we compare the total number of molecules predicted by the model with those derived from the observations (see Table~\ref{table:Nmol_literature}). To do this, we calculated the total number of molecules ($\mathcal{N}$) for each species by integrating our model abundances across the IR-emitting region (a gas temperature $>200$~K and above the dust photosphere at $14~\mathrm{\mu m}$).
For this, we used the following equation:
\begin{equation}
    \mathcal{N} = {\int_{R} N(r)~2\pi r~dr},
    \label{eq:tot-num-mol}
\end{equation}
\noindent where $N(r)$ is the vertically integrated column density as a function of the disk radius $r$, ranging from $0$ to $R$, where $R$ is the integration radius.

\begin{figure*}[ht]
\centering
\includegraphics[width=\linewidth]{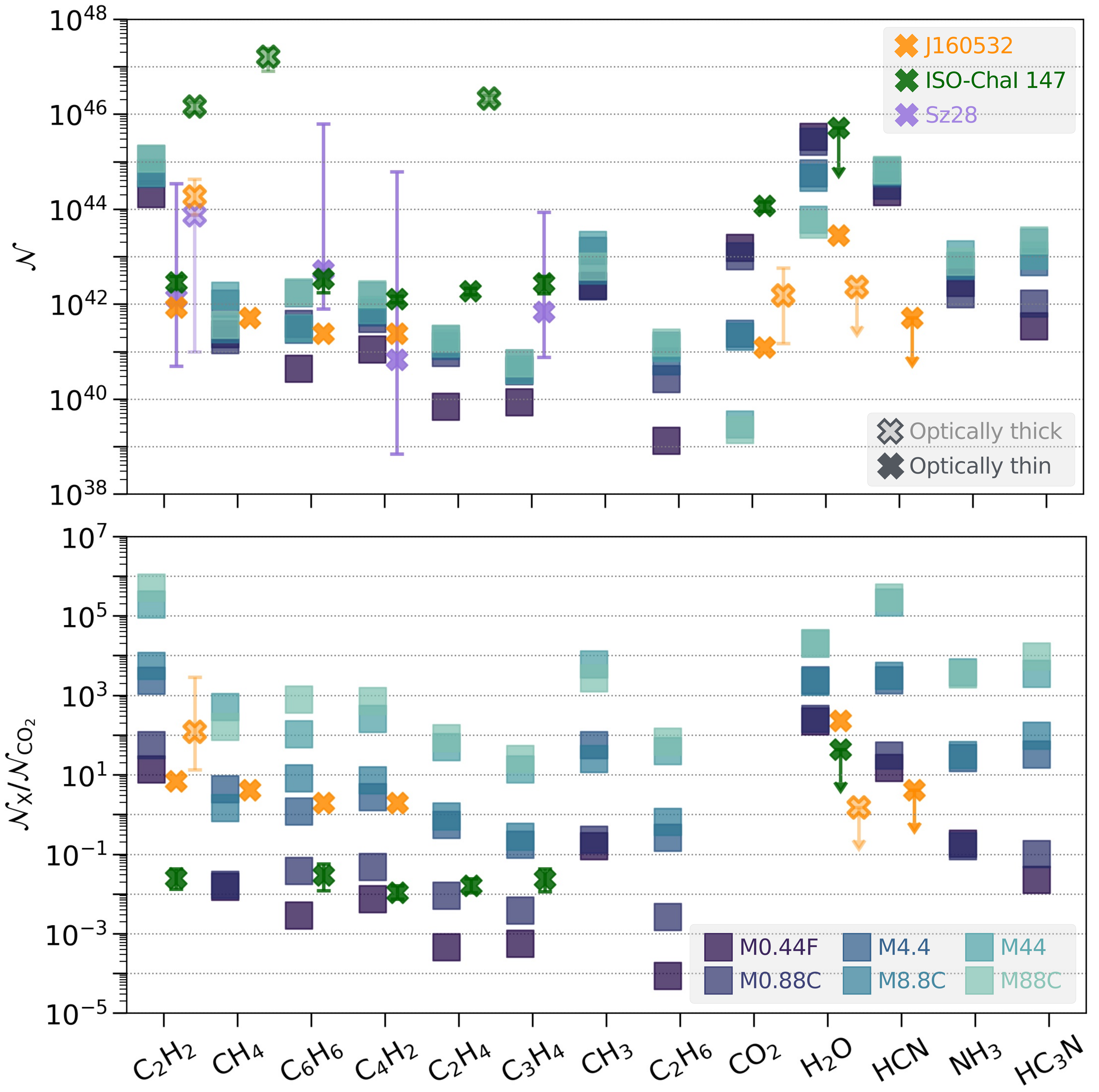}
\caption{
Top: total number of molecules of \ce{C2H2}, \ce{CH4}, \ce{C6H6}, \ce{C4H2}, \ce{C2H4}, \ce{C3H4}, \ce{CH3}, \ce{C2H6}, \ce{CO2}, \ce{H2O}, \ce{HCN}, \ce{NH3}, and \ce{HC3N}. 
The values estimated here are the integrated total number of molecules up to 10~au, integrating over the IR-emitting region.
Bottom: total number of molecule ratios of the same set of species with respect to \ce{CO2}.
The orange, green, and purple crosses represent the results from \cite{tabone_rich_2023}, \cite{arabhavi_abundant_2024}, \cite{kanwar_minds_2024}, and \citet{arabhavi_minds_2025} for J160532, ISO-ChaI~147, and Sz28, respectively. 
The filled and open crosses represent the thin and thick components for each species when available, respectively.
The shaded colored squares represent from dark to light the increase of the C/O ratio, i.e., the results from our chemical models with different C/O ratios.
The bottom panel shows the ratio of the number of molecules with respect to \ce{CO2}.
(The complete figure set (2 images) is available in the online article.)
}
\label{fig:TotalNmol_IRemitting}
\end{figure*}

Figure~\ref{fig:TotalNmol_IRemitting} shows the comparison between the model number of molecules for the IR-emitting region and those reported for the species observed toward J160532, ISO-ChaI~147, and Sz28 with orange, green, and purple crosses, respectively (top panel), and the ratios for the same set of species with respect to \ce{CO2} (bottom panel).
We used \ce{CO2} as our relative species because we find that it is more sensitive to the C/O ratio than others (e.g., \ce{C2H2}). 
Both optically thin and thick components are included if available.
The shaded squares represent the increasing C/O ratio value from dark to light, i.e., the different models tested in this work.
From the ratios in the bottom panel of Fig.~\ref{fig:TotalNmol_IRemitting}, it is clear that the ratio with respect to \ce{CO2} is a potentially powerful diagnostic of the global C/O in the infrared-emitting region of the disk.
Note that there is a spread of up to four orders of magnitude in the ratios with respect to \ce{CO2} for a spread of only two orders of magnitude in the C/O ratio.
For reference we present the column density ratio for \ce{C2H2}/\ce{CO2} as a function of radius in Fig.~\ref{fig:ColDensProf_ratio-Nmol_ratio-vs_int-radius}.

%%%%% Figure set for Figure 8 -- Begin
%\figsetstart
%\figsetnum{8}
%\figsettitle{Number of molecules and their ratios for the region above the $\tau=1$ surface at $14~\mu\mathrm{m}$ (disk atmosphere component) and full vertical disk extent down to the midplane (midplane component).}

%\figsetgrpstart
%\figsetgrpnum{8.1}
%\figsetgrptitle{Region above the $\tau=1$ surface at $14~\mu\mathrm{m}$}
%\figsetplot{Nmol_Nmolratio_atmosphere14um.pdf}
%\figsetgrpnote{
%Same as Fig.~\ref{fig:TotalNmol_IRemitting}, but for the integration down to the $\tau=1$ surface at $14~\mu\mathrm{m}$ (disk atmosphere component).}
%\figsetgrpend

%\figsetgrpstart
%\figsetgrpnum{8.2}
%\figsetgrptitle{Region above the midplane ($z=0$)}
%\figsetplot{Nmol_Nmolratio_midplane.pdf}
%\figsetgrpnote{
%Same as Fig.~\ref{fig:TotalNmol_IRemitting}, but for the integration down to the midplane (midplane component).}
%\figsetgrpend

%%%%% Figure set for Figure 8 -- End

From the top panel of Fig.~\ref{fig:TotalNmol_IRemitting} we see that the model reproduces (around the same order of magnitude) the total number of molecules for \ce{CH4} (thin component), \ce{C6H6}, \ce{C4H2}, \ce{CO2}, and \ce{H2O} but only for some disks and for different values of C/O, depending on the species.
For example, we can reproduce $\mathcal{N}$ of J160532 for \ce{CO2} but only for the moderately enhanced C/O models (M4.4 and M8.8C). 
This is also achieved for \ce{CH4} for the fiducial model (M0.44F) and cases that included carbon enrichment (M0.88C, M8.8C, and M88C) and for \ce{C6H6} for the carbon-enriched-only (M0.88C) and oxygen-depleted-only models (M4.4 and M44).
The number of molecules in the IR-emitting region for these three species up to 10~au varies differently with increasing C/O; it decreases for \ce{CO2} and oscillates for \ce{CH4} and \ce{C6H6}, with the carbon-rich models having a larger number of molecules than the equivalent oxygen-depleted models (see Figs.~\ref{fig:Nmol-C2Oratio_hydrocarbons} and \ref{fig:Nmol-C2Oratio_oxygen_nitrogen}).
All models overestimate the number of molecules in the thin component of \ce{C2H2}, reaching values of the order of magnitude of the thick component in J160532.
The fiducial model (M0.44F) is consistent with the upper limit reported for \ce{H2O} for ISO-ChaI~147, while the higher C/O ratio models (M44 and M88C) better fit that estimated for J160532.
For all other species (\ce{C2H4}, \ce{C3H4}, and \ce{HCN}), there is a difference of more than one order of magnitude between the models and observations for J160532, ISO-ChaI 147, and Sz28 from \cite{tabone_rich_2023}, \citet{arabhavi_abundant_2024}, and \citet{kanwar_hydrocarbon_2024}, with the observations generally suggesting higher values than the models (with the exception of HCN).
For the thin component of \ce{C2H2} and \ce{C3H4} in Sz28, the values obtained in the oxygen-rich and carbon-rich models, respectively, are consistent with the uncertainties reported for these species.
On the other hand, the observations for \ce{C4H2} in the same source do not constrain the C/O ratio, as all models fall within the corresponding error bars.

The bottom panel of Fig.~\ref{fig:TotalNmol_IRemitting} shows that the observed ratios for J160532 are consistent with a wide range of C/O ratios.
In particular, solar-like C/O ratios can reproduce the $\mathcal{N}$ ratios reported for \ce{C2H2}, HCN, and the oxygen-bearing species \ce{H2O}.
On the other hand, \ce{CH4}, \ce{C6H6}, and \ce{C4H2} are consistent with moderately enhanced C/O ratios ($\sim 4.4-8.8$).
However, all the equivalent values for ISO-ChaI~147 are consistent with solar C/O or moderately enhanced C/O, i.e., similar to 1 or of order a few ($\lesssim 4.4$).
We decided not to perform the same analysis for Sz28, because the values reported by \citet{kanwar_minds_2024} for \ce{CO2} are not a best fit to the available data and are used for illustration of the detection of \ce{CO2} only.
Taking the values from Table B.1 in \citet{kanwar_minds_2024} suggests a very high total number of molecules of \ce{CO2} ($\sim 10^{48}$), which is a factor $10^{4}$ higher than that extracted for ISO-ChaI~147.

Our results suggest that efficient production of hydrocarbon species does not always require a C/O~$>1$. 
Despite ISO-ChaI~147 appearing significantly more hydrocarbon-rich than J160532 through the number of species detected, the model ratios in Fig.~\ref{fig:TotalNmol_IRemitting} suggest that this source could be globally less carbon-rich than J160532.
We emphasize here that this is based on models assuming a single X-ray luminosity of $10^{29}~\mathrm{erg~s^{-1}}$, which is around the middle of the observed range for VLMSs. 
We cannot rule out any degeneracies in the results between X-ray luminosity and C/O, and we will explore this in future work.

In our definition of the IR-emitting region, we assume a lower temperature bound of 200~K. 
If this gas temperature constraint is relaxed and the integration to determine $\mathcal{N}$ is extended down to the $\tau=1$ surface at $14~\mu\mathrm{m}$ over the whole disk, most species show only small, model-dependent increases, because they reach their peak abundance well above this temperature threshold.
We find that \ce{CH4} and \ce{CO2} are the most strongly impacted, as they are more abundant than the other species considered here in the colder regions that would be now included in the integration.
Thus, the ratios with respect to \ce{CO2} that we present in Fig.~\ref{fig:TotalNmol_IRemitting} would be affected by a common factor.
For interest, versions of Fig.~\ref{fig:TotalNmol_IRemitting} for the disk atmosphere and midplane components are available online.

\begin{figure}[ht]
\centering
\includegraphics[width=\linewidth]{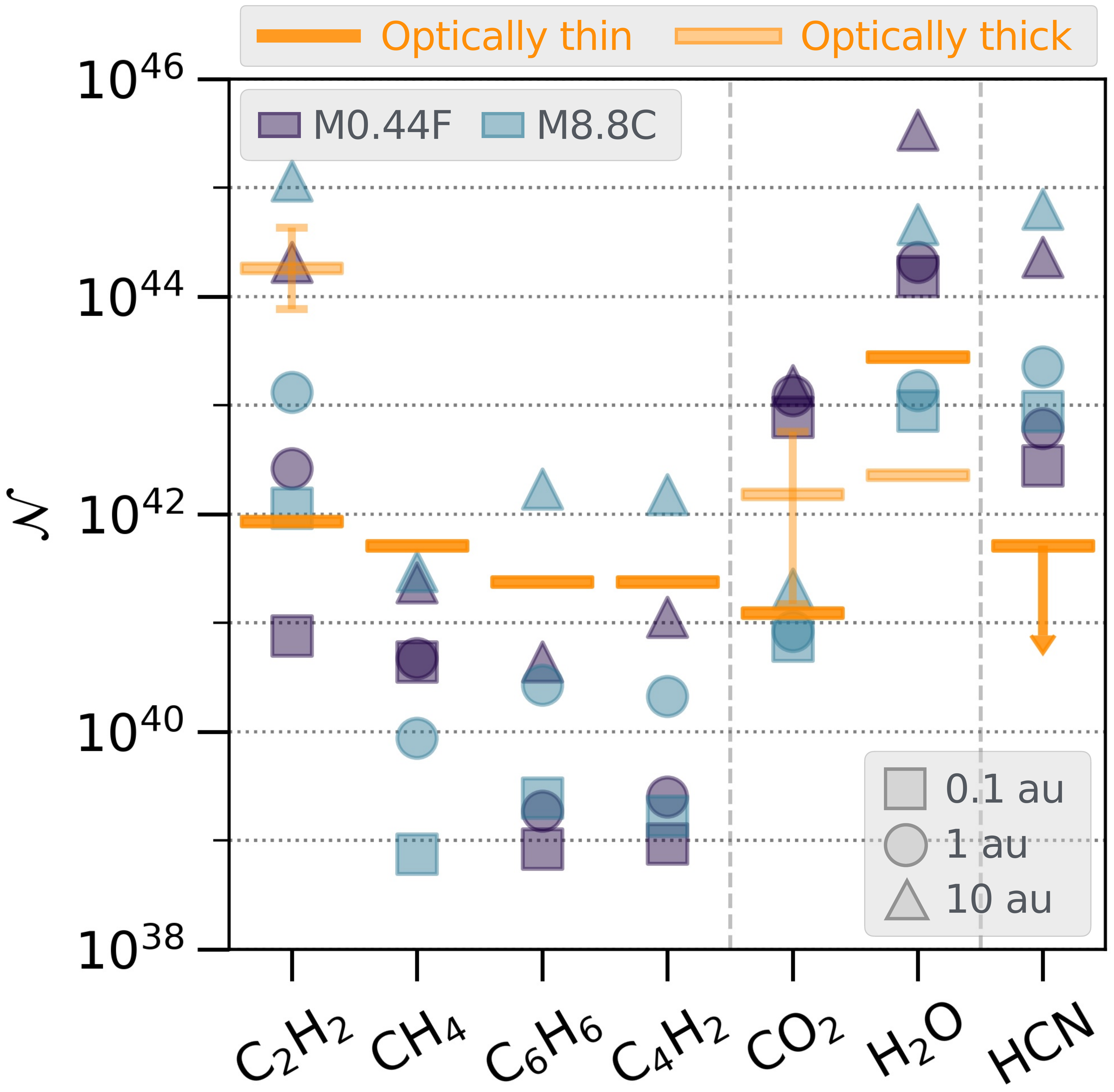}
\caption{
Total number of molecules in the IR-emitting region, estimated using eq.~\eqref{eq:tot-num-mol}, for J160532 \citep[orange horizontal lines;][]{tabone_rich_2023, arabhavi_minds_2025} and the fiducial (M0.44F) and carbon-rich (M8.8C) models (dark purple and light-blue shapes, respectively) for \ce{C2H2}, \ce{CH4}, \ce{C6H6}, \ce{C4H2}, \ce{CO2}, \ce{H2O}, and HCN.
The values estimated for our model are the result of an integration up to 0.1 (squares), 1 (circles), and 10~au (triangles).}
\label{fig:Nmol_IRemitting_M0.44F_M8.8C_J160}
\end{figure}

Although we obtain reasonable agreement with the optically thin component based on number of molecule ratios, the numbers of molecules estimated from our models do overestimate \ce{C2H2} and HCN (see top panel of Fig.~\ref{fig:TotalNmol_IRemitting}).
This motivated the question of whether it is possible to obtain a better agreement if we use a different radius of integration, given that we do not know the size of these compact disks.
To explore the sensitivity of the total number of molecules with the assumed IR-emitting area, we also estimate the total number of molecules integrated out to 0.1, 1, and 10~au, for one model (M8.8C) for which the number of molecules results have reasonable agreement with those for J160532.
Additionally, to test the possible degeneracy between disk size and C/O ratio, we include the $\mathcal{N}$ for the same sets of integration radius for the fiducial model (M0.44F).
Fig.~\ref{fig:Nmol_IRemitting_M0.44F_M8.8C_J160} presents these values and compares with the total number of molecules reported by \citet{tabone_rich_2023} and \citet{arabhavi_minds_2025} for the species presented (\ce{C2H2}, \ce{CH4}, \ce{C6H6}, \ce{C4H2}, \ce{CO2}, \ce{H2O}, and HCN).
For a solar-like C/O ratio (M0.44F), $\mathcal{N}$ for the optically thin component of \ce{C2H2} is better reproduced with a smaller radius of integration ($\sim 1$~au).
In the carbon-rich scenario (M8.8C) a radius of integration of $\sim0.1$~au better reproduces the number of molecules in the thin component.
Other hydrocarbons such as \ce{CH4}, \ce{C6H6}, and \ce{C4H2} are already well mimicked, with the full integration of the disk up to 10~au with the observed values lying close to the model values for both C/O ratios (e.g., \ce{CH4}) or in between the model predictions (e.g., \ce{C6H6} and \ce{C4H2}).

The models predict a low dependence of $\mathcal{N}_\mathrm{CO_2}$ on the radius of integration for both the M0.44F and M8.8C models.
The optically thin component of \ce{CO2} is well reproduced for all integration radii for C/O~$\sim8.8$, but overestimated for the oxygen-rich model.
The apparent convergence of $\mathcal{N}$ when increasing the integration radius suggests that the optically thick component of \ce{CO2} is not well reproduced for any of the models, being overestimated by M0.44F and underestimated by M8.8C; an intermediate C/O might achieve better agreement.
However, both models (M0.44F and M8.8C) and all integration radii are consistent with the reported uncertainties.
Figure~\ref{fig:TotalNmol_IRemitting} shows that $\mathcal{N}_\mathrm{CO_2}$ has a steplike behavior, only varying when O/H decreases, so C/O ratios of $\sim 0.88$ and $\sim 4.4$ will follow the M0.44F and M8.8C trends, respectively.
\ce{H2O} is overestimated by the fiducial model but better reproduced for C/O~$\sim8.8$ with an integration radius of $0.1-1$~au.
Species such as HCN are overproduced by at least one order of magnitude in all cases.
From these results, we see an overlap of the estimated $\mathcal{N}$ for the different radii of integration and C/O ratios.

%%%%%%%%%%%%%%%%%%%%%%%%%%%%%%%%%%%%%%%%%%%%%%%%%%%%%%%%%%%%%%%%%%%%%%

\subsection{Carbon, Oxygen, and Nitrogen Reservoirs}

Here we discuss the main carriers for carbon, oxygen, and nitrogen and how they vary when different integration radii and C/O ratios are considered. 
This is worth exploring to check whether there are additional carriers of C, O, and N that are not probed by the observations.
We focus on the effect of an increase of C/O by comparing the fiducial and solar-like model (M0.44F) with the first model that includes both oxygen depletion and carbon enrichment (M8.8C).

\begin{figure*}[ht]
\centering
\includegraphics[width=1\linewidth]{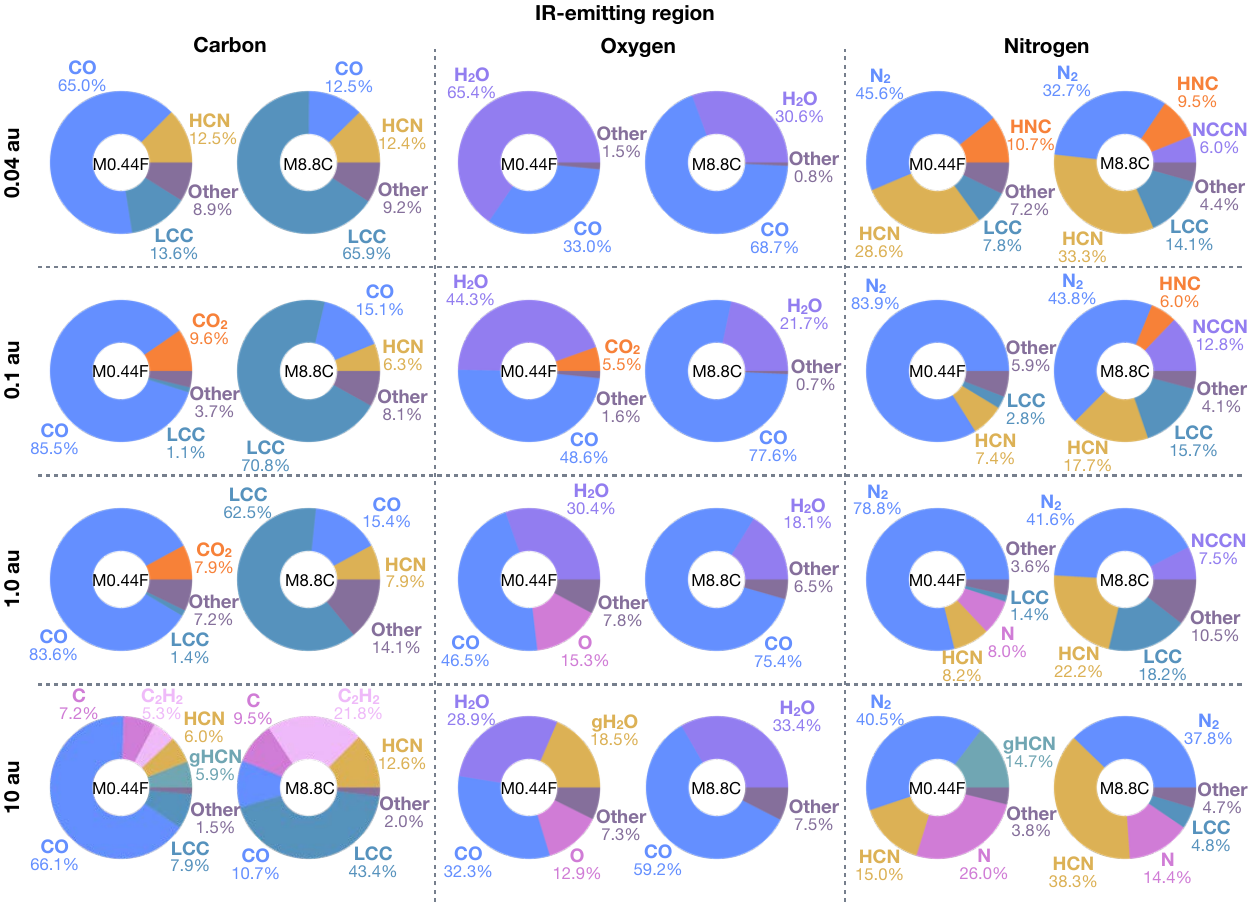}
\caption{Pie charts showing the contribution of species by molecular mass to the total budgets for carbon (left), oxygen (middle), and nitrogen (right) for two C/O ratios: $\sim 0.44$ (M0.44F) and $\sim 8.8$ (M8.8C). 
We show values at a radius of 0.04~au (innermost region of the disk) and for integration up to radii of 0.1, 1, and 10~au.
We present values for the \textit{atmosphere} only considering the IR-emitting region.
``LCC'' represents the summation of LCCs with $\geq 3$ carbon atoms with a contribution $\geq 1\%$ by molecular mass.
``Other'' represents the summation of all the species with a contribution $< 5$\%.
(The complete figure set (2 images) is available in the online article.)}
\label{fig:PieChart_IRemitting_massweighted}
\end{figure*}

Figure~\ref{fig:PieChart_IRemitting_massweighted} presents pie charts showing the percentage contribution of species to the total carbon (left), oxygen (middle), and nitrogen (right) budget for the IR-emitting region.
We consider averages over the different integration radii (0.1, 1, and 10~au) and include a comparison with the very innermost radius of our model grid (at 0.04~au).
We list individual species with more than a 5\% contribution by molecular mass.  
For species with a smaller contribution, we sum up their contribution in a category called ``Other.'' 
We have an additional category called ``LCC,'' which stands for ``long carbon chains'' (Fig.~\ref{fig:PieChart_LCC} presents the breakdown of this sector, showing in detail the most abundant species in the category). 
In many cases we find that a significant fraction of carbon is contained within multiple hydrocarbon species with three or more carbon atoms, and we group these together in a single category.

%%%%% Figure set for Figure 10 -- Begin
%\figsetstart
%\figsetnum{10}
%\figsettitle{Pie charts for the region above the $\tau=1$ surface at $14~\mu\mathrm{m}$ (disk atmosphere component) and full vertical disk extent down to the midplane (midplane component).}

%\figsetgrpstart
%\figsetgrpnum{10.1}
%\figsetgrptitle{Region above the $\tau=1$ surface at $14~\mu\mathrm{m}$}
%\figsetplot{PieChart_Atmosphere14um.pdf}
%\figsetgrpnote{
%Same as Fig.~\ref{fig:PieChart_IRemitting_massweighted}, but for the integration down to the $\tau=1$ surface at $14~\mu\mathrm{m}$ (disk atmosphere component).}
%\figsetgrpend

%\figsetgrpstart
%\figsetgrpnum{10.2}
%\figsetgrptitle{Region above the midplane ($z=0$)}
%\figsetplot{PieChart_Midplane.pdf}
%\figsetgrpnote{
%Same as Fig.~\ref{fig:PieChart_IRemitting_massweighted}, but for the integration down to the midplane (midplane component).}
%\figsetgrpend

%\figsetend

%%%%% Figure set for Figure 10 -- End

For a solar-like C/O, in the innermost region of the disk in the IR-emitting region (at 0.04~au), carbon is mostly in \ce{CO} ($\approx 65\%$), with smaller contributions from LCC and HCN (all $< 15$\%).
When we integrate up to 0.1~au, \ce{CO} remains as the main carrier ($86\%$) and the contribution from LCC drops to ($\approx 1\%$).
\ce{CO2} also emerges as a main carbon carrier here ($\approx 10$\%). 
The results out to 1~au are similar to those out to 0.1~au.
When integrating out to 10~au, because only regions where the gas temperature is higher than 200~K have been considered, the atomic carbon becomes a key carrier ($\approx 7\%$), along with \ce{C2H2}, HCN, and \ce{HCN} ice, all with a contribution lower than 6\%.
However, \ce{CO} remains the main carrier of carbon.

When C/O increases to $\sim8.8$, carbon is mainly contained in LCCs for all methods of calculation (i.e., at 0.04~au, and integrated out to 0.1, 1.0, and 10~au). 
\ce{CO} and \ce{HCN} show up as carbon carriers at 0.04~au and out to 10~au, but their individual contributions are lower than 16\%.

In the innermost region of the disk (at $0.04$~au) there is a high abundance of LCCs for both C/O~$<1$ and C/O~$>1$. 
This is also seen in \citet{kanwar_minds_2024}, who they report high column densities of hydrocarbons at small radii ($r<0.1$~au) regardless of the C/O adopted, showing that carbon enrichment may not be necessary for the efficient synthesis of hydrocarbons in the very innermost region of the disk.
Figures~\ref{fig:FAmap_LCCs_set1} and \ref{fig:FAmap_LCCs_set2} show the fractional abundance maps for some of the most abundant LCC species (\ce{C9H2}, \ce{C10H2}, \ce{CH2CCH2} ice, \ce{CH3CCH} ice, \ce{HC5N}, \ce{HC7N}, \ce{HC9N}, and \ce{HC3N} ice); see Fig.~\ref{fig:PieChart_LCC}.
The fractional abundance maps for other abundant LCC species, such as \ce{HC3N}, \ce{C3}, and \ce{C3H}, are shown in Figs.~\ref{fig:FAmap_nitrogen} and \ref{fig:FAmap_potentially_observed_set1}.
Note that the high abundance of large unsaturated hydrocarbon species is likely a carbon sink effect of the chemical network, with carbon piling up in the longest C-bearing species in the network.
This effect may be exacerbated by the fact that our network does not include destruction routes from LCCs back to \ce{C2H2} or smaller hydrocarbons.
We note here that the largest and most abundant LCCs in our models have not yet been detected, highlighting the need for future observations to test this scenario.

For oxygen, \ce{H2O} and CO are the main carriers in the inner disk ($\le 0.1~\mathrm{au}$).
In the IR-emitting region CO is the main carrier for both the oxygen-rich case (M0.44F) and the carbon-rich scenario (M8.8C).
Beyond 0.1~au, both models have different species as important oxygen carriers.
For the oxygen-rich case, atomic oxygen becomes a key main carrier when integrating up to 1 and 10~au.
Additionally, \ce{H2O} ice emerges as a main carrier when integrating over all disk extent.
None of this happens for the carbon-rich case M8.8C, which has \ce{H2O} and CO as key carriers across all methods of integration.
When integrating up to 0.1~au, \ce{CO2} emerges as an important carrier ($\sim 6\%$).

For nitrogen, in most cases \ce{N2} is the main carrier across the models.
Within 0.1~au, nitrogen is mainly in the form of \ce{N2}, HCN, HNC, and NCCN (in the carbon-rich model only), and \ce{HC5N} in the innermost disk (at 0.04~au) for M8.8C.
Beyond 0.1~au, HCN and \ce{N2} remain as main carriers, although atomic nitrogen becomes a key contributor to the nitrogen budget for the oxygen-rich case, as well as more complex molecules such as cyanopolynes (e.g., \ce{HC3N} and \ce{HC9N}) in the carbon-rich model.
For the outer-disk regions ($>1~\mathrm{au}$), nitrogen is mostly in its molecular and atomic form, with the appearance of HCN ice in the oxygen-rich model.
The main carriers of nitrogen between the fiducial case and carbon-rich case do not differ significantly in nature and usually differ only in percentage contribution. 
For example, HCN has an increased contribution in the inner disk ($\le 0.1$~au) in the carbon-rich case versus the oxygen-rich case ($\approx 33$\% vs. $\approx 29$\% at 0.04~au).

Our models predict a set of hydrocarbons and oxygen- and nitrogen-bearing species as key and main carriers for the C, O, and N reservoirs.
In particular, CO, \ce{H2O}, and \ce{N2} remain as key carriers for all radii of integration for both values of C/O.
From these species, only \ce{H2O} has been detected in J160532 and ISO-ChaI~147.
Other species such as \ce{C2H2}, \ce{CO2}, and HCN also become key carriers under specific conditions. 
For example, \ce{C2H2} is a key carbon carrier when the full disk extent is considered (up to 10~au) for both M0.44F and M8.8C models; however, when C/O increases from 0.44 to 0.88, its contribution to the carbon budget increases from $\approx5\%$ to $\approx22\%$, which suggests that an enhanced C/O ratio is needed to explain the high abundances of \ce{C2H2} in the three disks.
However, \ce{CO2}, an oxygen-bearing species detected in J160532, ISO-ChaI~147, and Sz28, is only a key carrier for a solar-like C/O and smaller radius of integration ($\sim0.1-1$~au).
Finally, our results suggest that other carbon carriers such as NCCN and LCCs (see Fig.~\ref{fig:PieChart_LCC}) would be good targets to look for in the IR spectra, as they are key carriers of carbon and not probed by current observations.
Additionally, these species are likely to be good tracers of a an enhanced C/O ratio.
For example, \ce{C9H2} appears as a key contributor to LCCs when C/O~$\sim0.44$, and increases to $\approx 52\%-69\%$ when C/O increases to $\sim 8.8$ (when integrating up to 1~au). 
We also note with caution here that the chemistry in our network does not go to saturation for species with three or more carbons; thus, the unsaturated LCC species act as carbon sinks in the network.

%%%%%%%%%%%%%%%%%%%%%%%%%%%%%%%%%%%%%%%%%%%%%%%%%%%%%%%%%%%%%%%%%%%%%%

\subsection{Effects of C/O on Nitrogen-bearing Species}
In the different scenarios that we have explored, we only vary the C/O ratio. 
However, from the fractional abundance maps (Fig.~\ref{fig:FAmap_nitrogen}), column density profiles (Fig.~\ref{fig:ColdensProfile_IRemitting}) and number of molecules versus C/O ratio profiles of HCN, \ce{NH3}, and \ce{HC3N} in Fig.~\ref{fig:Nmol-C2Oratio_oxygen_nitrogen}, we can see that nitrogen-bearing species do respond to C/O variations even though N/H is constant across the different models. 

Overall, the total number of molecules of \ce{NH3} increases when O/H decreases, and it slightly decreases when C/H increases. 
For all C/O ratios, the number of \ce{NH3} molecules varies by less than one order of magnitude.
For the case of \ce{N2}, carbon enrichment and/or an oxygen depletion result in a constant $\mathcal{N}$ with respect to the fiducial model (M0.44F).

\cite{wei_effect_2019} suggest that the \ce{HCN} abundance is a good tracer of the effect of carbon grain destruction within the soot line. 
From the fractional abundance maps in Fig.~\ref{fig:FAmap_nitrogen} we see that the \ce{HCN} midplane abundance increases when C/H increases, which is consistent with the results from \cite{wei_effect_2019}. 
This increase in HCN with carbon enhancement is also evident in Figs.~\ref{fig:ColdensProfile_IRemitting} and \ref{fig:Nmol-C2Oratio_oxygen_nitrogen}; however, as found for many species, the biggest relative change in HCN total number of molecules comes in the first two perturbations, i.e., for a C/O of 0.47 and 0.87.

\ce{NH3} is a potentially observable molecule, perhaps detected toward Sz28 in the reanalysis of the data by \citet{kaeufer_disentangling_2024}. 
Additionally, to compare with other published literature, albeit for T~Tauri stars, \cite{pontoppidan_nitrogen_2019} present a comparison between the \ce{NH3}/\ce{H2O} and \ce{HCN}/\ce{H2O} ratios in the inner disk of the protoplanetary disks around three solar-mass stars. 
They show that the \ce{HCN} abundance is expected to be between one and two orders of magnitude larger than the \ce{NH3} abundance.
Potential future diagnostics of C/O in the inner regions of disks should include \ce{NH3}.
Further, the models suggest that the ratio of \ce{NH3} to nitriles may depend on the X-ray ionization rate in the disk, with more carbon driven into nitriles for a higher X-ray ionization rate.
We intend to explore these potential diagnostics in future work.

%%%%%%%%%%%%%%%%%%%%%%%%%%%%%%%%%%%%%%%%%%%%%%%%%%%%%%%%%%%%%%%%%%%%%%

\subsection{Chemical Pathways}
In general, we find the same active chemical pathways as discussed in \citet{walsh_molecular_2015} and \citet{kanwar_hydrocarbon_2024}.
Note that we use the full {\sc{rate12}} version of the UMIST database for astrochemistry, whereas \cite{kanwar_hydrocarbon_2024} adopt a subset of those reactions in their version of the network.
We see that the main chemistry routes between an oxygen-rich and carbon-rich scenario are not so different.
We noted that some formation and destruction routes are different, for example, oxygen-related reactions are not as efficient in the carbon-rich model as in the oxygen-rich one, due to the high oxygen depletion by a factor of 100.
However, other pathways are preferred to form and destroy the same species.
\citet{kanwar_hydrocarbon_2024} reached a similar conclusion albeit for a C/O ratio of 2 when comparing with a model assuming C/O~=~0.45.
To better illustrate this, see Figs.~4 and 5 from \citet{walsh_molecular_2015} and Figs.~3 and 6 in \citet{kanwar_hydrocarbon_2024}.
The main variation we see is in the percentage of the dominance of the different reactions, but there are few new chemical pathways for formation or destruction at higher C/O ratios.

As an example, Fig.~\ref{fig:chemnetwork_C6H6_C4H2_M0.44F_M8.8C} illustrates the main formation pathways for \ce{C6H6} and \ce{C4H2} for both the M0.44F and M8.8C models.
In this, we see a bottom-up scheme, where long hydrocarbons, such as \ce{C6H6} are formed starting with the reaction between atomic carbon and \ce{H3+}.
Note that we only list reactions that have a contribution of at least 10\% to the rate of formation of the different species.
As mentioned before, the main chemistry routes do not change significantly as the C/O increases.
In the oxygen-rich scenario (M0.44F), chemistry involving CO is important for the formation of \ce{C+}, whereas in the carbon-rich case (M8.8C) this reaction is replaced by reactions involving \ce{CH3} and \ce{C3}.

%%%%%%%%%%%%%%%%%%%%%%%%%%%%%%%%%%%%%%%%%%%%%%%%%%%%%%%%%%%%%%%%%%%%%%

\subsection{Physical Scenarios That Can (Re)set C/O in the Inner Disk and Implications for Planet Formation}\label{sec:physical-scenarios}

We have shown that the observations toward J160532 suggest a higher C/O ratio in the inner disk of this source than that for ISO-ChaI~147 when estimations of $\mathcal{N}$ from observations are available for both sources (e.g., \ce{C6H6} and \ce{C4H2}; see bottom panel of Fig.~\ref{fig:TotalNmol_IRemitting}).
In particular, we found that ISO-ChaI~147 could be consistent with solar, or a moderate enhancement in C/O.
However, the picture for J160532 is not as simple, and it strongly depends on the species considered, ranging from solar-like values to highly enhanced C/O ratios.
We reiterate again, that our results apply only for a model that has an assumed stellar X-ray luminosity of $10^{29}$~erg~s$^{-1}$, which is in the middle of the observed range for VLMSs \citep{preibisch_evolution_2005}.
Here we briefly discuss the different mechanisms and events that could lead to an increase in the C/O ratio, due to both carbon grain destruction releasing carbon into the gas phase and oxygen depletion. 

Several processes have been proposed to increase the C/O ratio as a consequence of the disk physical and chemical evolution and planet formation, which in synergy increase C/H and decrease O/H.
One possible scenario is the presence of the soot line, which is a region within which polycyclic aromatic hydrocarbons (PAHs) and refractory carbon grains are destroyed via thermal reactions, resulting in the sublimation of these species, which injects excess carbon into the gas phase \citep{kres_soot_2010, vant_hoff_carbon_2020}. 
Note that PAHs toward disks around VLMSs have not been detected with JWST. 
This is likely related to the weaker UV radiation field in these sources that is unable to excite PAHs, compared with T~Tauri and Herbig Ae/Be stars \citep{geers_C2D_2006, stum_JWST_2024}.

Mechanisms to inject excess carbon into the gas phase are (1) carbon being released from the grains to the gas phase by the collisions of hot oxygen atoms with carbonaceous dust grains \citep{lee_solar_2010}, which mainly happens in the disk atmosphere, but the effect is spread through the disk by vertical mixing; (2) oxidation of carbon dust \citep{finocchi_chemical_1997, gail_radial_2001, gail_radial_2002}, which is the reaction between carbon-rich grains and oxygen-bearing species, such as \ce{OH}; (3) carbon released from PAHs through destruction by X-ray and high-energy photons \citep{siebenmorgen_destruction_2010}; and (4) carbon-rich grains being photochemically destroyed by far-UV photons \citep{atala_vacuum_2015, anderson_destruction_2017}. 
The sublimation of refractory carbon is proposed to take place in regions with gas temperatures $ \gtrsim 500~\mathrm{K}$ \citep{li_earth_2021}.
Reactions of carbon grains with atomic H can also liberate carbon into the gas phase \citep[see][]{lenzuni_dust_1995}.
In our M~dwarf physical model (see Fig.~\ref{fig:disk-structure}) the gas temperature in the innermost region ($\le 0.1$~au) is $\gtrsim 500~\mathrm{K}$, as is that in the inner-disk atmosphere ($z/r \ge 0.1$).
Hence, although we assume that carbon grain sublimation has occurred globally in the disk for simplicity, the temperature at which this effect should take place is also aligned with the region of enhanced hydrocarbon formation. 

However, there are mechanisms that may destroy carbonaceous grains over a larger spatial area than suggested by the current gas temperature.
Episodic accretion events in the early stages of disk formation and evolution due to an increased accretion rate will increase both the gas and dust temperature, pushing outward the snowline and the soot line \citep{Audard_episodic_2014, vant_hoff_carbon_2020}. 
Carbon grain destruction is an irreversible process; hence, carbon grains could have been destroyed in an earlier evolutionary phase when episodic accretion is active.
That the inner disks could be affected by such processes is hinted at in recent JWST results.
Studies comparing JWST and Spitzer observations of disks around T~Tauri stars have signs of highly dynamic inner disks, with the observations suggesting a temporal variability in the molecular line emission of species such as \ce{C2H2}, \ce{HCN}, \ce{H2O}, \ce{OH}, and \ce{CO2} \citep{schwarz_reveals_2024, romero-mirza_spectroscopy_2024}. 
A similar conclusion was made by \citet{kospal_jwst_2023} when comparing the strength of the molecular emission during the quiescent epoch of EX Lup with the outburst event studied by \citet{banzatti_exlupi_2012}.

\cite{wei_effect_2019} investigated this process of carbon grain destruction in a chemical model of a protoplanetary disk around a T~Tauri star and proposed that the midplane abundances of HCN, \ce{c-C3H2}, and LCCs at 1~au are significantly enhanced when using an increased C/O of 1.7 owing to injection of carbon via carbon grain destruction. 
Other species, such as \ce{CH4} and \ce{C2H2}, also show a positive response to carbon injection, although the contrast between the oxygen-rich and carbon-rich model is less pronounced.
This is accompanied by a decrease in \ce{CO2}, which is a similar trend to that found here (see Figs.~\ref{fig:Nmol-C2Oratio_hydrocarbons} and \ref{fig:Nmol-C2Oratio_oxygen_nitrogen}).
\cite{tabone_rich_2023}, \cite{van_dishoeck_diverse_2023}, and \cite{arabhavi_abundant_2024} proposed this scenario to explain the J160532 and ISO-ChaI~147 observations, suggesting that the increase in \ce{C2H2} abundance is occurring within the location of the ``soot line," where hydrocarbon grains are destroyed by the sublimation front. 
Here we tested this scenario by adopting the same assumption of carbon enrichment as \cite{wei_effect_2019}. 
We mimicked the effect of carbon grain destruction in the cases M0.88C, M8.8C, and M88C. 
However, in our models, we find that carbon grain destruction is not always necessary to explain the high number of molecules of many hydrocarbons in the inner regions of the disk and that oxygen-depletion-only models are also a possible explanation (see below).
This is supported by the fact that some $\mathcal{N}$ ratios, such as \ce{H2O}/\ce{CO2} and HCN/\ce{CO2}, show that solar and oxygen depletion models (M0.44, M4.4, and M44) have similar results to their counterparts including carbon enrichment (M0.88C and M8.8).
Note that this might also be driven by the strong sensitivity of \ce{CO2} to O/H.

A second mechanism to increase C/O is oxygen depletion via the trapping of water-ice-coated pebbles beyond the water snowline.  
Models have shown that disk substructures such as gaps can trap icy pebbles in the outer disk, resulting in a decrease of the O/H ratio in the inner region of the disk \citep{kalyaan_linkin_2021, kalyaan_effect_2023}.
This mechanism was also proposed by \cite{tabone_rich_2023}, \cite{kamp_chemical_2023}, and \cite{van_dishoeck_diverse_2023} to reproduce the high \ce{C2H2}/\ce{H2O} and \ce{HCN}/\ce{H2O} ratios inferred for disks around VLMSs by \citet{tabone_rich_2023}, attributing this to a low water abundance in the inner disk. 
We tested this scenario by depleting the available oxygen in the models by factors of $10$ (M4.4 and M8.8C) and $100$ (M44 and M88C).
We find that our $\mathcal{N}$ are also sensitive to oxygen depletion, with a particularly strong trend seen in the \ce{C6H6}/\ce{CO2} and \ce{C4H2}/\ce{CO2} ratios, suggesting oxygen depletion factors as high as 10, with a preference for a depletion of $\sim 10$ for J160532 and no oxygen depletion for ISO-ChaI~147 (see Fig.~\ref{fig:TotalNmol_IRemitting}).
However, again, this strongly depends on the species we are focusing on and the sensitivity of our model results to the assumed stellar X-ray luminosity.
For example, \ce{C2H2}/\ce{CO2} suggests that neither oxygen depletion nor carbon enrichment is necessary to fit the estimations for J160532.
Icy pebble trapping in the outer disk may explain these high oxygen depletion values for J160532.
However, note that the new estimations for the \ce{H2O} column density in J160532 by \citet{arabhavi_minds_2025} suggest a ``less water-depleted'' inner disk, with \ce{C2H2}/\ce{H2O} and \ce{HCN}/\ce{H2O} ratios one order of magnitude higher than the upper limit inferred by \citet{tabone_rich_2023}.

In \citet{kanwar_minds_2024} models of the inner-disk composition are presented for two scenarios: one for their canonical C/O ratio of 0.45, and another in which oxygen has been depleted by a factor of $\approx 4$ (C/O~$=2$).
They calculated both the maximum column density reached for species of interest and the column density at 0.1~au, taking into consideration the full disk column.
They find that \ce{C4H2} is the only maximum column density that increases significantly when oxygen is depleted, with an increase of a factor of $\sim 300$.
They also find that their maximum column density of \ce{OH} decreases by a factor of $\sim 50$.
The maximum column densities of all other species increase or decrease by a factor of a few at most.
However, \citet{kanwar_minds_2024} do find that the column density at 0.1~au is very sensitive to C/O.
The column densities of \ce{CH3}, \ce{C2H2}, \ce{C2H6}, \ce{C4H2}, \ce{C3H4}, \ce{C6H6}, and \ce{HCN} all increase by several orders of magnitude when oxygen is depleted.
In this work, we do not find the same sensitivity for the maximum column densities and the column densities at 0.1~au for the hydrocarbons, with the canonical and C/O~$=4.37$ results differing by no more than a factor of 10.
Differences in the adopted disk structure could explain the discrepancies. 
It is also possible that differences are due to the adopted chemistry (e.g., we use a gas-grain model) and how it is calculated (e.g., we calculate the chemistry in a time-dependent manner as opposed to assuming a steady state).
In addition, we see that smaller hydrocarbons are more rapidly driven into the LCC reservoir in our model, which is likely related to the carbon sink effect.

The scenarios discussed above can result in a high C/O ratio in the inner region of the disk. 
However, the C/O ratio (and C/H ratio) can increase as a consequence of additional physics in the disk. 
\citet{booth_chemical_2017} and \citet{booth_pebble_2019} have shown that high metallicities and, in particular, C/O ratios can be reached as a consequence of gas accretion and the growth and radial drift of icy pebbles in the disk \citep[see also][]{kalyaan_linkin_2021, kalyaan_effect_2023}.
\citet{mah_close-in_2023} also suggest that the C/O ratio in the inner region of disks around VLMSs can increase from substellar to superstellar more rapidly than (after $\sim 2~\mathrm{Myr}$), and up to higher values than, their higher-mass counterparts, T~Tauri and Herbig~Ae disks. 
This is a consequence of the close-in snowlines of water and \ce{CO2} in the disks around low-mass stars coupled with the short distances and faster travel speeds for the superstellar C/O material to travel into the inner disk. 
These scenarios are mimicked by models M8.8C and M88C, which include both carbon enrichment and oxygen depletion.
However, our results suggest that the trends in the observations can be reproduced with oxygen depletion or carbon enrichment alone.
Furthermore, we see that for ISO-ChaI~147, the \ce{C2H2}/\ce{CO2} ratio in Fig.~\ref{fig:ColDensProf_ratio-Nmol_ratio-vs_int-radius} suggests that C/O~$>1$ is not always necessary, since the M0.44F and M0.88C results still reproduce observations in the innermost disk ($\lesssim0.1~\mathrm{au}$) and when only regions with gas temperature $>200$~K are included.

Another mechanism to increase the C/O ratio is the destruction of gas-phase CO by dissociative ionization with \ce{He+}, combined with radial drift and gas advection, which was recently studied by \citealt{sellek_chemical_2025}.
In this work, they find that this process would deplete O and liberate enough C to be incorporated into hydrocarbons (see \citet{sellek_chemical_2025} for more details about the reactions).
They find that this is more efficient and more likely to happen in a disk with a high ionization rate ($\gtrsim 10^{-17}~\mathrm{s^{-1}}$).
Another feature of the disk that would increase the chances of CO depletion is its size; compact disks are more likely to have efficient dust and ice trapping \citep[e.g.,][]{banzatti_JWST_2023}, which, as we discussed before, will also increase the C/O ratio.
Future studies on the effects of the ionization rate in the formation of hydrocarbons in disks around VLMSs are needed to test these scenarios.

Regarding planet formation, the atmospheric composition of exoplanets is directly linked to the place where they are formed \citep[e.g.,][]{oberg_astrochemistry_2021,oberg_protoplanetary_2023}. 
Planets can be formed early or late in the disk lifetime. 
If they form early in the inner disk, most of the material available to be accreted will have pristine compositions, similar to the host star (i.e., initial abundances of our models), whereas if they form late, the available material will be more similar to the abundances of our results. 
Therefore, understanding the dominating chemistry within the planet-forming region is key to also comprehending the diversity of chemical compositions of exoplanetary atmospheres. 
The scenarios explored in this work, mimicking the effects of carbon enrichment and oxygen depletion, will have a direct impact on the atmospheric composition of planets forming in this region.

%%%%%%%%%%%%%%%%%%%%%%%%%%%%%%%%%%%%%%%%%%%%%%%%%%%%%%%%%%%%%%%%%%%%%%

\section{Summary} \label{sec:conclusion}

Recent JWST observations of disks around the low-mass stars reported unexpectedly high abundances of hydrocarbons species in the inner disk \citep[e.g.,][]{tabone_oh_2021, arabhavi_abundant_2024, kanwar_minds_2024}, results that are not fully consistent with the currently available models. 
Motivated by these observations, we revisited the M~dwarf chemical models by \cite{walsh_molecular_2015} and explored the different physical scenarios proposed by \cite{tabone_rich_2023}, \cite{van_dishoeck_diverse_2023} and \cite{arabhavi_abundant_2024} that can lead to oxygen depletion (e.g., icy pebble trapping beyond the snowline) and carbon enrichment (e.g., destruction of carbon grains) and that would increase gas-phase C/O to possibly explain the observations. 
The C/O ratios ranged from solar values ($\mathrm{C/O}=0.44$) to an extreme case combining carbon enrichment and extreme oxygen depletion ($\mathrm{C/O}= 87.47$). 
Our models indicate that reaching high abundances and numbers of hydrocarbon molecules generally requires a carbon enrichment of at least a factor of 2 and/or oxygen depletion by a factor of 10 or more.
In general, we find that the peak abundance, column density, and number of molecules reached by both hydrocarbons (\ce{C2H2}, \ce{CH4}, \ce{C6H6}, \ce{C4H2}, \ce{C2H4}, \ce{C3H4}, \ce{C2H6}, and \ce{CH3}) and nitriles (HCN, \ce{HC3N}, and \ce{CH3CN}) are sensitive to the C/O ratio, with both oxygen depletion and carbon enrichment increasing the values of the peak abundance and column density in the IR-emitting region.
However, the largest relative increases in column density usually occur at moderately enhanced C/O ratios (up to $4.4$). 
Increasing the C/O ratio to higher values (up to $87.47$) does not always result in further enhancement.
Hence, the chemistry becomes limited by C/H, rather than O/H, and higher depletion factors for oxygen are unlikely to change the results. 
This may not be true for higher C/H values, although we are already assuming that all carbon that is present in the disk is available in the gas phase by assuming that all carbon grains have been destroyed. 
To increase C/H further would require some additional mechanism to preferentially deliver carbon, and carbon alone, to the inner disk.

The peak abundance and column densities and number of molecules of O-bearing species (e.g., OH and \ce{H2O}) scale almost linearly with oxygen depletion, and their behavior is not as sensitive to C/H.
This is true even for the carbon-bearing species CO and \ce{CO2}, whose chemistry is limited more by O/H than C/H.

Overall, we find that the inner-disk chemistry of oxygen- and carbon-bearing species in disks around M~dwarf stars can be strongly affected by the C/H and O/H ratios and that nitrogen-bearing molecules are also affected, even though we have used a constant N/H over all models.

Based on the total number of molecules, $\mathcal{N}$, no single model (i.e., value of C/O) completely fits the trends in the results reported in \citet{tabone_rich_2023}, \citet{arabhavi_abundant_2024}, \citet{kanwar_minds_2024}, and \citet{arabhavi_minds_2025}.
However, some molecules do achieve $\mathcal{N}$ close to those inferred from the observations. 
We find that using the ratio of number of molecules with respect to \ce{CO2} may allow a better discrimination of the underlying C/O in the inner disk.
Considering the ratios calculated for the IR-emitting region, we find that a wide range of C/O values are required to reproduce the observed ratios depending on species and source.
However, a comparison of the trends in the modeled and observed ratios of total number of hydrocarbons relative to \ce{CO2} suggests that the C/O ratio might not be the same for the inner disk of J160532 and ISO-ChaI~147.
This suggests that the systems observed so far show a range of C/O and metallicities even if they present similar species detected in their spectra.
Given that the sources discussed here are not unique, our findings indicate that an enhanced C/O could explain the observations of not only J160532, ISO-ChaI~147, and Sz28 but also the successful detection of abundant hydrocarbons in the disks around other VLMSs \citep[e.g.,][]{arabhavi_very_2025, grant_transition_2025}.
Note that we use a generic inner-disk model around an M~dwarf star with a single value for X-ray luminosity at the middle of the observed range ($10^{29}~\mathrm{erg~s^{-1}}$).
Hence, we cannot rule out any degeneracies that may arise in our results between the assumed accretion rate and X-ray luminosity and C/O ratio. 
We plan to investigate this in future work. 

Finally, as the material available in disks will determine the atmospheric composition of planet formation therein, it is expected that the scenarios explored in this work are setting the initial conditions for planet formation.

%%%%%%%%%%%%%%%%%%%%%%%%%%%%%%%%%%%%%%%%%%%%%%%%%%%%%%%%%%%%%%%%%%%%%%

%\begin{acknowledgments}
\section*{Acknowledgments}

J.K.D.-B. acknowledges support from the Science and Technology Facilities Council via a doctoral training grant (grant No. ST/Y509711/1).

C.W.~acknowledges financial support from the Science and Technology Facilities Council and UK Research and Innovation (grant Nos. ST/X001016/1 and MR/T040726/1).

EvD acknowledges funding from the European Research Council (ERC) under the European Union’s Horizon 2020 research and innovation program (grant agreement No. 101019751
MOLDISK) and the Danish National Research Foundation through the Center of Excellence “InterCat” (Grant agreement No. DNRF150).
%\end{acknowledgments}

\bibliography{bibliography.bib}{}
\bibliographystyle{aasjournal}

\appendix
\counterwithin{figure}{section}

%%%%%%%%%%%%%%%%%%%%%%%%%%%%%%%%%%%%%%%%%%%%%%%%%%%%%%%%%%%%%%%%%%%%%%

\section{Caveats}\label{appendix:caveats}
There are some model assumptions and caveats we need to take into consideration when interpreting our results. 
First of all, our disk physical model is not self-consistent. 
When we change the C/O ratio, we assume a constant temperature structure across our models.
We assume this for simplicity, so we are considering only the chemical changes with increasing C/O rather than more complex changes that may occur in the balance between heating and cooling when the metallicity of the gas changes.
Note that this assumption is also made in the models presented in \citet{kanwar_minds_2024}, where they explore C/O ratios of 0.45 and 2.
Additionally, we are considering a smooth disk, without gas or dust substructures. 
However, in the models adopting different values of C/O, we are implicitly mimicking the effects that those substructures have, i.e. dust trapping, on the chemistry.
We also do not assume that the disk model has an inner rim.  
If a warm inner rim is the origin of the mid-IR line emission as suggested for the case of EX~Lupi in \citet{woitke_2d_2024}, the inner rim would be exposed directly to radiation from the central star and may reveal more of the warm molecular reservoir here, which could lead to the high number of molecules estimated from observations.
Additionally, our model has a temperature inversion due to viscous heating in the midplane, which increases the abundances of hydrocarbons in that region (see Figs.~\ref{fig:FAmap_hydrocarbon_set1}, \ref{fig:FAmap_hydrocarbon_set2}, and \ref{fig:Nmol-C2Oratio_hydrocarbons}). 
However, note that our main analysis focuses only on the IR-emitting region.
Nonetheless, it is intended to test the inner-disk chemistry in a passively heated disk in future work.

In this work, we use a generic M~dwarf disk model, which means that we are not modeling specifically the star-disk systems studied by \cite{tabone_rich_2023}, \cite{arabhavi_abundant_2024}, \cite{kanwar_minds_2024}, or \cite{kaeufer_disentangling_2024}. 
However, our M~dwarf model properties are similar enough to the stellar and physical properties of J160532, ISO-ChaI~147, and Sz28 (see Tables \ref{table:model-properties} and \ref{table:Observed_sources_properties}) to allow a comparison with the data toward these sources. 
Nevertheless, some differences may be responsible for some of the discrepancies between the model total number of molecules and those derived from the observations.
For example, perhaps the disk mass of our model is too low, and that may explain why we underpredict $\mathcal{N}$ in the IR-emitting region. 
However, the total disk gas mass in our physical model is almost $0.6~M_\mathrm{Jup}$ out to 10~au.
The disks around the low-mass stars J160532, ISO-ChaI~147, and Sz28 have gas masses of $0.2~M_\mathrm{Jup}$ \citep{tabone_rich_2023}, $1.05~M_\mathrm{Jup}$ \citep{arabhavi_abundant_2024}, and $0.08~M_\mathrm{Jup}$ \citep{kanwar_minds_2024, kaeufer_disentangling_2024}, respectively.
\citet{franceschi_minds_2024} estimate a total warm \ce{H2} gas mass of J160532 of $2.3\times 10^{-5}~M_\mathrm{Jup}$, confined to the emitting area within a radius of 0.033~au.
Using our disk surface density at 0.04~au and assuming an emitting area of 0.04~au, we calculated a warm gas mass of $4\times 10^{-5}~M_\mathrm{Jup}$, which is only a factor of two higher than that inferred for J160532.
Hence, our model disk has a higher mass than either J160532 and Sz28. 
In addition, the mass distribution in these two low-mass sources may be different, with more mass concentrated in the innermost disk than the model we have adopted here.

A potential explanation for the discrepancy, in particular, in the difference in $\mathcal{N}$ of HCN between our estimations and observations in J160532 is that we could be overestimating the X-ray ionization rate throughout the disk.
An increase in the X-ray ionization rate would have the effect of driving more carbon into the nitriles \citep{walsh_molecular_2015, kanwar_hydrocarbon_2024}.
Note that our results only apply to a source with a single value of X-ray luminosity at the middle of the observed range for VLMSs, and any further conclusion would require new models to test these scenarios. 
\citet{pascucci_atomic_2013} estimate  $L_\mathrm{X}\sim6.3\times10^{28}~\mathrm{erg~s^{-1}}$ for Sz28, which is a factor of only $\approx 1.6$ weaker than the X-ray luminosity of our model ($\sim10^{29}~\mathrm{erg~s^{-1}}$).
However, it was not possible to compare our model results with $\mathcal{N}$ ratios for that disk, as there is no estimation of $\mathcal{N}_\mathrm{CO2}$ available.
Our model numbers of molecules achieve reasonable agreement with those derived for J160532 and ISO-ChaI~147 (depending on the species), which could be due to the similar X-ray luminosity for both sources ($\approx 6.3\times 10^{28}~\mathrm{erg~s^{-1}}$ and $\lesssim 1.1\times 10^{29}~\mathrm{erg~s^{-1}}$, respectively) to that assumed in our model ($10^{29}~\mathrm{erg~s^{-1}})$.
Our preliminary tests indicate that even lower X-ray luminosities increase hydrocarbon abundances.
A full investigation of these effects will be explored in future work.

The chemical network used in this work is extensive, includes both grain- and gas-phase reactions, and has been supplemented with reactions important in hot dense gas, e.g., collisional processes, neutral-neutral reactions that possess a large barrier, and reactions with excited \ce{H2}.
The inclusion of gas-grain chemistry is important in disks around such low-mass stars because they are colder than T~Tauri disks and their snowlines will lie much closer to the star. 
We can see evidence of the importance of grain-surface chemistry in the depletion region that emerges around the \ce{H2O} snow surface and that is also affected by the C/O ratio.

However, the chemistry follows a bottom-up scheme in that we can only build larger hydrocarbons from smaller species. 
Most of the column densities and numbers of molecules calculated for the IR-emitting region are lower than those derived from observations (e.g., \ce{CH4}, \ce{C6H6}, \ce{C2H4}, and \ce{C3H4}) which could be explained with the inclusion of a top-down chemical scheme, i.e., allowing the injection of carbon and hydrocarbon fragments from the breakdown of larger species, e.g., PAHs.

Finally, we are comparing our model results with numbers of molecules typically estimated using 0D slab models \citep{tabone_oh_2021, kanwar_minds_2024}.
Additionally, as discussed by \citet{van_dishoeck_diverse_2023}, \citet{kamp_chemical_2023}, and \citet{arulanantham_jdisc_2025}, 0D slab models are a simplified approximation for the abundance retrieval, assuming a constant temperature and column density for each molecule. 
However, the physical structure from our model and the abundance map for the different species show that the gas temperature and density of the region where each molecule is abundant are unlikely to be represented by one value only, with emission likely arising over a radial and vertical gradient of the disk \citep{romero-mirza_retrieval_2024}. 
Thus, this is another caveat to take into account when comparing between our model results with the column densities and number of molecules reported by the literature based on observational data.
For instance, the emitting area is well constrained for optically thick emission of \ce{C2H2} and \ce{CO2}, which means that it is likely not appropriate to compare these observed values with the results over the full radial range since the disk-averaged column density results are likely not representative. 
For this reason, for further comparison, we include the number of molecules as a function of C/O for the innermost region of the disk model at 0.4, 0.1, and 1~au in Fig.~\ref{fig:Nmol_vs_CO2_innermost}, which may better represent the emitting area estimated from the observations.

However, retrieval models are becoming more complex. 
For example, \citet{kaeufer_disentangling_2024} use Bayesian analysis and multiple 1D emission components in their abundance retrieval method and estimate higher column density values than those reported by \citet{kanwar_minds_2024} for the same source (Sz28). 
Future more sophisticated methods for abundance retrieval from observations using 2D models may use our models for comparison.
Alternatively, generating directly the IR spectra from our 2D models, as done with other chemical models such as \texttt{ProDiMo} \citep[e.g.,][]{kanwar_hydrocarbon_2024} and \texttt{DALI} \citep[e.g.,][]{tabone_ohIII_2024}, would provide another approach for interpreting observations.

%%%%%%%%%%%%%%%%%%%%%%%%%%%%%%%%%%%%%%%%%%%%%%%%%%%%%%%%%%%%%%%%%%%%%%

\section{Vertical Distributions}\label{appendix:vertical_distribution}
\begin{figure*}[ht]
\centering
\includegraphics[width=1\linewidth]{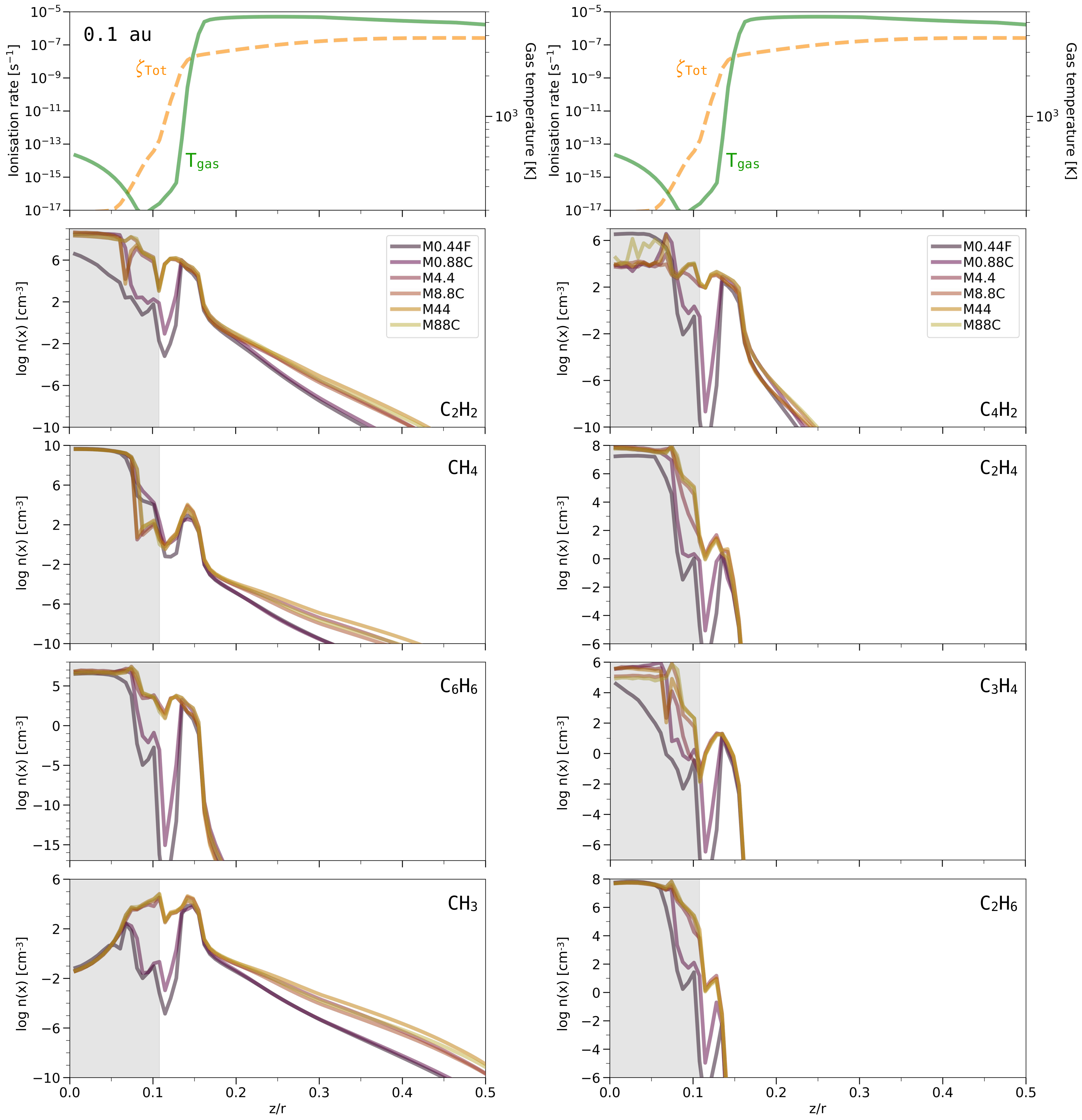}
\caption{
First row: vertical distribution of the ionization rate (orange dashed curve) and gas temperature (green solid curve) at 0.1~au.
Second to fifth rows: vertical distribution of the total abundance of the hydrocarbons: \ce{C2H2}, \ce{CH4}, \ce{C6H6}, \ce{CH3}, \ce{C4H2}, \ce{C2H4}, \ce{C3H4}, and \ce{C2H6}.
The different colors represent the different models.
The shaded region represents the $z/r$ at which the dust photosphere at $14~\mu\mathrm{m}$ is defined at that radius $r$.}
\label{fig:TA_Tgas_ionizationRate_0.1au_hydrocarbons}
\end{figure*}

\begin{figure*}[ht]
\centering
\includegraphics[width=1\linewidth]{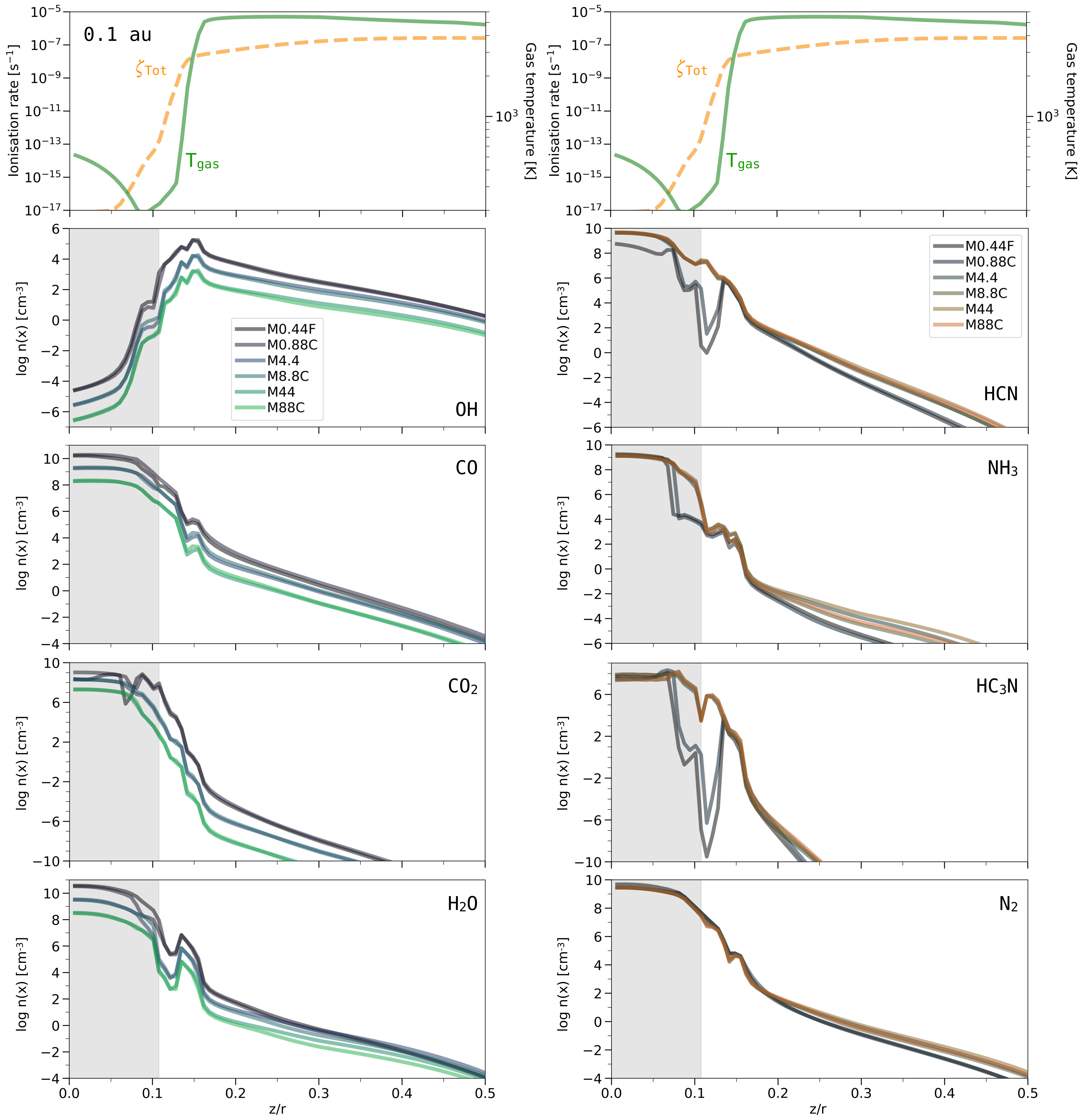}
\caption{Same as Fig.~\ref{fig:TA_Tgas_ionizationRate_0.1au_hydrocarbons}, but for the oxygen (OH, CO, \ce{CO2}, and \ce{H2O}) and nitrogen-bearing species (HCN, \ce{NH3}, \ce{HC3N}, and \ce{N2}) at 0.1~au.}
\label{fig:TA_Tgas_ionizationRate_0.1au_nitrogen_oxygen}
\end{figure*}

\begin{figure*}[ht]
\centering
\includegraphics[width=1\linewidth]{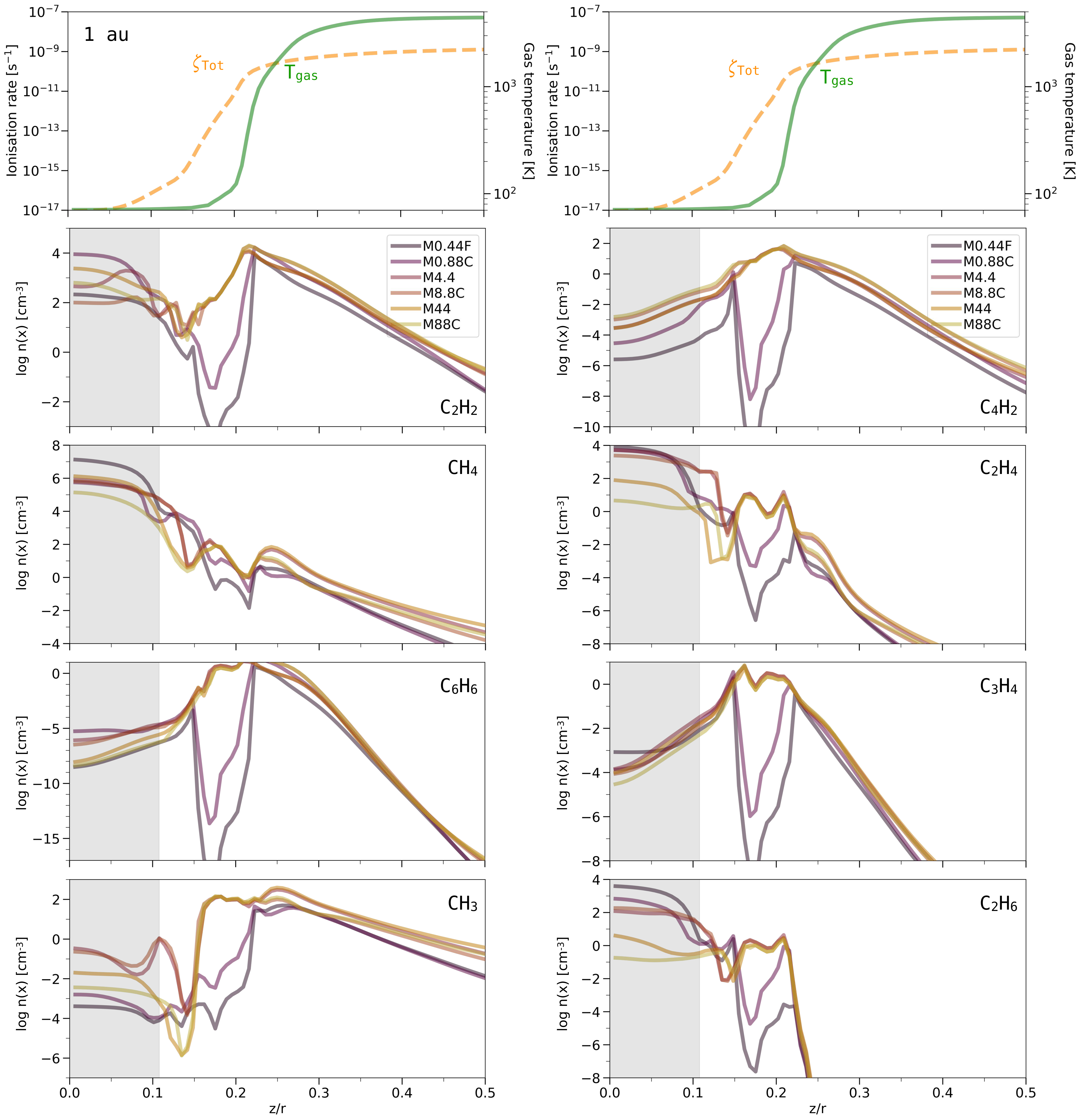}
\caption{Same as Fig.~\ref{fig:TA_Tgas_ionizationRate_0.1au_hydrocarbons}, but at 1~au.}
\label{fig:TA_Tgas_ionizationRate_1.0au_hydrocarbons}
\end{figure*}

\begin{figure*}[ht]
\centering
\includegraphics[width=1\linewidth]{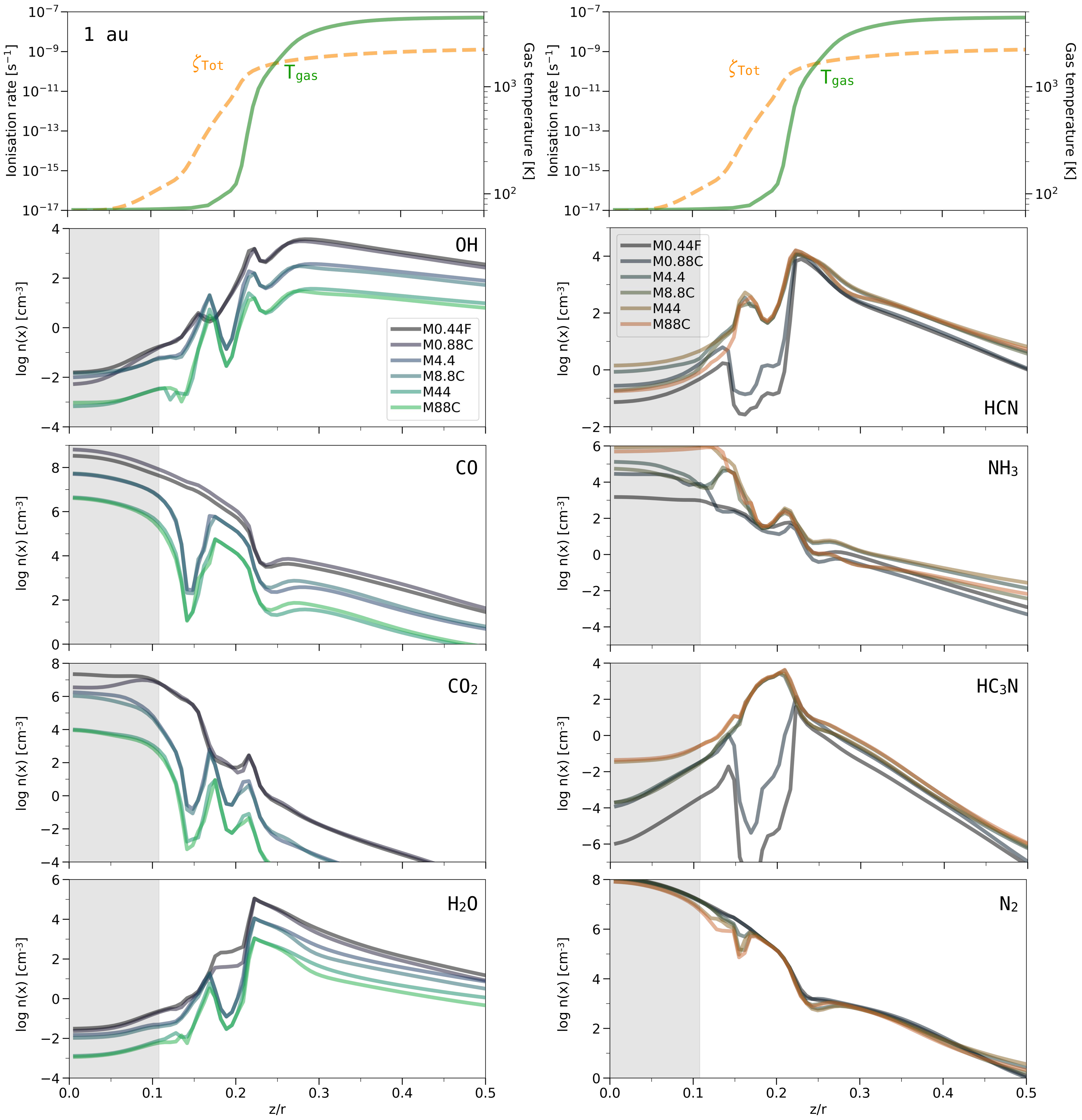}
\caption{Same as Fig.~\ref{fig:TA_Tgas_ionizationRate_0.1au_nitrogen_oxygen}, but at 1~au.}
\label{fig:TA_Tgas_ionizationRate_1.0au_nitrogen_oxygen}
\end{figure*}

Figures \ref{fig:TA_Tgas_ionizationRate_0.1au_hydrocarbons}, \ref{fig:TA_Tgas_ionizationRate_0.1au_nitrogen_oxygen}, \ref{fig:TA_Tgas_ionizationRate_1.0au_hydrocarbons}, \ref{fig:TA_Tgas_ionizationRate_1.0au_nitrogen_oxygen} show the vertical distribution of the ionization rate, gas temperature, and total abundance of a set of carbon-, nitrogen-, and oxygen-bearing species.
The vertical distributions for two radii of the disk (0.1 and 1~au) are presented.

%%%%%%%%%%%%%%%%%%%%%%%%%%%%%%%%%%%%%%%%%%%%%%%%%%%%%%%%%%%%%%%%%%%%%%

\section{Quantitative Descriptions of the Abundance and Column Density Variations for $\mathrm{C}_x\mathrm{H}_y$ with C/O}\label{appendix:hydrocarbons_CxHy}

In the main text, we present a summary of the behavior of $\mathrm{C}_x\mathrm{H}_y$ with increasing C/O because the members of this group of species can exhibit similar responses to the change in metallicity. 
Here we provide a more quantitative description of how each individual species responds to the change in C/O.

\paragraph{\ce{C2H2}}
The abundance distribution of \ce{C2H2} follows the general pattern as described in Section~\ref{sec:results} with two reservoirs.  
The fractional abundance is high in the upper layers, where the temperature ranges from $300$ to $3000~\mathrm{K}$, reaching a peak of $\sim 10^{-5}$ to $10^{-4}$ depending on the model.
However, as we will later see in most of the species' abundances, there is a correlation of the spatial extent over which the peak abundance is reached in the IR-emitting region with O/H and C/H separately, but not necessarily with C/O. 
Figures~\ref{fig:FAmap_hydrocarbon_set1}, \ref{fig:TA_Tgas_ionizationRate_0.1au_hydrocarbons}, and \ref{fig:TA_Tgas_ionizationRate_1.0au_hydrocarbons} show that from model M0.44F (the fiducial model) to M0.88C (carbon enrichment), in which C/O increases from 0.44 to 0.88, the vertical width of the region over which \ce{C2H2} reaches its peak abundance expands. 
For the model in which we deplete oxygen by a factor of 10 (M4.4), leading to a C/O of 4.37, we also see that the peak abundance region is broader in vertical extent than that for the fiducial model.
However we do see \ce{C2H2} reaching its peak abundance over a greater vertical extent and/or reaching a value $\sim10$ times higher in the IR-emitting region for the carbon-enriched cases (M0.88C, M8.8C, and M88C) than for the models with oxygen depletion alone (M0.44F, M4.4, and M44).
In conclusion, it appears that excess carbon and an increase in the C/H ratio lead to more efficient synthesis of \ce{C2H2} in the IR-emitting region.

The radial extent of the midplane reservoir of \ce{C2H2} is also affected by the increase in C/O. 
Whilst the peak fractional abundance here is not as high as reached in the disk atmosphere ($\sim 10^{-6}$ to $10^{-5}$), increasing the C/O ratio first by carbon enhancement (M0.88C) extends the midplane reservoir from $\lesssim 0.1$ to $\lesssim 0.2$~au.  
An increase in C/O through oxygen depletion alone (M4.4) leads to the midplane reservoir extending out to $\lesssim 0.3$~au.  
The combined model (M8.8C) and further increases in C/O, do not significantly extend this reservoir further. 
Hence, an increase in C/O in the disk midplane does allow more efficient synthesis of \ce{C2H2}, but only out to a radius of $\sim 0.3$~au.

The behavior with C/O for \ce{C2H2} is reflected in the calculated column density shown in Fig.~\ref{fig:ColdensProfile_IRemitting}. 
The column density profile of \ce{C2H2} in the IR-emitting region is generally smooth, with a gentle increase in column density within $\sim 1$~au, by up to one order of magnitude, reaching a peak at the innermost point in our model (0.04~au).  

The biggest increase in the column density in the IR-emitting region in the inner disk comes in the first three perturbations (enhancing C, C/O$~=0.87$; depleting O, C/O$~=4.37$; and both, C/O$~=8.75$).
In the outer disk, the behavior is different.
Here we see the column density increasing by a factor of $\sim 2$ when enhancing carbon (M0.88C, M8.8C, and M88C) with respect to their counterparts with oxygen depletion only (M0.44F, M4.4, and M44, respectively). 
This suggests that C/H becomes the limiting factor in setting the amount of \ce{C2H2} that can be synthesized in the disk.

\paragraph{\ce{C6H6}}
The peak abundance of \ce{C6H6} in the disk atmosphere is lower than that for the hydrocarbons considered thus far, reaching a peak value of $\sim 10^{-8}$ in the disk atmosphere and midplane in the fiducial model (M0.44F). 
The spatial distribution exhibits the same double-component morphology as \ce{C2H2}. 
The abundance resides in a narrow layer in the disk atmosphere (see Fig.~\ref{fig:FAmap_hydrocarbon_set1}), which becomes wider in vertical extent when C/O increases from 0.44 (M0.44F) to 0.87 (M0.88C). 
The response of the \ce{C6H6} abundance to initial abundances is nonlinear with C/O, similar to that found for \ce{C2H2}. 
When an oxygen depletion factor of $10$ is applied (M4.4), the \ce{C6H6} peak abundance region moves downward vertically compared to that for M0.44F; however, if the oxygen depletion factor increases to $100$ (M44), the layer does not vary in vertical extent compared to M4.4 but remains broader than the fiducial model (M0.44F).
This effect is also observed in the M8.8C and M88C scenarios, where the \ce{C6H6} peak fractional abundance increases compared with the fiducial model, and over a broader layer, in response to carbon enrichment. 
However, the two more extreme C/O ratios considered here (44 and 88) do not lead to the synthesis of significantly more \ce{C6H6} than found in the two moderate cases (4.4 and 8.8). 
For the midplane component, again, the first perturbation has the biggest effect: the midplane reservoir of \ce{C6H6} increases in radial extent from $\lesssim 0.2$~au in the fiducial model (M0.44F) out to $\lesssim 0.3$~au for all other cases.

In the column density profiles for \ce{C6H6} in Fig.~\ref{fig:ColdensProfile_IRemitting}, the initial abundances have an impact across the disk. 
The column density integrated over the IR-emitting region in the outer disk,  $\gtrsim 1$~au, ranges from $\ll 10^{12}$~cm$^{-2}$ (M0.44F) to $\lesssim10^{13}$~cm$^{-2}$ (M88C).
In the inner disk, the column density reaches a peak of $\sim10^{15}$~cm$^{-2}$ in all models within $0.05$~au.
Beyond 0.1~au, oxygen-rich models drop by almost two to three orders of magnitude (M0.44F and M0.88C, respectively).
On the other hand, the column densities in the carbon-rich models decrease moderately by an order of magnitude and drop again beyond 1~au.

\paragraph{\ce{C4H2}} 
\ce{C4H2} exhibits the same abundance pattern as \ce{C2H2} and \ce{C6H6} (see Fig.~\ref{fig:FAmap_hydrocarbon_set2}). 
The fractional abundance in the fiducial model (M0.44F) peaks with a value $\sim10^{-8}$ in the upper layers of the disk, where the temperature exceeds around $1000~\mathrm{K}$. 
The midplane reservoir reaches a similar peak value and extends out to $\sim 0.2$~au in the fiducial model.
For higher C/O ratios, the layer where the abundance peaks extends vertically, similar to that found for \ce{C2H2}, \ce{C6H6}, and \ce{CH3}. 
The cases in which C/H increases and O/H decreases (M8.8C and M88C) show a more vertically extended peak abundance region compared with the M0.88C model, where only carbon enrichment is considered.
The molecular reservoir in the midplane also varies in shape and extent as C/O increases.
There is a radial expansion from $\sim0.2~\mathrm{au}$ (M0.44F) to $\sim0.7~\mathrm{au}$ (M88C) combined with a drop in the peak abundance in the midplane. 
In these scenarios, it appears that increasing carbon and/or depleting oxygen increases the destruction rate (or slows down the production rate) of \ce{C4H2} in the disk midplane, with the conditions present therein likely sequestering carbon into other carriers. 
As we will discuss later, both \ce{C2H4} and \ce{C2H6} show the opposite behavior, lending credence to this theory.

The column density profiles in Fig.~\ref{fig:ColdensProfile_IRemitting} reflect these variations in \ce{C4H2} abundance across the full radial extension.

\paragraph{\ce{C2H4}}
The behavior of \ce{C2H4} is similar to \ce{C2H2} in its abundance distribution and is similar to \ce{CH4} in that the midplane component reaches a larger peak fractional abundance than that in the disk atmosphere.
The fractional abundance of \ce{C2H4} in the innermost region of the disk within $\sim0.2~\mathrm{au}$ and below $z/r \approx 0.1$, reaches values of $\sim 10^{-7}$ in the fiducial model. 
On the other hand, the fractional abundance in the region where the temperature is $\sim500-1500~\mathrm{K}$ reaches values of $\sim10^{-10}$ within $\sim7$~au, and then increases almost two orders of magnitude in the outermost region.
As the C/O ratio increases, this region expands vertically and the peak abundance increases by around one order of magnitude (model M0.88C).
The peak abundance in the innermost midplane region also radially expands from $\sim0.2$~au (M0.44F) to $\sim0.5~\mathrm{au}$ (M4.4 and higher C/O models).

We see for the M0.44F model in Fig.~\ref{fig:ColdensProfile_IRemitting}, that the column density over the IR-emitting region peaks at $\sim10^{15}~\mathrm{cm^{-2}}$ at $\sim0.04$~au but is $\ll 10^{12}$~cm$^{-2}$ beyond 0.8~au.
As the C/O ratio increases, the column density profiles reach higher values through the innermost disk (up to $10^{16}~\mathrm{cm^{-2}}$ for M88C) within $\sim0.1$~au and then decrease by around four orders of magnitude beyond 0.8~au.

\paragraph{\ce{C3H4}}
The fractional abundance of \ce{C3H4} (propyne) follows a similar behavior to \ce{C2H4}, with a peak abundance in the atmosphere of $\sim10^{-9}$. 
In the oxygen-rich models (M0.44F and M0.88C), \ce{C3H4} is abundant again only in a very narrow layer that expands in vertical extent with increasing C/O.  
For the C/O scenarios above unity, the peak abundance value does not vary. 
On the other hand, the abundance in the midplane component increases by at least an order of magnitude when C/O increases.

In general, the column density across the IR-emitting region increases with increasing C/O, reaching values higher than $10^{12}~\mathrm{cm^{-2}}$ between $\sim0.05$ and $\sim 0.1$~au for C/O~$>8.8$.
Regarding the column density profiles, the differences between the cases where oxygen depletion is included with and without carbon enhancement (e.g., M4.4 and M8.8C or M44 and M88C) vary depending on the case.
For example, for the moderate C/O models (M4.4 and M8.8C), we see that in the innermost region M8.8C reaches higher column density values than M4.4 by at least an order of magnitude.
On the other hand, for the more extreme C/O cases (M44 and M88C), the difference between those results is very subtle, similar to what we see in the other hydrocarbon cases. 

\paragraph{\ce{C2H6}}
\ce{C2H6} is also similar to \ce{CH4} in that its peak abundance is reached in the midplane (see Fig.~\ref{fig:FAmap_hydrocarbon_set2}).
\ce{C2H6} reaches a peak between $10^{-6}$ and $10^{-5}$ (depending on the model) in the innermost region of the disk midplane within a radius of $\approx 0.4$~au, where the gas temperature is $150-500$~K across all models (see Fig.~\ref{fig:FAmap_hydrocarbon_set2}).  
The very innermost region (within 0.07~au) is depleted in \ce{C2H6}, with the carbon preferring to be in \ce{C2H2} and \ce{CH4} (see Fig.~\ref{fig:FAmap_hydrocarbon_set2}). 
When C/O increases, the abundance of \ce{C2H6} also increases in the disk atmosphere around $z/r\sim 0.2-0.4$, reaching a peak abundance of $\sim 10^{-7}$ in the outer-disk region ($\gg 0.1$~au).

This is reflected in the calculated column density shown in Fig.~\ref{fig:ColdensProfile_IRemitting}, where the peak column density in the IR-emitting region increases from $\sim10^{14}~\mathrm{cm^{-2}}$ in the fiducial model to $\sim10^{16}~\mathrm{cm^{-2}}$ in the model M4.4 (and in models with higher C/O ratios) within $\sim0.2~\mathrm{au}$. 
Beyond $\sim0.2$~au, there is no distinguishable variation in the column density, with values $\lesssim10^{12}~\mathrm{cm^{-2}}$ in all models.

%%%%%%%%%%%%%%%%%%%%%%%%%%%%%%%%%%%%%%%%%%%%%%%%%%%%%%%%%%%%%%%%%%%%%%

\section{Variations in Midplane Reservoir Size}\label{appendix:midplane}

\begin{figure}[ht]
\centering
\includegraphics[width=1\linewidth]{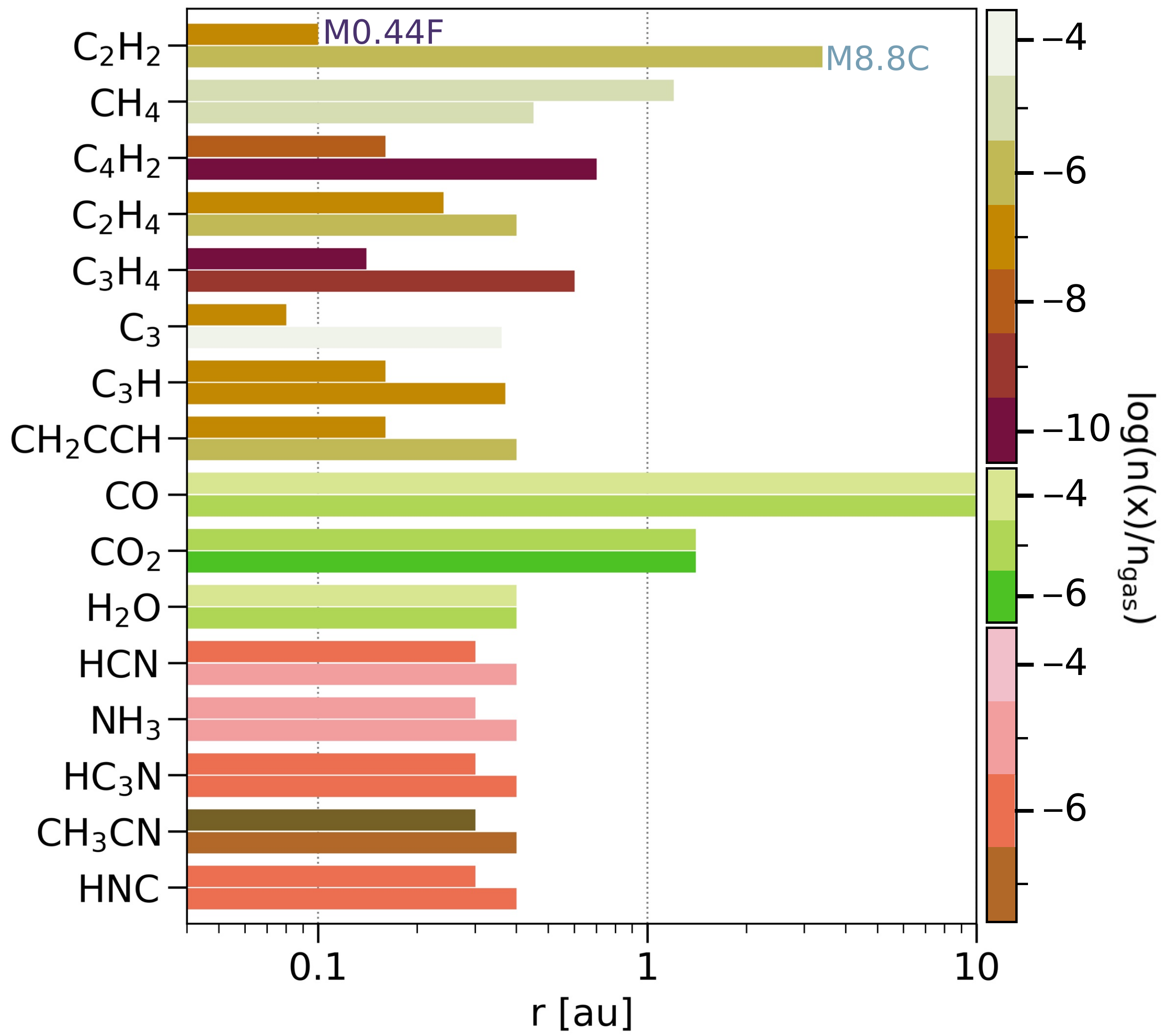}
\caption{
Midplane radial ranges for those species that show a variation in the size and abundance in the midplane reservoir, using similar color palettes to those in Figs.~\ref{fig:FAmap_hydrocarbon_set1}, \ref{fig:FAmap_oxygen}, and \ref{fig:FAmap_nitrogen} for hydrocarbons and oxygen and nitrogen carriers, respectively.
The two bars represent the midplane radial extension and abundance for M0.44F (top) and M8.8C (bottom), respectively.
}
\label{fig:midplane-size}
\end{figure}

Figure~\ref{fig:midplane-size} illustrates the midplane sizes for those molecules that have a midplane reservoir, which varies radially and/or in the peak abundance reached when C/O increases. 
We show the results for two values of C/O only, 0.44 and 8.75, to illustrate the behavior and give a qualitative idea of the response of the extent and abundance of the midplane reservoir to the C/O variation.
The midplane size was estimated by eye, based on where the abundance varies by a factor of 10 or more.
For simplicity, we consider only the outer radius of the reservoir, which means that we do not capture radial variations in this simple illustration.

As discussed previously, there are two common variations for hydrocarbons: a radial expansion and an increase in the peak abundance value as C/O increases. 
\ce{CH4} does not follow the same behavior, and the peak abundance in the midplane distribution instead shrinks from $\sim 1.2$~au to $\approx 0.45$~au. 
In addition, the peak abundance of \ce{C4H2} drops by a factor of $\sim 10$ as C/O increases, despite an increase in the radial range over which the peak value is reached. 
For oxygen carriers such as CO, \ce{CO2}, and \ce{H2O}, the radial extent does not vary as C/O increases; however, the peak abundance of all three species decreases by a factor of $10-100$ as C/O changes from 0.44 to 8.75.
Finally, for nitrogen-bearing species, all midplane reservoirs expand radially by a small amount, $\approx 0.1$~au. 
The peak abundance for HCN and \ce{CH3CN} also increases with increasing C/O.
That for \ce{NH3}, \ce{HC3N}, and \ce{HNC} have a similar peak abundance for both values of C/O.
Note that, as mentioned in Appendix~\ref{appendix:caveats}, our model has a temperature inversion in the midplane owing to viscous heating. 
However, these results are useful to compare with radial drift and accretion models, which are typically confined to the midplane \citep[e.g.,][]{sellek_chemical_2025}.

%%%%%%%%%%%%%%%%%%%%%%%%%%%%%%%%%%%%%%%%%%%%%%%%%%%%%%%%%%%%%%%%%%%%%%

\section{Potentially Observable Species}\label{appendix:potentially_observable}
\begin{figure*}[ht]
\centering
\includegraphics[width=1\linewidth]{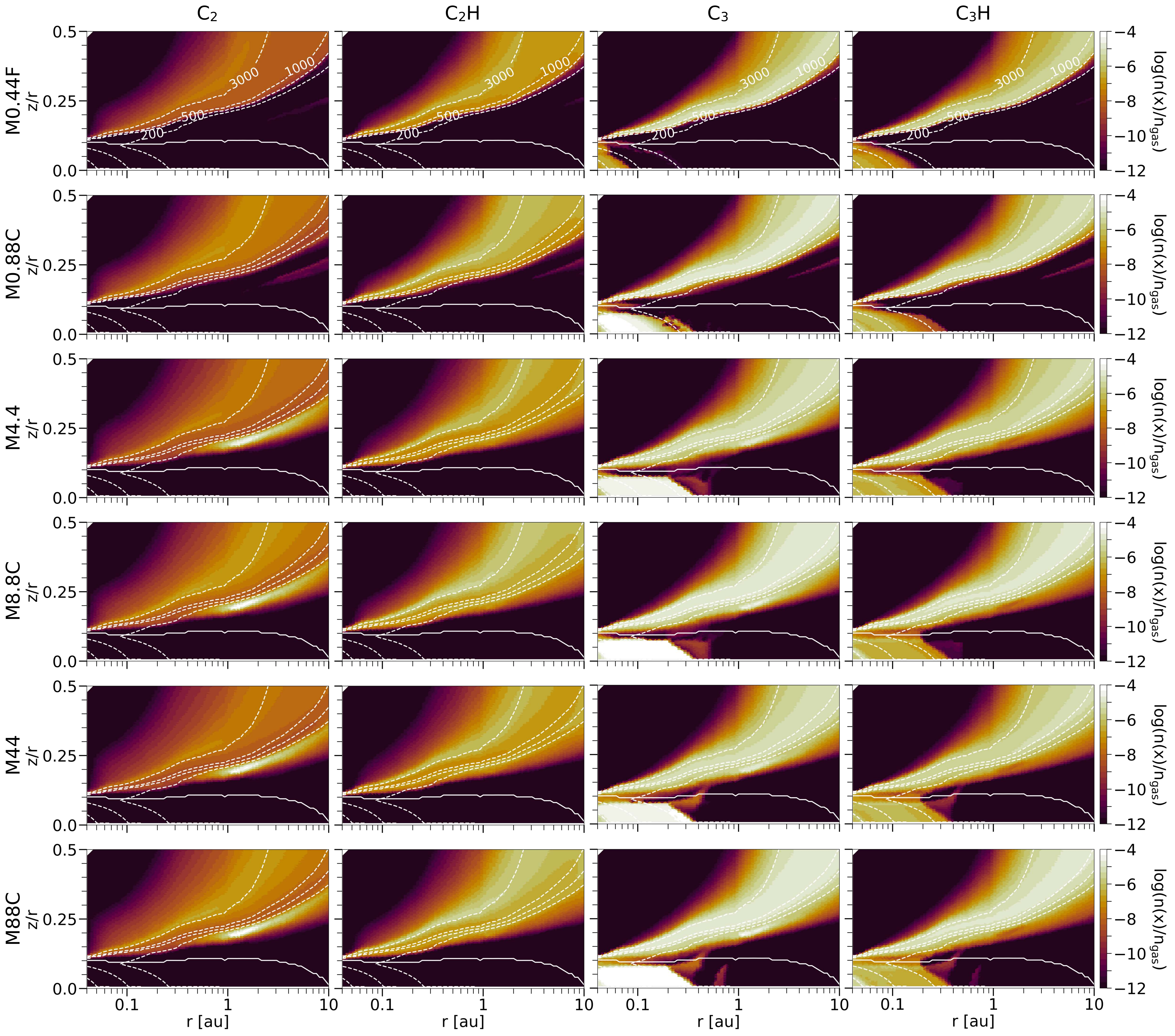}
\caption{Same as Fig.~\ref{fig:FAmap_hydrocarbon_set1}, but for the potentially observable species \ce{C2}, \ce{C2H}, \ce{C3}, and \ce{C3H}.
}
\label{fig:FAmap_potentially_observed_set1}
\end{figure*}

\begin{figure*}[ht]
\centering
\includegraphics[width=1\linewidth]{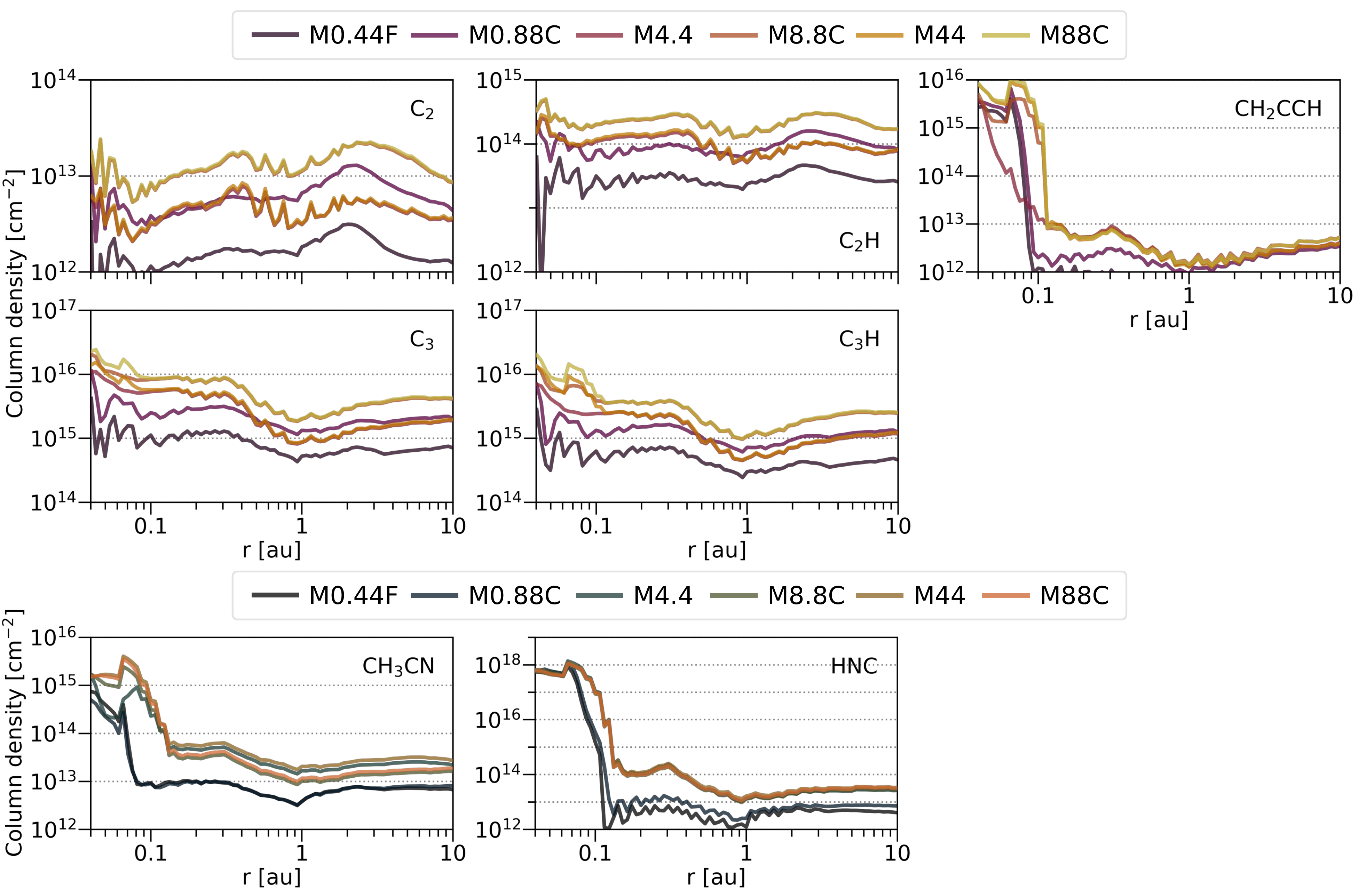}
\caption{Same as Fig.~\ref{fig:ColdensProfile_IRemitting}, but for the potentially observable species \ce{C2}, \ce{C2H}, \ce{C3}, \ce{C3H}, \ce{CH2CCH}, \ce{CH3CN}, and HNC.
(The complete figure set (2 images) is available in the online article.)
}
\label{fig:ColdensProf_potentially_observed_IRemitting}
\end{figure*}

\begin{figure*}[ht]
\centering
\includegraphics[width=1\linewidth]{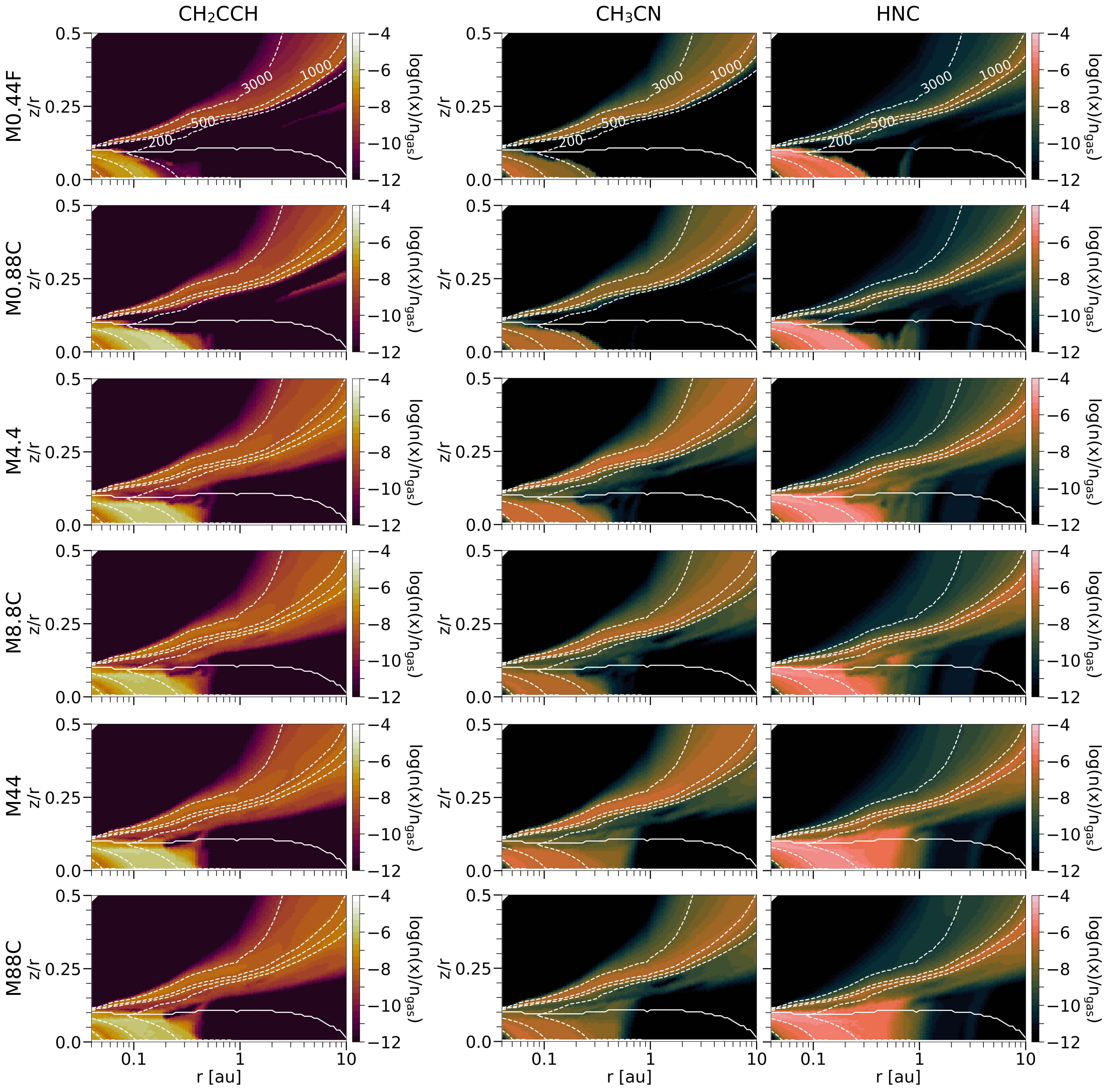}
\caption{Same as Fig.~\ref{fig:FAmap_hydrocarbon_set1}, but for the potentially observable species \ce{CH2CCH}, \ce{CH3CN}, and \ce{HNC}.
}
\label{fig:FAmap_potentially_observed_set2}
\end{figure*}

Other hydrocarbons and nitrogen-bearing species have been proposed to be potentially observable in the inner region of the disk. 
Figures \ref{fig:FAmap_potentially_observed_set1} and \ref{fig:FAmap_potentially_observed_set2} show the fractional abundance maps for \ce{C2}, \ce{C2H}, \ce{C3}, \ce{C3H}, \ce{CH2CCH}, \ce{CH3CN}, and \ce{HNC}.
Figure~\ref{fig:ColdensProf_potentially_observed_IRemitting} show the column density profiles for the same set of species.
Figure~\ref{fig:Nmol-C2Oratio-others} shows the total number of molecules integrated over the full disk model (out to 10~au) as a function of the C/O ratio.

As shown in Figs.~\ref{fig:FAmap_potentially_observed_set1} and \ref{fig:ColdensProf_potentially_observed_IRemitting}, the abundances and column densities of \ce{C3} are particularly enhanced when C/O increases, for example, the abundance is $\sim10^{-4}$ in both the disk midplane and the IR-emitting region when C/O is greater than 0.44.
One caveat of the chemical network used here is that the chemistry does not include processes that lead to full saturation of species that contain three or more carbon atoms.
Whilst the predictions for the IR-emitting region are likely robust, it is recommended to treat with caution the predictions for the disk midplane, where the temperatures and pressures should tend to move the chemistry toward equilibrium.
The same is true for species with a higher number of carbon atoms ($>3$).
Figures \ref{fig:FAmap_potentially_observed_set1} and \ref{fig:Nmol-C2Oratio-others} also show that the \ce{C2} and \ce{C2H} abundance distribution is confined to the IR-emitting region, similar to that for \ce{CH3} and \ce{OH}.

%%%%% Figure set for Figure E.2 -- Begin
%\figsetstart
%\figsetnum{E.2}
%\figsettitle{Column density profiles for potentially observed species for the region above the $\tau=1$ surface at $14~\mu\mathrm{m}$ (disk atmosphere component) and full vertical disk extent down to the midplane (midplane component).}

%\figsetgrpstart
%\figsetgrpnum{E.2.1}
%\figsetgrptitle{Region above the $\tau=1$ surface at $14~\mu\mathrm{m}$}
%\figsetplot{ColumnDensityProfiles_potentially_atmosphere14um.pdf}
%\figsetgrpnote{
%Same as Fig.~\ref{fig:ColdensProf_potentially_observed_IRemitting}, but for the integration down to the $\tau=1$ surface at $14~\mu\mathrm{m}$ (disk atmosphere component).}
%\figsetgrpend

%\figsetgrpstart
%\figsetgrpnum{E.2.2}
%\figsetgrptitle{Region above the midplane ($z=0$)}
%\figsetplot{ColumnDensityProfiles_potentially_midplane.pdf}
%\figsetgrpnote{
%Same as Fig.~\ref{fig:ColdensProf_potentially_observed_IRemitting}, but for the integration down to the midplane (midplane component).}
%\figsetgrpend

%\figsetend

%%%%% Figure set for Figure E.2 -- End

\section{Total Number of Molecules with Respect to C/O}\label{appendix:Nmol_vs_C2O}

\begin{figure*}[ht]
\centering
\includegraphics[width=1\linewidth]{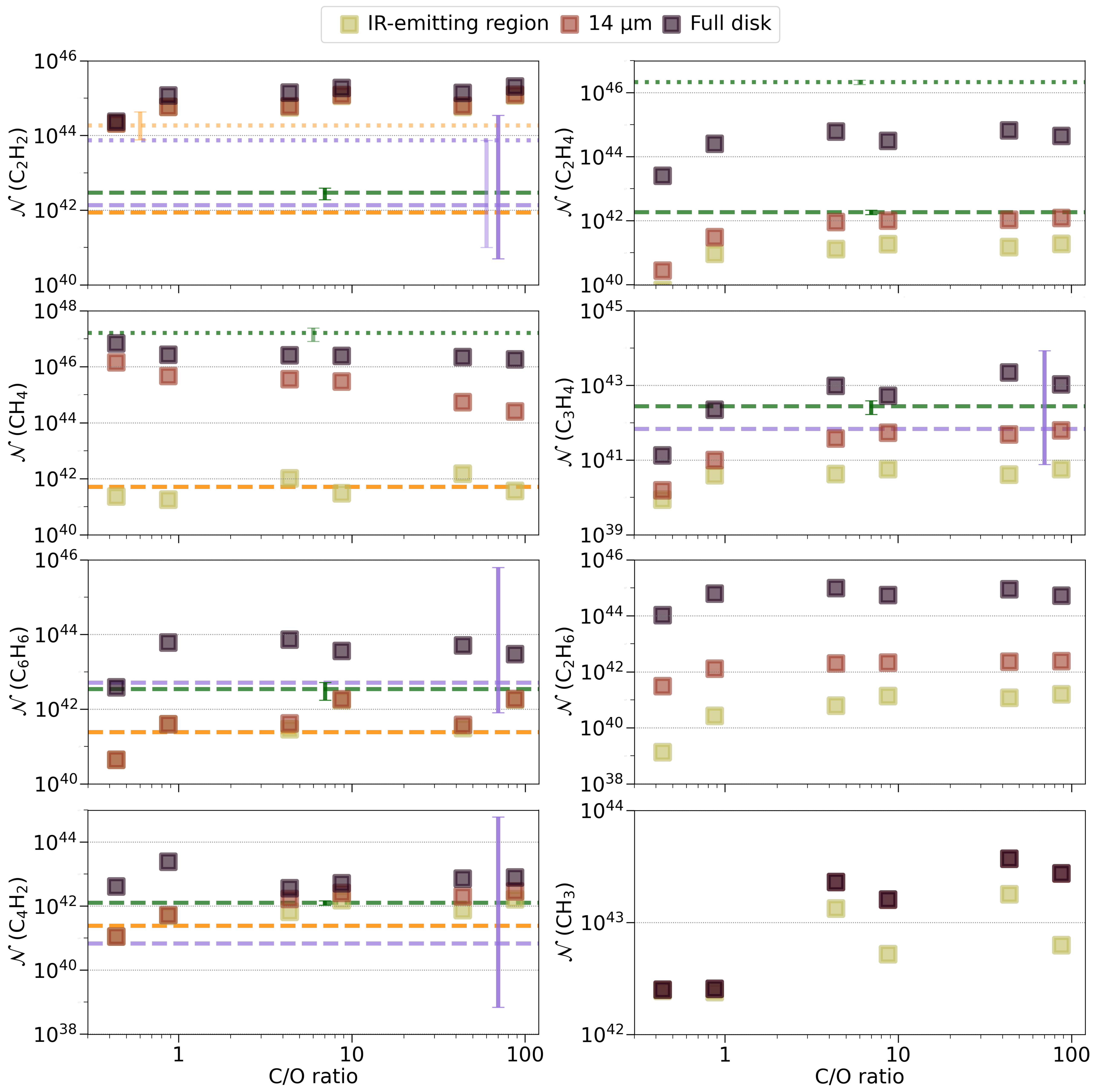}
\caption{
Number of molecules (integrated over 10~au) for the hydrocarbon species (\ce{C2H2}, \ce{CH4}, \ce{C6H6}, \ce{C4H2}, \ce{C2H4}, \ce{C3H4}, \ce{CH3}, and \ce{C2H6}) as a function of C/O.
The different colors from light to dark represent the integration over the IR-emitting region, down to the dust photosphere at 14~$\mu$m, and over the full disk.
The orange, green, and purple horizontal dashed lines represent the number of molecules reported for J160532, ISO-ChaI~147, and Sz28, respectively, for the optically thin component.
If available, the optically thick component is represented by horizontal dotted lines for the same disks.
}
\label{fig:Nmol-C2Oratio_hydrocarbons}
\end{figure*}

\begin{figure*}[ht]
\centering
\includegraphics[width=1\linewidth]{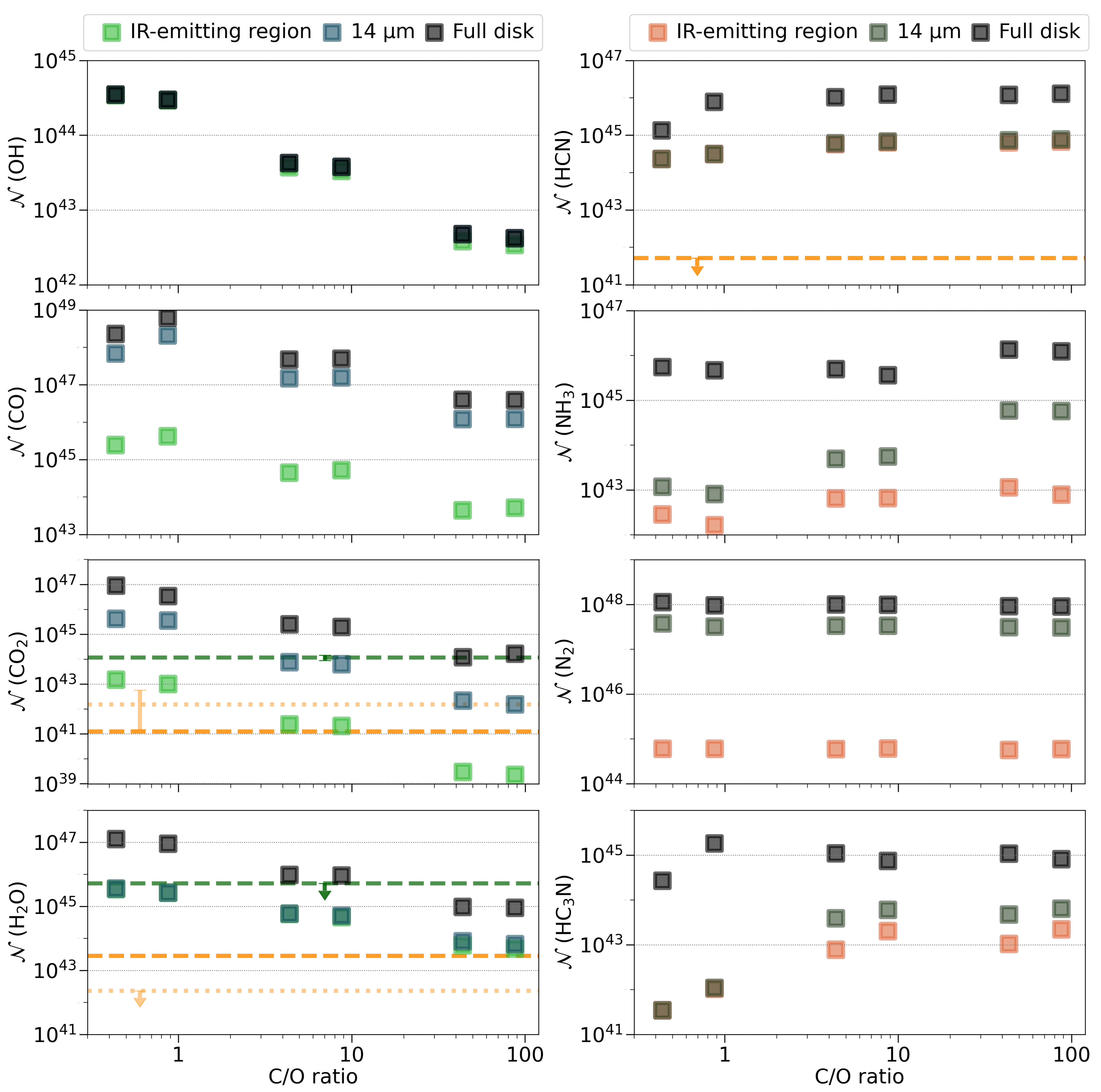}
\caption{Same as Fig.~\ref{fig:Nmol-C2Oratio_hydrocarbons}, but for the oxygen- and  nitrogen-bearing species (\ce{OH}, \ce{CO}, \ce{CO2}, \ce{H2O}, \ce{HCN}, \ce{NH3}, \ce{N2}, and \ce{HC3N}).
}
\label{fig:Nmol-C2Oratio_oxygen_nitrogen} 
\end{figure*}

\begin{figure*}[ht]
\centering
\includegraphics[width=1\linewidth]{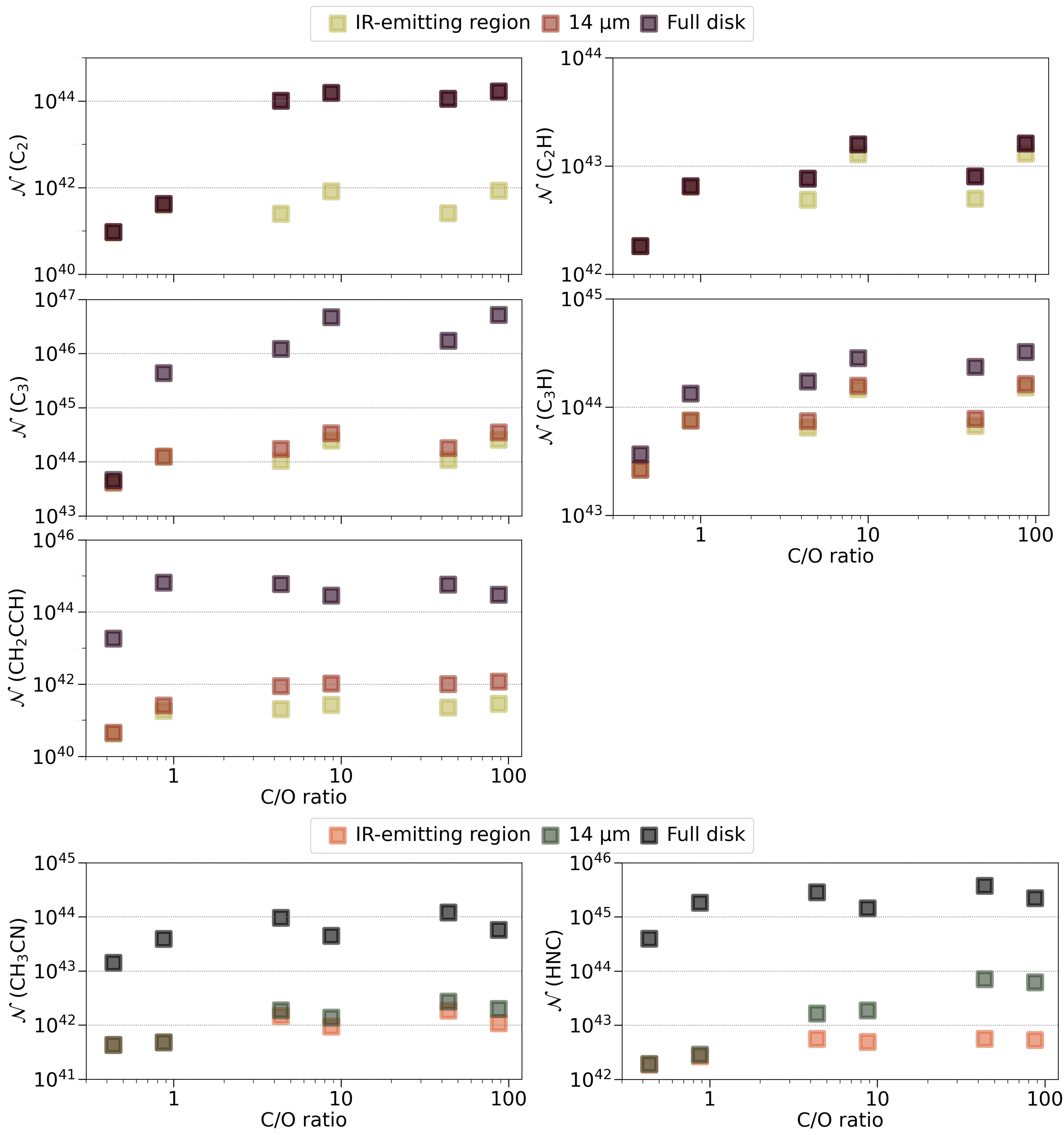}
\caption{Same as Fig.~\ref{fig:Nmol-C2Oratio_hydrocarbons}, but for the potentially observable species \ce{C2}, \ce{C2H}, \ce{C3}, \ce{C3H}, \ce{CH2CCH}, \ce{CH3CN}, and \ce{HNC}.
}
\label{fig:Nmol-C2Oratio-others}
\end{figure*}

\begin{figure*}[ht]
\centering
\includegraphics[width=\linewidth]{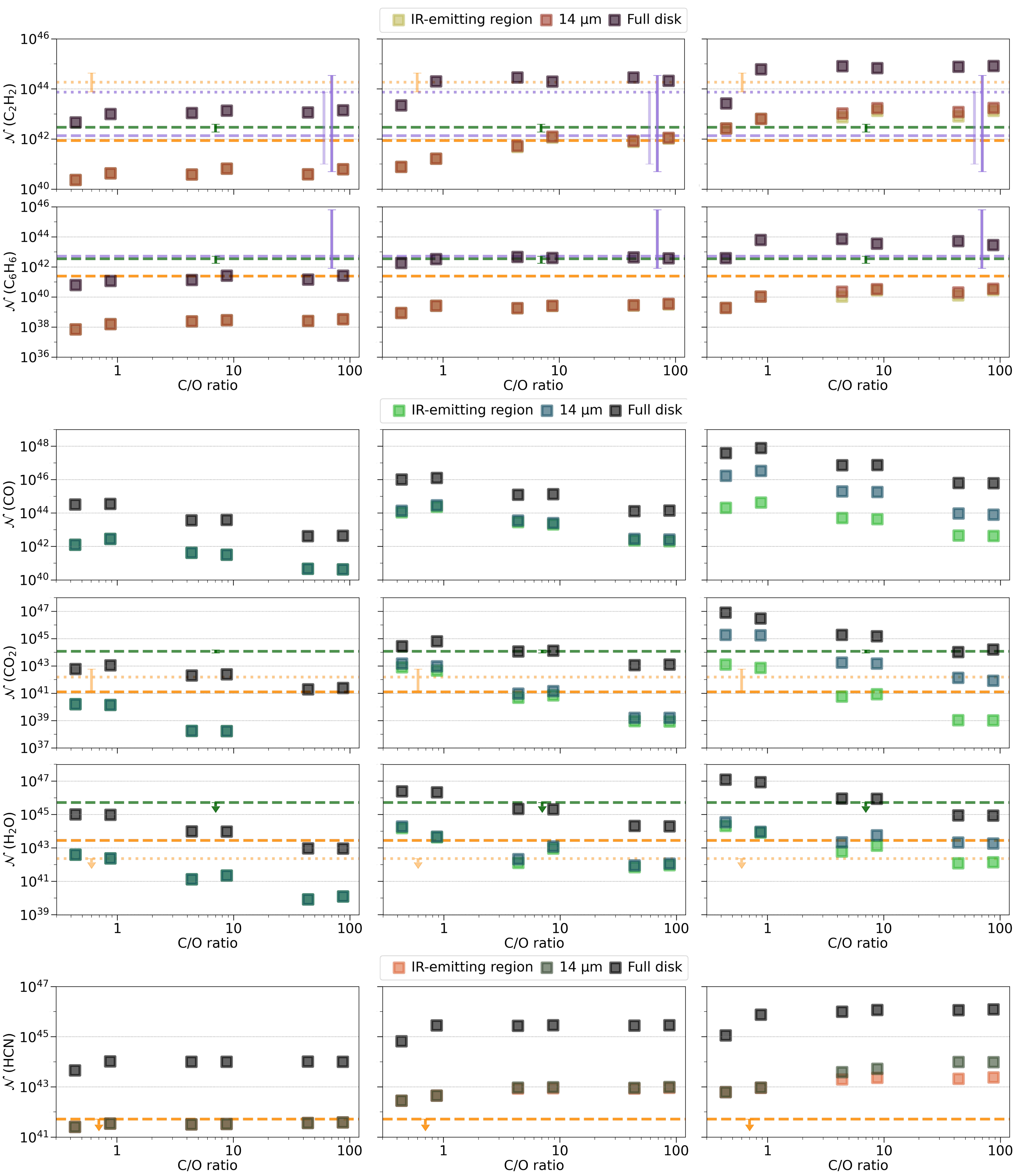}
\caption{
Same as Figure~\ref{fig:Nmol-C2Oratio_hydrocarbons}, but at 0.04~au (left) and averaged over 0.1 (middle) and 1~au (right) for \ce{C2H2}, \ce{C6H6}, \ce{CO}, \ce{CO2}, \ce{H2O}, and HCN (each row).
}
\label{fig:Nmol_vs_CO2_innermost}
\end{figure*}

Figures~\ref{fig:Nmol-C2Oratio_hydrocarbons}, \ref{fig:Nmol-C2Oratio_oxygen_nitrogen}, and \ref{fig:Nmol-C2Oratio-others} show the total number of molecules over the full disk model out to 10~au for all the key species discussed in this work and potentially observed species, as a function of C/O ratio.  
These plots show nicely the general trends in total number of molecules, $\mathcal{N}$, with C/O as discussed in the text.
The horizontal dashed and dotted lines represent the derived number of molecules for the optically thin and thick components, respectively, for J160532, ISO-ChaI~147, and Sz28 \citep{tabone_rich_2023,arabhavi_abundant_2024,kanwar_minds_2024,kaeufer_disentangling_2024,arabhavi_minds_2025}.
Figure~\ref{fig:Nmol_vs_CO2_innermost} shows the same estimation at 0.04, 0.1, and 1~au of a subset of species (\ce{C2H2}, \ce{C6H6}, \ce{CO}, \ce{CO2}, \ce{H2O}, and HCN).
We also include the $\mathcal{N}$ calculations for the integration down to the dust photosphere at $14~\mu\mathrm{m}$.
This is because in some cases (e.g., \ce{CH4} and \ce{C2H6}) the abundance in the region below $T_\mathrm{gas}=200~\mathrm{K}$ and above the dust photosphere at $14~\mu\mathrm{m}$ is not negligible.
However, for other species, such as \ce{C6H6} and \ce{C2H4}, $\mathcal{N}$ values are more similar to the estimations that include only the IR-emitting region.

%%%%%%%%%%%%%%%%%%%%%%%%%%%%%%%%%%%%%%%%%%%%%%%%%%%%%%%%%%%%%%%%%%%%%%

\section{Summary of M~dwarf Systems Observed with JWST}\label{appendix:JWST_observations}

Table~\ref{table:Nmol_literature} summarizes the inferred number of molecules for the same sources reported by \citet{tabone_rich_2023}, \citet{arabhavi_abundant_2024}, \citet{kanwar_minds_2024}, and \citet{kaeufer_disentangling_2024}.
Table~\ref{table:Observed_sources_properties} presents the star-disk system physical properties for the J160532, ISO-ChaI~147, and Sz28 disks. 

\begin{deluxetable*}{@{\extracolsep{10pt}}lccc}
\tablecaption{Inferred number of molecules for J160532, ISO-ChaI~147, and Sz28.}
\tablehead{
    \colhead{Molecule} & \multicolumn{3}{c}{Number of Molecules ($\mathcal{N}$)}\\
                     & \colhead{J160532 [1]} & \colhead{ISO-ChaI~147 [3]} & \colhead{Sz28 [4]}
    }
    \startdata
    \hline
    \multicolumn{4}{c}{Hydrocarbons} \\
    \hline
    \hline
     \ce{C2H2}* & $1.84^{+2.45}_{-1.07}\times 10^{44}$     & $1.45\pm 0.34\times 10^{46}$  & $7.40_{-7.39}\times 10^{43}$ \\
     \ce{C2H2} & $8.61\times 10^{41}$      & $2.90\pm0.99\times 10^{42}$  & $1.35^{+343}_{-1.30}\times 10^{42}$ \\
     \ce{C6H6} & $2.41\times 10^{41}$      & $3.45\pm1.72\times 10^{42}$  & $5.10^{+6240}_{-4.30}\times 10^{42}$ \\
     \ce{C4H2} & $2.41\times 10^{41}$      & $1.27\pm0.20\times 10^{42}$  & $6.75^{+6.08}_{-6.68}\times 10^{40}$ \\
     \ce{CH4}*  & --                      & $1.62\pm0.81\times 10^{47}$   & --                 \\
     \ce{CH4}  & $5.17\times 10^{41}$      & Not detected          & Detected           \\
     \ce{C2H4}* & Not detected            & $2.14\pm0.34\times 10^{46}$  & Not detected        \\
     \ce{C2H4} & Not detected             & $1.87\pm0.30\times 10^{42}$  & Not detected        \\
     \ce{C3H4} & Not detected             & $2.76\pm1.10\times 10^{42}$  & $6.86^{+848}_{-6.10}\times 10^{41}$ \\
     \ce{C2H6} & Not detected             & Detected              & Detected           \\
     \ce{CH3}  & Not detected             & Detected              & Detected           \\
    \hline
    \multicolumn{4}{c}{Oxygen-bearing Species} \\
    \hline
    \hline
     \ce{H2O}*  & $\leq 2.30\times 10^{42}$   & -- & Not detected          \\
     \ce{H2O}  & $2.81\times 10^{43}$ [2]   & $<5.12\times 10^{45}$ & Not detected          \\
     \ce{CO2}*  & $1.53^{+4.21}_{-1.38}\times 10^{42}$     & --  & Detected              \\
     \ce{CO2}  & $1.24\times 10^{41}$     & $1.17\pm0.27\times 10^{44}$  & Detected              \\
    \hline
    \multicolumn{4}{c}{Nitrogen-bearing Species} \\
    \hline
    \hline
     \ce{HCN}  & $\leq 5.17\times 10^{41}$ & Detected              & Detected              \\
     \ce{NH3}  & Not detected             & Not detected            & Not detected          \\
     \ce{HC3N} & Not detected             & Detected              & Detected              \\
    \hline
    \enddata
\label{table:Nmol_literature}
\tablecomments{[1] \citet{tabone_rich_2023}; [2] \citet{arabhavi_minds_2025}; [3] \citet{arabhavi_abundant_2024}; and [4] \citet{kanwar_minds_2024}.
Optically thick values are indicated by the asterisk on the species named.}
\end{deluxetable*}

\begin{deluxetable*}{@{\extracolsep{10pt}}lcccc}[htb]
\tablecaption{Properties of the Low-mass Stars, J160532, ISO-ChaI~147, and Sz28.}
\tablehead{
    \colhead{Parameter} & \colhead{Description} & \multicolumn{3}{c}{Value}\\
                        &                       & \colhead{J160532} & \colhead{ISO-ChaI~147} & \colhead{Sz28}
    }
    \startdata
    \hline  
     ${M}_\star~(M_\odot)$          & Stellar mass         & $0.14$ & $0.11$ & $0.12$\\
     ${L}_\star~(L_\odot)$          & Stellar luminosity   & $0.04$ & $0.03$ & $0.04$\\
     ${\dot{M}}~(M_\odot~\mathrm{yr^{-1}})$          & Accretion mass rate  & $7.9\times 10^{-10}$ & $7\times 10^{-12}$ & $1.8\times 10^{-11}$\\
     $\mathrm{Age}_\star~(\mathrm{Myr})$        & Stellar age          & $2.6\pm 1.6$ &  $1-2$ & $3.5$\\
     $\mathrm{SpT}$              & Spectral type        & M4.75 &  M5.5 & M5.5\\
     ${M}_\mathrm{grain}~(M_\oplus)$ & Dust grain disk mass & $0.75$ &  $<0.72$ & $0.48$\\
     ${M}_\mathrm{gas}~(M_{Jup})$   & Gas disk mass        & $<0.2$ & $<1.05$ & $0.08$\\
    \hline
    \enddata
\label{table:Observed_sources_properties}
\tablecomments{Physical properties of the J160532 \citep{pascucci_atomic_2013, tabone_rich_2023}, ISO-ChaI~147 \citep{arabhavi_abundant_2024}, and Sz28 star-disk systems \citep{kanwar_minds_2024}.}
\end{deluxetable*}

%%%%%%%%%%%%%%%%%%%%%%%%%%%%%%%%%%%%%%%%%%%%%%%%%%%%%%%%%%%%%%%%%%%%%%

\section{Temperature Distribution of Molecules in the IR-emitting Region}\label{appendix:temp_distribution}

\begin{figure*}[ht]
\centering
\includegraphics[width=\linewidth]{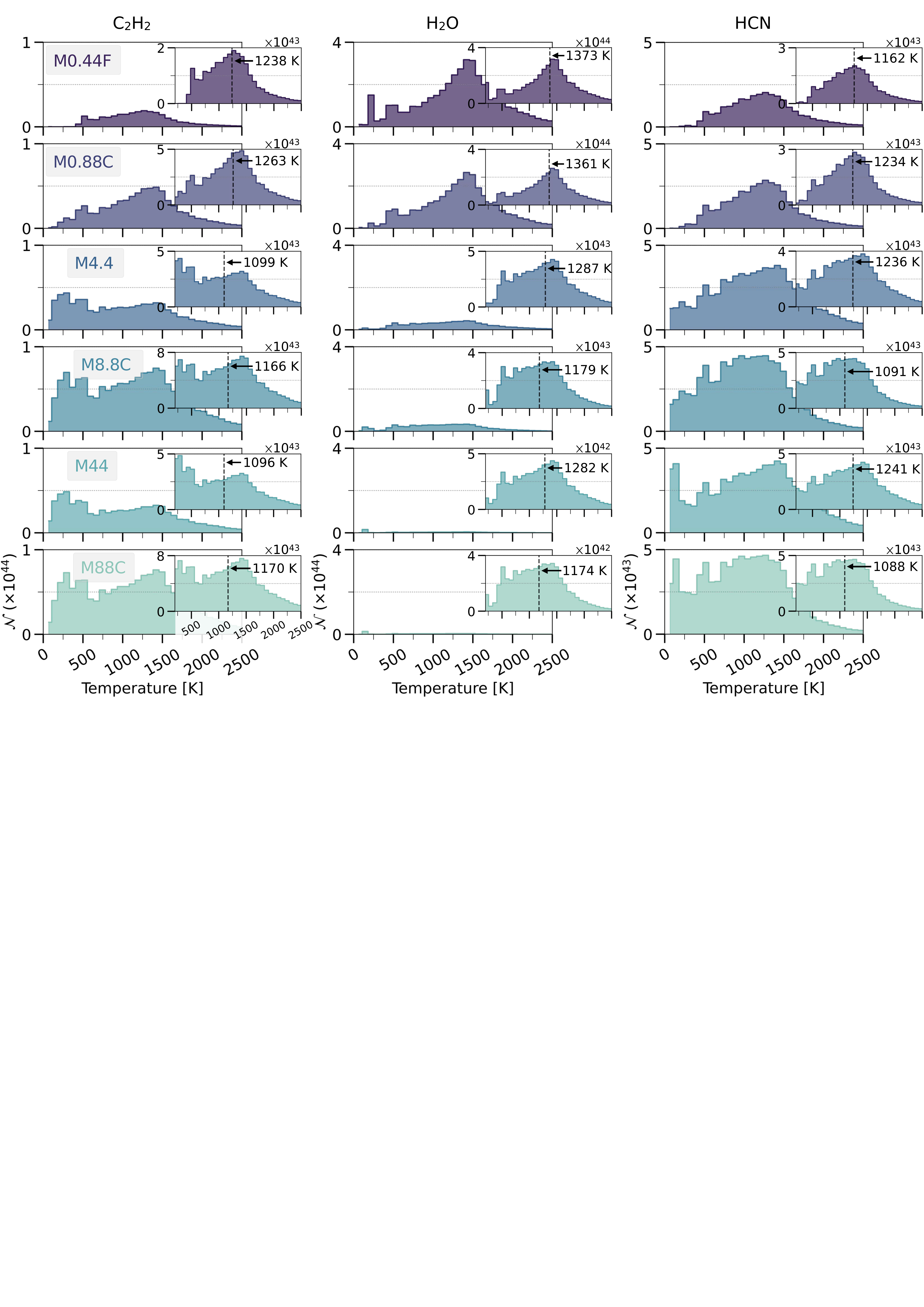}
\caption{
Histogram of the number of molecules as a function of gas temperature for \ce{C2H2}, \ce{H2O}, and \ce{HCN} (columns).
From top to bottom, each row represents a different model, with increasing C/O ratio from 0.44 (M0.44F, top) to 88 (M88C, bottom).
In each column, the left panel shows the histogram across the full gas temperature range, and the right panel presents a zoom-in to better display the vertical spread. 
In these cases, a gas temperature cut of $>200$~K is applied and the $x$-axis ranges from 200 to 2500~K.
The black dashed lines represent the average temperature of the gas for regions of the disk $>200~\mathrm{K}$.
}
\label{fig:all_histograms}
\end{figure*}

In Section~\ref{sec:observational_trends} we discussed the observational trends of the excitation temperatures reported for the species detected in each disk.
In this appendix, we focus on the results of our models regarding the gas temperature distribution for each molecule.
We conducted this analysis to check and compare whether or not the temperature of the region in which the molecules are mainly present in the disk is consistent with the excitation temperature derived from the observations.
Note that the gas temperature structure is not recalculated for different elemental abundances (see Appendix~\ref{appendix:caveats}); therefore, the temperature variation with C/O discussed below should be taken with caution.
Figure~\ref{fig:all_histograms} presents histograms of the number of molecules as a function of gas temperature for \ce{C2H2}, \ce{H2O}, and \ce{HCN} according to the C/O ratio.
We choose these species because they have slightly different behavior in their temperature distribution with the C/O variations.
Note that we do not plot the equivalent data for \ce{CO2} because our model predicts that the majority of \ce{CO2} molecules reside at low temperatures, less than 250~K.
That the observations derive excitation temperatures for \ce{CO2} generally higher than this requires further investigation.

\paragraph{\ce{C2H2}}
From Fig.~\ref{fig:all_histograms}, we see that the bulk of \ce{C2H2} has a higher peak gas temperature between 1000 and 1750~K in the oxygen-rich cases (M0.44F and M0.88C) than in the carbon-rich cases, where a large fraction of \ce{C2H2} molecules also reside at low gas temperatures ($\lesssim500$~K).
When the C/O increases to $>1$, two distinct components at different temperatures appear.
The colder component peaks between 200 and 600~K with a number of molecules of $\sim 5\times 10^{43}$.
The peak temperature in this component remains constant for higher C/O ratio.
On the other hand, the peak temperature of the hotter component oscillates when increasing C/O ratios ($\sim 100$~K between models). 
The number of \ce{C2H2} molecules in the same component also varies depending on the model; for oxygen-depleted-only models (M4.4 and M44), $\mathcal{N}$ is $\lesssim 5\times 10^{43}$~K, while for the carbon-enriched and oxygen-depleted models (M8.8C and M88C), it is nearly two times higher.
From Fig.~\ref{fig:Nmol_Tex_literature}, the excitation temperature of \ce{C2H2} is 325~K in ISO-ChaI~147, to 425~K in Sz28, and 400 and $\sim 525$~K for the thin and thick components of J160532, respectively. 
Comparing with our gas temperature distributions, only for the carbon-rich cases is there a large fraction of \ce{C2H2} sitting in the colder region of the disk, which is more consistent with the excitation temperature from the observations.
This is complementary evidence for the disks having a C/O~$>1$ in the IR-emitting regions of the disk.
Note that the oxygen-depleted-only models (M4.4 and M44) have fewer molecules residing at high temperature ($>500$~K) than those with carbon enrichment, which suggests that the models with carbon enrichment and oxygen depletion (M8.8C and M88C) may overestimate the abundance of \ce{C2H2} present in the hot molecular layer.

\paragraph{\ce{H2O}}
The gas temperature distribution for water suggests that this molecule is more abundant at high temperatures, ranging from 1250 to 1750~K for the fiducial and carbon-enriched-only models (M0.44F and M0.88C). 
Once C/O exceeds 1, the shape of the histogram changes to a skewed Gaussian-like shape, peaking at $\sim1500$~K. 
Compared with \ce{C2H2} the overall temperature distribution of \ce{H2O} does not significantly change with increasing C/O and only the total number of molecules is affected.
From Fig.~\ref{fig:Nmol_Tex_literature}, the excitation temperature of \ce{H2O} ranges from 300~K in ISO-ChaI~147 to 525~K in J160532.
In general, our models predict higher temperatures for \ce{H2O}.

\paragraph{HCN}
The HCN gas distribution has a similar behavior to the hydrocarbon \ce{C2H2}.
For C/O~$<1$, the histogram has a Gaussian-like shape, peaking between 1000 and 1750~K.
For C/O~$>1$, the histogram changes to a skewed Gaussian-like shape, with M8.8C peaking between 750 and 1500~K, and M4.4 at $1250-1500$~K.
For the most extreme C/O values ($\sim 43.72$ and $\sim 87.47$), two components appear.
The colder component peaks at the lower end of the gas temperature, around $50-75$~K. 
The hotter component for M44 and M88C remains similar in distribution to that for M4.4 and M8.8C, respectively.
From Fig.~\ref{fig:Nmol_Tex_literature}, the excitation temperature of HCN is around 400~K in J160532.
Our models predict a colder and hotter component; however, neither of those peaks at $\sim 400$~K.

\paragraph{Other species}
The histograms for other species detected in J160532, ISO-ChaI~147, or Sz28, such as \ce{C2H4}, \ce{C4H2}, \ce{C6H6}, \ce{C3H4}, and \ce{CH4}, are not presented because their gas temperature distributions are somewhat similar to those presented before. 
For example, \ce{C6H6} shows a similar distribution to \ce{C2H2}, with a peak in the number of molecules at higher temperatures in the M8.8C model.
Other molecules, namely \ce{C2H4}, \ce{C4H2}, \ce{C3H4}, and \ce{CH4}, are mostly abundant in colder regions ($\lesssim 250$~K), similar to the cold component of HCN in the oxygen-poor models.
Additionally, we find that as C/O increases \ce{C6H6}, \ce{C4H2}, \ce{CH4}, and \ce{C3H4} have an increasing fraction of molecules at higher temperatures ($\sim500-2000$~K) in the carbon-enriched cases, whereas \ce{C2H4} does not.
The temperatures predicted for C/O~$<1$ are higher than those estimated from the observations, which in most cases remains true for carbon-rich scenarios.
However, as C/O increases, the average temperature of \ce{C2H4} in the IR-emitting region reaches similar values to those observed in ISO-ChaI~147.
For \ce{CH4}, when C/O increases owing to oxygen depletion (M4.4 and M44), the average temperature in the IR-emitting region is almost 400~K higher than their counterpart’s models with carbon enrichment (M8.8C and M88C).

%%%%%%%%%%%%%%%%%%%%%%%%%%%%%%%%%%%%%%%%%%%%%%%%%%%%%%%%%%%%%%%%%%%%%%

\section{Column Density Ratios and Radial Integration Sensitivity}\label{appendix:ColDensProfile_ratio}

\begin{figure*}[ht]
\centering
\includegraphics[width=\linewidth]{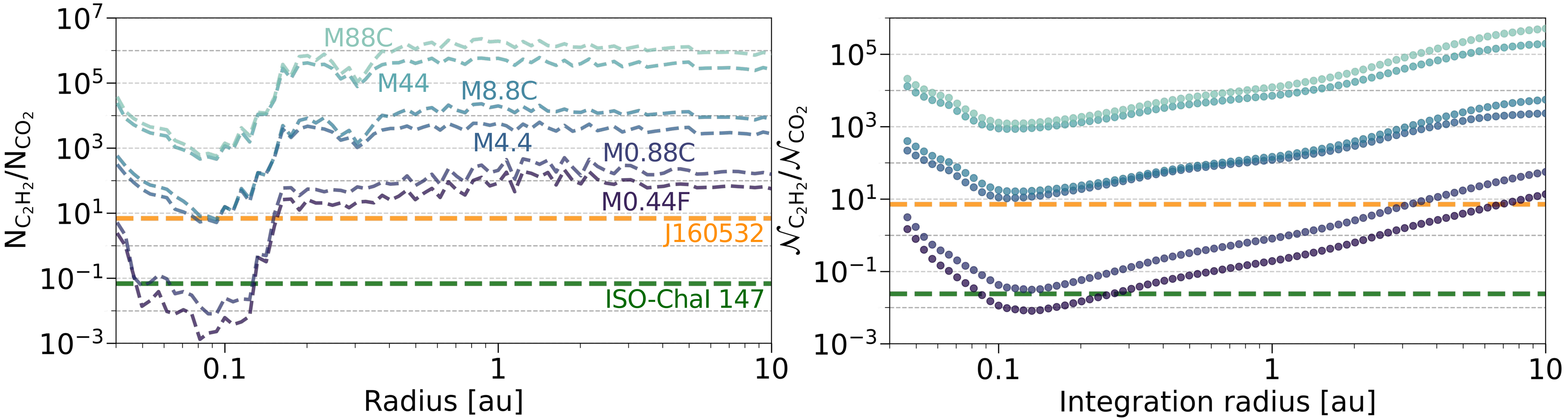}
\caption{
Left: column density ratio for \ce{C2H2}/\ce{CO2} as a function of radius.
Right: total number of molecule ratios up to a fixed integration radius (shown on the $x$-axis), where each circle represents an integration step.
Both panels consider the IR-emitting region.
The different colors represent the different C/O ratios.
The horizontal dashed orange and green lines represent the number of molecule ratios from the results reported by \cite{tabone_rich_2023} and \cite{arabhavi_abundant_2024}, respectively.
}
\label{fig:ColDensProf_ratio-Nmol_ratio-vs_int-radius}
\end{figure*}

Figure~\ref{fig:ColDensProf_ratio-Nmol_ratio-vs_int-radius} presents the \ce{C2H2}/\ce{CO2} ratio as function of radius (top panel) and the $\mathcal{N}$ ratio  as a function of radius, when the integration radius varies (bottom panel).

%%%%%%%%%%%%%%%%%%%%%%%%%%%%%%%%%%%%%%%%%%%%%%%%%%%%%%%%%%%%%%%%%%%%%%

\section{Long Carbon Chains in the IR-emitting Region}\label{appendix:PieChart_LCC}

\begin{figure*}[ht]
\centering
\includegraphics[width=0.76\linewidth]{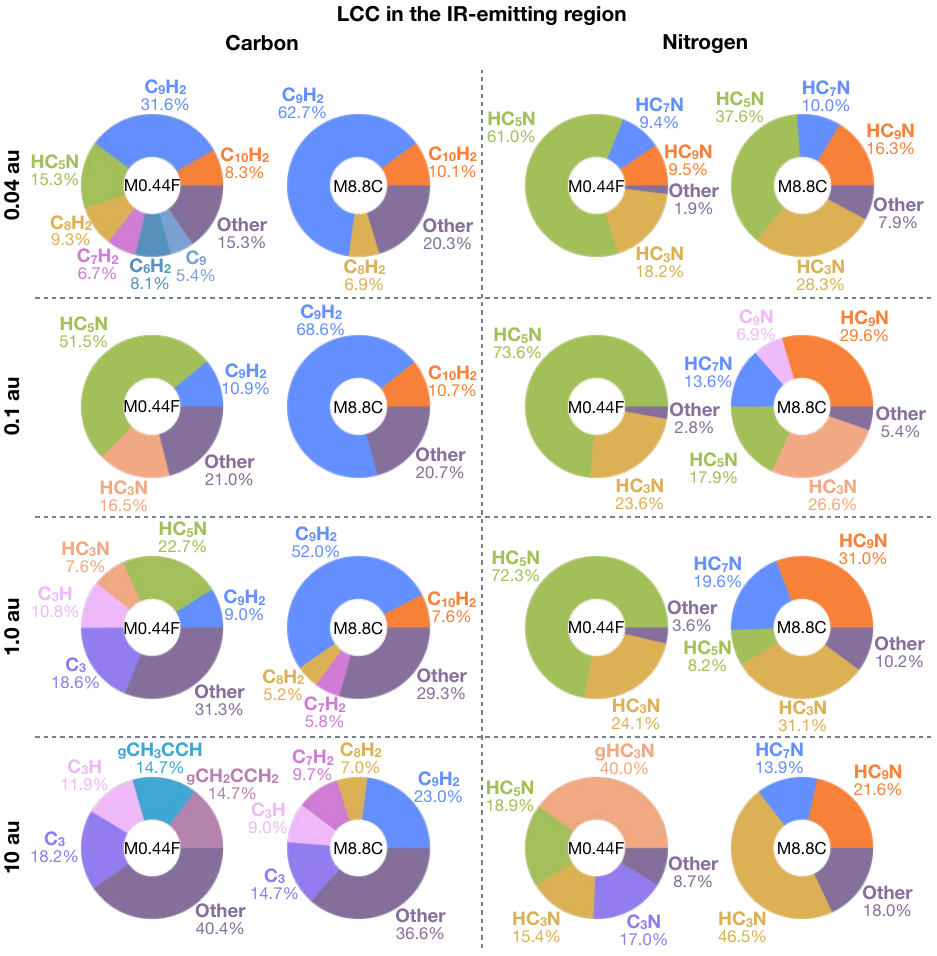}
\caption{
Pie charts showing the contribution of LCCs by molecular mass to the LCC category in the carbon (left) and nitrogen (right) budgets for the same models (M0.44F and M8.8C) and integrations radii (0.04, 0.1, 1.0, and 10~au) as those in Fig.~\ref{fig:PieChart_IRemitting_massweighted}.
}
\label{fig:PieChart_LCC}
\end{figure*}

\begin{figure*}[ht]
\centering
\includegraphics[width=\linewidth]{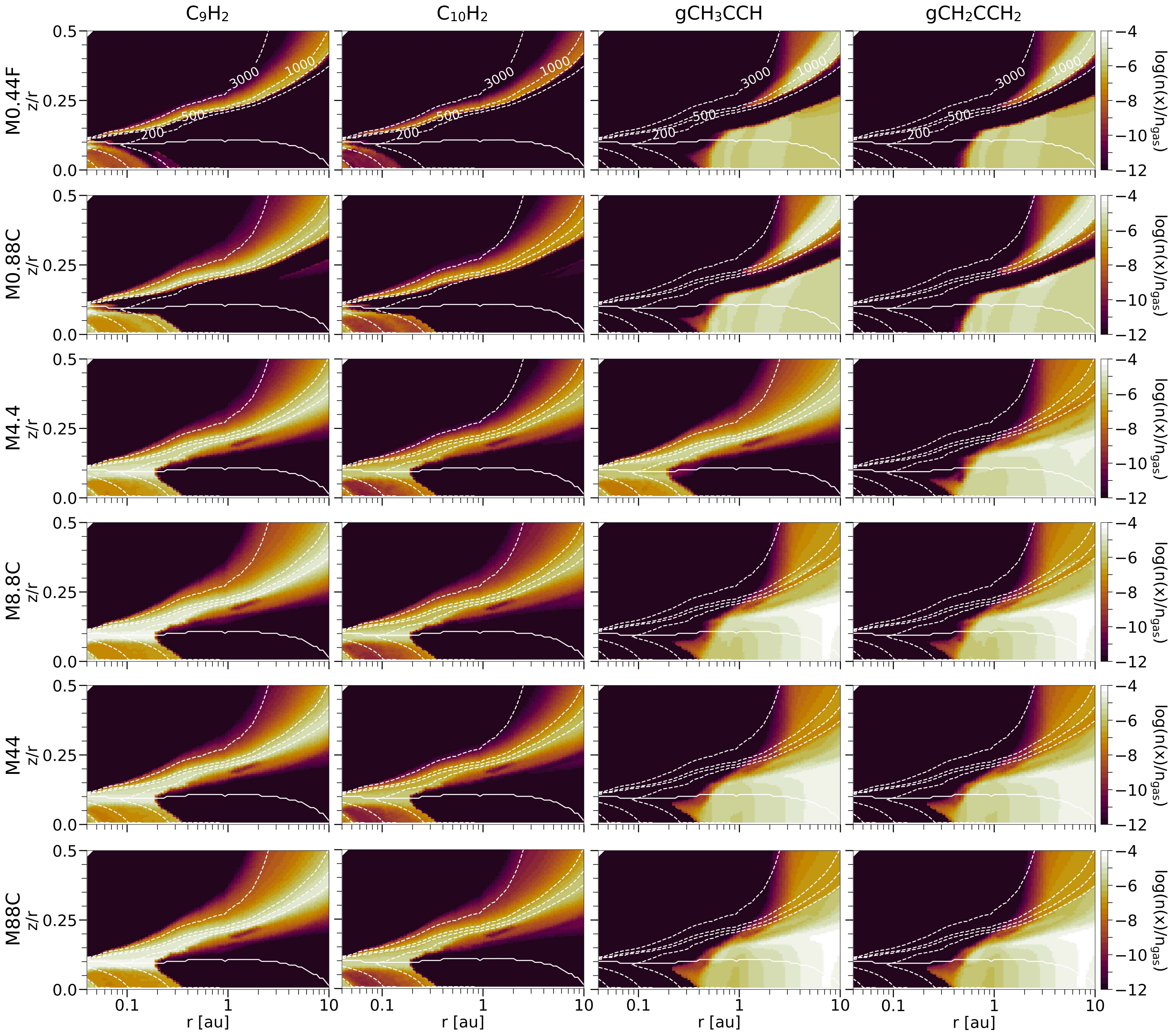}
\caption{
Same as Fig.~\ref{fig:FAmap_hydrocarbon_set1}, but for the LCCs \ce{C9H2}, \ce{C10H2}, \ce{CH3CCH} ice, and \ce{CH2CCH2} ice.
}
\label{fig:FAmap_LCCs_set1}
\end{figure*}

\begin{figure*}[ht]
\centering
\includegraphics[width=\linewidth]{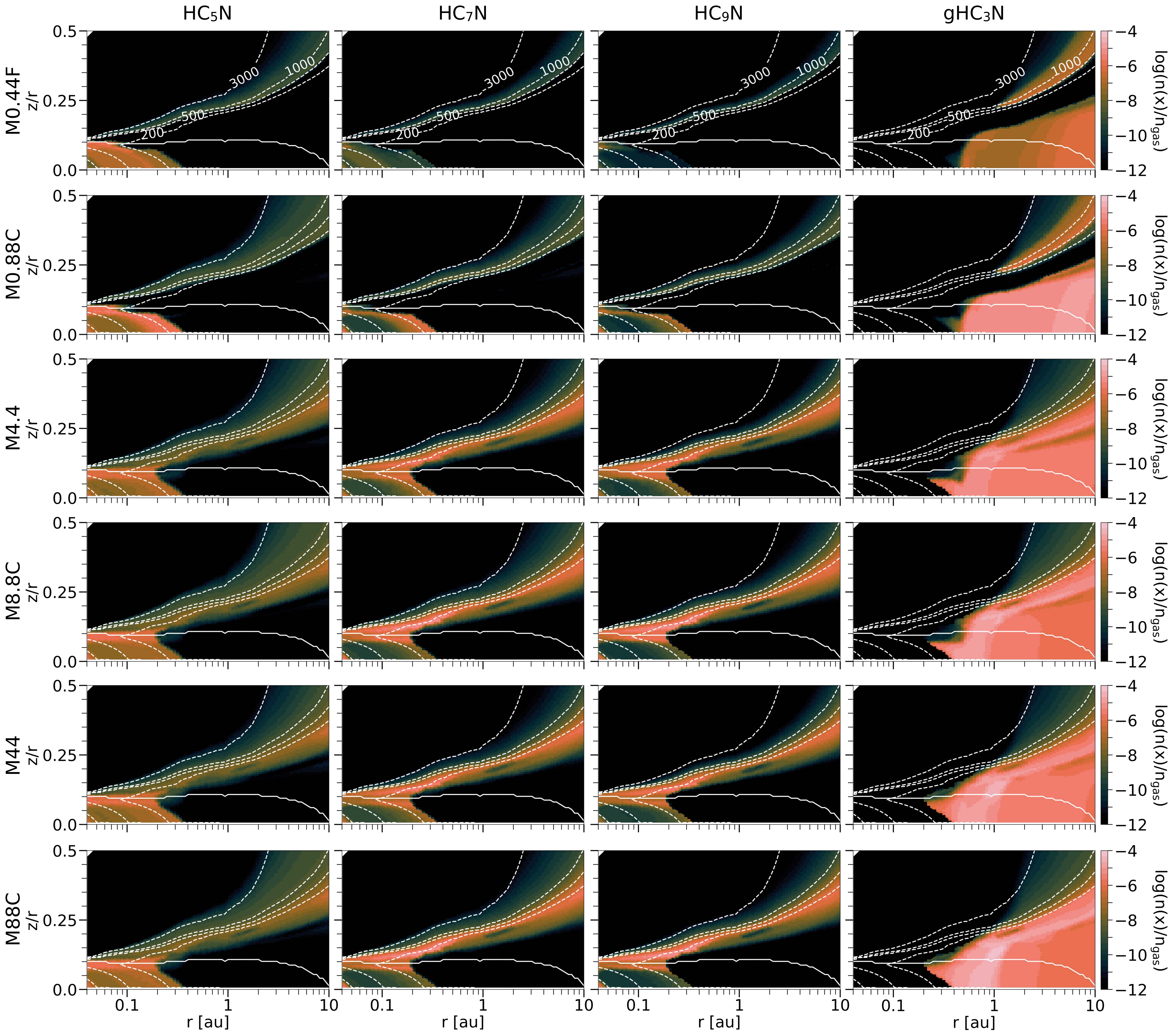}
\caption{
Same as Fig.~\ref{fig:FAmap_hydrocarbon_set1}, but for the LCCs \ce{HC5N}, \ce{HC7N}, \ce{HC9N}, and \ce{HC3N} ice.
}
\label{fig:FAmap_LCCs_set2}
\end{figure*}

Figure~\ref{fig:PieChart_LCC} presents pie charts showing the percentage contribution of long-carbon chains (LCC) to the total LCCs reservoir in the carbon budget for the IR-emitting region.
Figures~\ref{fig:FAmap_LCCs_set1} and \ref{fig:FAmap_LCCs_set2} show the fractional abundance maps for the set of LCCs with higher abundance in this category (\ce{C9H2}, \ce{C10H2}, \ce{CH3CCH} ice, \ce{CH2CCH2} ice, \ce{HC5N}, \ce{HC7N}, \ce{HC9N}, and \ce{HC3N} ice).

%%%%%%%%%%%%%%%%%%%%%%%%%%%%%%%%%%%%%%%%%%%%%%%%%%%%%%%%%%%%%%%%%%%%%%

\section{Formation Pathways for \ce{C6H6} and \ce{C4H2}}

\begin{figure*}[ht]
\centering
\includegraphics[width=1\linewidth]{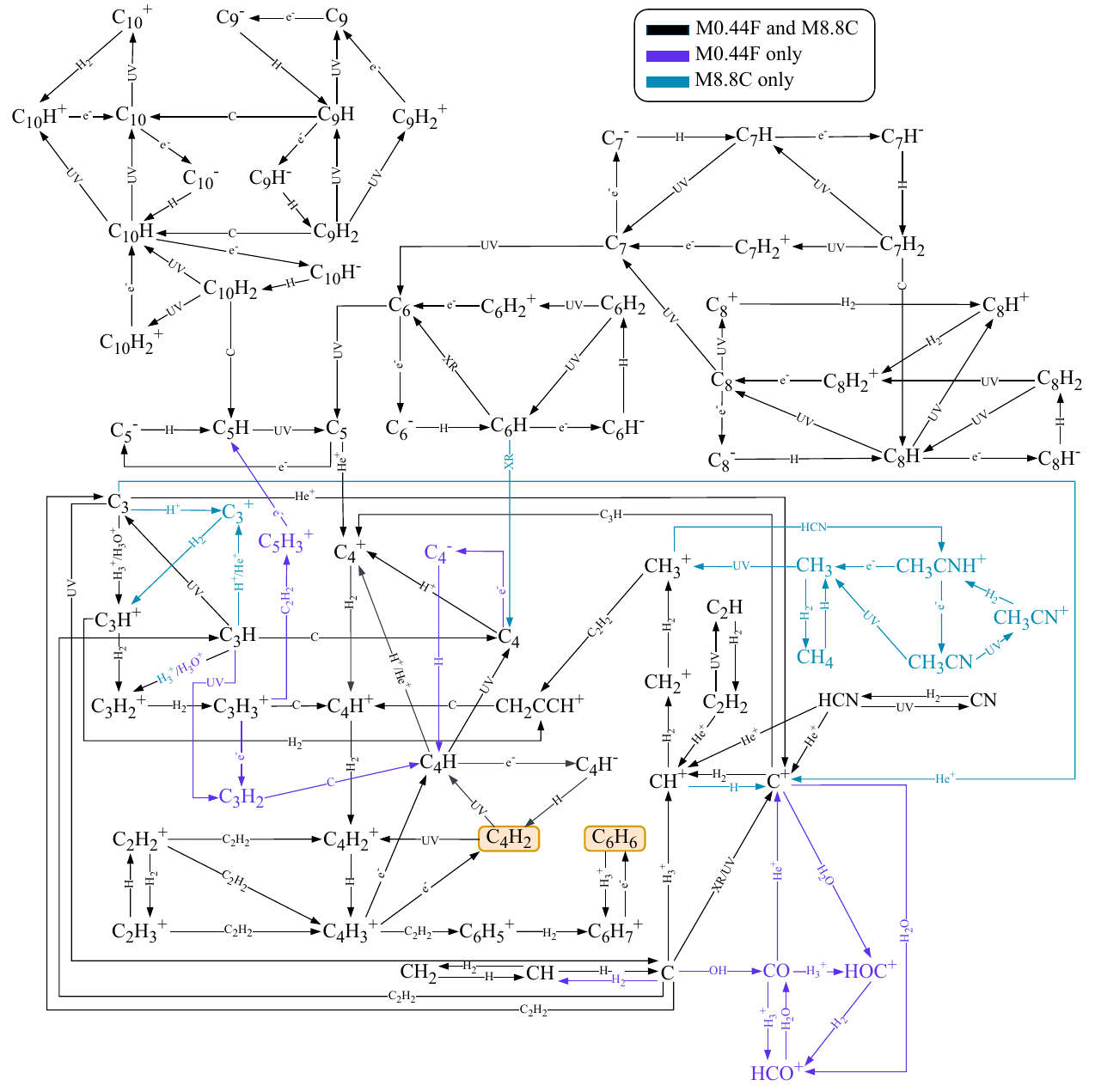}
\caption{Main gas-phase formation pathways for \ce{C6H6} and \ce{C4H2} at $r=1~\mathrm{au}$ and $z/r=0.23$ in the oxygen-rich and carbon-rich models, M0.44F and M8.8C, respectively. 
Purple and turquoise lines represent those reactions only dominant in the M0.44F and M8.8C models, respectively. Black lines represent the main formation reactions for both models.
}
\label{fig:chemnetwork_C6H6_C4H2_M0.44F_M8.8C}
\end{figure*}

Figure~\ref{fig:chemnetwork_C6H6_C4H2_M0.44F_M8.8C} illustrates the dominant gas-phase reactions controlling the formation of \ce{C6H6} and \ce{C4H2} at $r=1~\mathrm{au}$ and $z/r=0.23$ in the models M0.44F and M8.8C.

\end{document}